\documentclass[preprint,amssymb,aps]{revtex4-1}%
\usepackage{amsfonts}
\usepackage{amsmath}
\usepackage{amssymb}
\usepackage{graphicx}%
\providecommand{\U}[1]{\protect\rule{.1in}{.1in}}
\begin{document}

\title{Groundstate and Collective Modes of a Spin-Polarized Dipolar Bose-Einstein
Condensate in a Harmonic Trap.}
\author{I. Sapina, T. Dahm and N. Schopohl \footnote{corresponding author:
nils.schopohl@uni-tuebingen.de}\\Institut f\"{u}r Theoretische Physik and Center for Collective Quantum
Phenomena, Universit\"{a}t T\"{u}bingen, Auf der Morgenstelle 14, D-72076
T\"{u}bingen, Germany}
\date{}

\begin{abstract}
We report new results for the Thomas-Fermi groundstate of a spin- polarized
dipolar interacting Bose-Einstein condensate for the case when the external
magnetic field $\mathbf{B}$ is not orientated parallel to a principal axis,
but is aligned parallel to a symmetry plane of a harmonic anisotropic trap.
For a dipole interaction strength parameter $\varepsilon_{D\ }\neq0\ $the
release energy of the condensate depends on the trap orientation angle
$\vartheta_{T}$ between the principal axis $\mathbf{e}_{z,T}$ of the trap and
the field $\mathbf{B}$. From the quasiclassical Josephson equation of
macroscopic quantum physics we determine the low-lying eigenfrequencies of
small amplitude collective modes of the condensate density for various trap
frequencies $\omega_{a}$ and trap orientation angles $\vartheta_{T}$. For the
special case of a \emph{spherical} harmonic trap with trap frequency $\omega$
it is rigorously shown for $-\frac{1}{2}<\varepsilon_{D\ }<1$, that a pure
$s$-wave symmetry breather excitation of the condensate density exists, that
oscillates at a constant frequency $\Omega_{s}=\sqrt{5}\omega$ around the
groundstate cloud, despite the well known fact, that the shape of the
groundstate cloud of a spin-polarized dipolar condensate is for $\varepsilon
_{D\ }\not =0$ not isotropic. For $\vartheta_{T}\neq0$ the small amplitude
modes of the particle density with isotropic and quadrupolar symmetry consist
of two groups. There exist four modes that are combinations of basis functions
with $s$-wave, $d_{x^{2}-y^{2}}$ - and $d_{z^{2}\ }$-wave, and also $d_{xz}%
$-wave symmetry, and two modes that are combinations of basis functions with
$d_{yz}$- and $d_{xy}$-wave symmetry. A characteristic difference in the
dependence of the frequencies of these six collective modes on the dipole
interaction strength parameter $\varepsilon_{D\ }$ for prolate and oblate
harmonic tri-axial traps, respectively, is suggested to be used as an
experimental method to measure the $s$-wave scattering length $a_{s}$ of the atoms.

\end{abstract}
\maketitle

\affiliation{Institut f\"ur Theoretische Physik and Center for Collective Quantum Phenomena, Universit\"at T\"ubingen, Auf der Morgenstelle 14, D-72076 T\"ubingen, Germany}

\affiliation{Institut f\"{u}r Theoretische Physik , Center for Collective Quantum Phenomena
and their Applications, Eberhard Karls-Universit\"{a}t T\"{u}bingen, Auf der
Morgenstelle 14, 72076 T\"{u}bingen, Germany}

\section{Introduction}

\bigskip Experiments with trapped, extremely dilute gas clouds, consisting of
identical atoms with mass $m^{\star}$, and forming at ultracold temperatures a
quantum degenerate Bose-Einstein condensate (BEC), are nowadays a research
focus in many laboratories. Early on it has been realized that cold atom
clouds do not form an ideal Bose gas, but experience in the low energy sector
of the system \emph{isotropic} interaction forces that can be well described
by a microscopic $s$-wave scattering length $a_{s}$ \cite{Pethick and Smith}.
While the size of an ideal Bose gas confined inside a \emph{harmonic} trap
with trap frequency $\omega$ is determined by the width $a_{\omega}%
=\sqrt{\frac{\hbar}{m^{\star}\omega}}$ of the groundstate wavefunction of a
single particle, the size of an \emph{interacting} cold atom cloud consisting
of a large number $N>>1$ of condensed Bose atoms may increase to much larger
distances $\Lambda_{TF}=a_{\omega}\left(  \frac{4\pi Na_{s}}{a_{\omega}%
}\right)  ^{\frac{1}{5}}$. Fortunately, the necessary requirement $\frac{4\pi
Na_{s}}{a_{\omega}}\gg1$ for observing a BEC in a harmonic trap can be
realized simultaneously with the condition of a small diluteness parameter
$n_{0}a_{s}^{3}\ll1$, so that the mean field theory of Ginzburg and Pitaevskii
for interacting Bose systems is applicable for a wide range of parameters
$a_{\omega}$ and $a_{s}$. As the length $\Lambda_{TF}$ increases with
increasing $N$ the kinetic energy $E_{K}\simeq\frac{\hbar^{2}}{2m^{\star
}\Lambda_{TF}^{2}}$ of the interacting particles in the groundstate eventually
becomes much smaller than the potential energy $V_{T}$ $\simeq\frac{m^{\star}%
}{2}\omega^{2}\Lambda_{TF}^{2}$ of the particles, because the density inside
the BEC becomes a smooth and slowly varying function of position. In the
Thomas-Fermi approximation the kinetic energy term for the particles in the
groundstate of the BEC is neglected altogether. This is justified when the
chemical potential $\mu$ of the interacting system is much larger than the
chemical potential $\sim\frac{3}{2}\hbar\omega$ of the non interacting Bose
gas. So for $\frac{4\pi Na_{s}}{a_{\omega}}\gg1$ the dominant balance required
for mechanical equilibrium of a trapped BEC is between the repulsive
interactions of the atoms and the confinement forces of the trap.

New interesting physics can be observed when in addition to the usual $s$-wave
contact interaction the atoms get influenced by long ranged dipole-dipole
forces \cite{Santos I}. This occurs, for example, for Bose atoms with nuclear
spin $I=0$ and integer (electronic) spin $S$, thus giving rise to a multiplet
$-S\leq M_{S}\leq S$ of atomic magnetic dipole moments with $z$-component
$2\mu_{B}M_{S}$. A transition metal atom like chromium $^{52}Cr$ has $I=0$ and
$S=3$. On the other hand, Alkali atoms like $^{87}Rb$ carry (nuclear) spin
$I=\frac{3}{2}$ and $S=\frac{1}{2}$, thus coupling to a total spin $F=1$ in
the lowest energy state. The first experimental study of magnetic
dipole-dipole interactions in a BEC was realized with $^{52}Cr$-atoms
\cite{Lahaye et al} carrying a large magnetic moment $\left\vert \left\langle
\mathbf{M}\right\rangle \right\vert =6\mu_{B}$. A quantum degenerate $F=1$
spinor BEC was synthezised successfully with $^{87}Rb$-atoms\cite{Vengalattore
I}. Recently, intrinsically anisotropic BEC systems with electric
dipole-dipole interactions between polar molecules have been studied
experimentally \cite{Ospelkaus}. New research directions are concerned with
magnetic quantum gases consisting of heavy rare earth atoms like
Thulium\cite{Thulium} with $\left\vert \left\langle \mathbf{M}\right\rangle
\right\vert =4\mu_{B}$ , Erbium\cite{Erbium} with $\left\vert \left\langle
\mathbf{M}\right\rangle \right\vert =7\mu_{B}$, and
Dysprosium\cite{Dysprosium} with $\left\vert \left\langle \mathbf{M}%
\right\rangle \right\vert =10\mu_{B}$.

In the ensuing considerations we study spin-polarized Bose atom clouds. When
the magnetic dipole moments of the atoms are $100\%$ polarized under a
homogeneous external magnetic induction field $\mathbf{B}=B^{\left(
ext\right)  }\mathbf{e}_{z}$ , so that all atoms in the cold gas cloud carry
the identical effective magnetic moment, say $\left\langle \mathbf{M}%
\right\rangle =-2\mu_{B}S\ \mathbf{e}_{z}$ , it is still possible to describe
the Bose condensed groundstate $\Psi$ of $N$ interacting atoms by a
\emph{scalar} Hartree ansatz
\begin{equation}
\Psi\left(  \mathbf{r}^{\left(  1\right)  },...,\mathbf{r}^{\left(  N\right)
}\right)  =\psi(\mathbf{r}^{\left(  1\right)  })\cdot\psi(\mathbf{r}^{\left(
2\right)  })\cdot\cdot\cdot\psi(\mathbf{r}^{\left(  N\right)  })
\label{Hartree Groundstate}%
\end{equation}
The expectation value of the many body Hamiltonian $H$ , evaluated with such a
trial wave function $\Psi$ consisting of a product of $N$ identical
one-particle wave functions $\psi(\mathbf{r})$, is then minimized with respect
to variations of that one-particle wave function $\psi(\mathbf{r})$. The
\emph{optimal} one-particle wavefunction $\psi(\mathbf{r})$ so found is a
solution to the Gross-Pitaevskii equation \cite{Pethick and Smith}:
\begin{align}
& \label{GP}\\
\left[  -\frac{\hbar^{2}}{2m^{\star}}\nabla_{\mathbf{r}}^{2}+V_{T}%
(\mathbf{r})-\mu+\left(  N-1\right)  \int_{\mathbb{R}^{3}}d^{3}r^{\prime
}U\left(  \mathbf{r},\mathbf{r}^{\prime}\right)  \left\vert \psi\left(
\mathbf{r}^{\prime}\right)  \right\vert ^{2}\right]  \psi(\mathbf{r})  &
=0\nonumber
\end{align}
Here, $V_{T}(\mathbf{r})$ denotes the potential of the trap, and $U\left(
\mathbf{r},\mathbf{r}^{\prime}\right)  $ describes the interaction potential
between two Bosons. The chemical potential $\mu$ is a Lagrange parameter
connected to the particle number $N$ in the condensate by the constraint:
\begin{equation}
\int_{\mathbb{R}^{3}}d^{3}r^{\prime}\ \left\vert \psi\left(  \mathbf{r}%
^{\prime}\right)  \right\vert ^{2}=1 \label{normalisation  I}%
\end{equation}

\section{Thomas-Fermi Theory of Spin-Polarized Dipolar Bose-Einstein
Condensate}

In the following we investigate the macroscopic quantum degenerate groundstate
of a spin-polarized system of interacting Bose atoms carrying a magnetic
dipole moment $\left\vert \left\langle \mathbf{M}\right\rangle \right\vert $.
The interaction potential
\begin{equation}
U\left(  \mathbf{r},\mathbf{r}^{\prime}\right)  =U_{0}\left(  \mathbf{r}%
,\mathbf{r}^{\prime}\right)  +U_{md}\left(  \mathbf{r},\mathbf{r}^{\prime
}\right)  \label{interaction  I}%
\end{equation}
between two atoms, one at position $\mathbf{r}$ and the other at
$\mathbf{r}^{\prime}$ , consists of two contributions, the short ranged
isotropic $s-$wave interaction pseudopotential
\begin{align}
U_{0}\left(  \mathbf{r},\mathbf{r}^{\prime}\right)   &  =\ g_{s}%
\ \delta^{\left(  3\right)  }\left(  \mathbf{r}-\mathbf{r}^{\prime}\right)
\label{interaction  Is}\\
g_{s}  &  =\frac{4\pi\hbar^{2}}{m^{\star}}a_{s}\nonumber
\end{align}
$\ $, and the long ranged \emph{magnetic} dipole-dipole interaction potential:%
\begin{align}
U_{md}\left(  \mathbf{r},\mathbf{r}^{\prime}\right)   &  =\frac{g_{md}}{4\pi
}\left[  \frac{1}{\left\vert \mathbf{r}-\mathbf{r}^{\prime}\right\vert ^{3}%
}-\frac{3\left(  r_{z}-r_{z}^{\prime}\right)  ^{2}}{\left\vert \mathbf{r}%
-\mathbf{r}^{\prime}\right\vert ^{5}}\right] \label{interaction Imd}\\
g_{md}  &  =\mu_{0}\left\vert \left\langle \mathbf{M}\right\rangle \right\vert
^{2}\nonumber
\end{align}
Here the external magnetic induction field $\mathbf{B}$ is orientated parallel
to the Cartesian unit vector $\mathbf{e}_{z}\ $in the laboratory frame, so
that the magnetic moments of two interacting atoms, one at positions
$\mathbf{r}$ and the other at $\mathbf{r}^{\prime}$, are both aligned parallel
to $\mathbf{e}_{z}$.

Using well known identities
\begin{align}
\frac{3\left(  r_{z}-r_{z}^{\prime}\right)  ^{2}}{\left\vert \mathbf{r}%
-\mathbf{r}^{\prime}\right\vert ^{5}}-\frac{1}{\left\vert \mathbf{r}%
-\mathbf{r}^{\prime}\right\vert ^{3}}  &  =\frac{\partial^{2}}{\partial
r_{z}^{2}}\frac{1}{\left\vert \mathbf{r}-\mathbf{r}^{\prime}\right\vert
}-\frac{1}{3}\cdot\nabla_{\mathbf{r}}^{2}\frac{1}{\left\vert \mathbf{r}%
-\mathbf{r}^{\prime}\right\vert }\label{identity I}\\
& \nonumber\\
-\nabla_{\mathbf{r}}^{2}\frac{1}{\left\vert \mathbf{r}-\mathbf{r}^{\prime
}\right\vert }  &  =4\pi\delta^{\left(  3\right)  }\left(  \mathbf{r}%
-\mathbf{r}^{\prime}\right) \nonumber
\end{align}
, and introducing the dimensionless parameter \cite{O'Dell}, \cite{Griesmaier
et al}
\begin{equation}
\varepsilon_{D}=\frac{g_{md}}{3g_{s}}=\frac{\mu_{0}\left\vert \left\langle
\mathbf{M}\right\rangle \right\vert ^{2}}{\frac{12\pi\hbar^{2}}{m^{\star}%
}a_{s}} \label{dipole strength}%
\end{equation}
as a measure of relative strength of magnetic dipole interaction forces, the
interaction potential between two atoms in the gas cloud may be rewritten in
the guise:%
\begin{equation}
U\left(  \mathbf{r},\mathbf{r}^{\prime}\right)  =g_{s}\left[  \left(
1-\varepsilon_{D}\right)  \delta^{\left(  3\right)  }\left(  \mathbf{r}%
-\mathbf{r}^{\prime}\right)  \ -3\varepsilon_{D}\frac{\partial^{2}}{\partial
r_{z}^{2}}\frac{1}{4\pi}\ \frac{1}{\left\vert \mathbf{r}-\mathbf{r}^{\prime
}\right\vert }\right]  \label{interaction  II}%
\end{equation}

In the Thomas-Fermi approximation the particle density profile in the
groundstate of the trapped BEC
\begin{equation}
n\left(  \mathbf{r}\right)  =\left\vert \sqrt{N}\psi\left(  \mathbf{r}\right)
\right\vert ^{2} \label{density}%
\end{equation}
is a solution to the integral equation
\begin{equation}
\left(  1-\varepsilon_{D}\right)  n_{TF}\left(  \mathbf{r}\right)
-3\varepsilon_{D}\frac{\partial^{2}}{\partial r_{z}^{2}}\ \frac{1}{4\pi}%
\int_{\mathbb{D}_{TF}}d^{3}r^{\prime}\frac{1}{\left\vert \mathbf{r}%
-\mathbf{r}^{\prime}\right\vert }n_{TF}\left(  \mathbf{r}^{\prime}\right)
=\frac{\mu-V_{T}\left(  \mathbf{r}\right)  }{g_{s}}
\label{TF integral equation I}%
\end{equation}
This is actually a non linear problem, because the solution of the integral
equation is sought inside the Thomas-Fermi cloud%
\begin{equation}
\mathbb{D}_{TF}=\left\{  \ \mathbf{r}\in\mathbb{R}^{3}|\ n_{TF}(\mathbf{r}%
)\geq0\right\}  \label{TF domain}%
\end{equation}
, which region is not known a priori. The determination of the shape of the
cloud $\mathbb{D}_{TF}$ , or its boundary $\partial\mathbb{D}_{TF}$ , is part
of the problem.

Eberlein et al. \cite{O'Dell} have found for a dipolar interacting BEC
confined inside a \emph{harmonic} trap, that despite the non local anisotropic
dipole-dipole interaction term, the domain $\mathbb{D}_{TF}$ always maintains
the shape of an ellipsoid, as in the case $\varepsilon_{D}=0$, but with
different semi axes. We confirm this finding and present new results for the
case, when the external magnetic field $\mathbf{B}$ is not in alignment with
the principal axis $\mathbf{e}_{z,T}$ of the trap.

Consider a harmonic \emph{anisotropic} trap potential $V_{T}(\mathbf{r})$ with
its minimum at position $\mathbf{r}=\mathbf{0}$ , and with the principal axis
$\mathbf{e}_{z,T}$ of the trap not in alignment with the field $\mathbf{B}$ :%
\begin{equation}
V_{T}(\mathbf{r})=\frac{m^{\star}}{2}\left(  \omega_{x}^{2}r_{x,T}^{2}%
+\omega_{y}^{2}r_{y,T}^{2}+\omega_{z}^{2}r_{z,T}^{2}\right)
\label{harmonic anisotropic trap}%
\end{equation}
For $\omega_{x}\neq\omega_{y}$ , $\omega_{y}\neq\omega_{z}$ and $\omega
_{z}\neq\omega_{x}\ $surfaces of constant trap potential $V_{T}(\mathbf{r}%
)=V_{T}>0$ have the geometrical shape of a tri-axial ellipsoid. Three mutually
orthogonal Cartesian unit vectors $\mathbf{e}_{x,T}\ ,\mathbf{e}_{y,T}\ $and
$\mathbf{e}_{z,T}$ determine the orientation of the principal axes of such a
trap. The magnetic field $\mathbf{B}$ is then in general a linear combination
of \emph{all }three principal axis vectors: $\mathbf{B=}B_{x,T}\mathbf{e}%
_{x,T}+$ $B_{y,T}\mathbf{e}_{y,T}+B_{z,T}\mathbf{e}_{z,T}$. For simplicity we
restrict our considerations in the following to the special case, when the
magnetic field $\mathbf{B}$ and the principal axis $\mathbf{e}_{z,T}$ of the
trap span a symmetry plane of the trap, say the plane $r_{y}=0$. Then we have
in (\ref{harmonic anisotropic trap}):
\begin{align}
& \label{trap coordinates I}\\
r_{x,T}  &  =r_{x}\left(  \vartheta_{T}\right)  =\cos\left(  \vartheta
_{T}\right)  r_{x}+\sin\left(  \vartheta_{T}\right)  r_{z}\nonumber\\
r_{y,T}  &  =r_{y}\left(  \vartheta_{T}\right)  =r_{y}\nonumber\\
r_{z,T}  &  =r_{z}\left(  \vartheta_{T}\right)  =-\sin\left(  \vartheta
_{T}\right)  r_{x}+\cos\left(  \vartheta_{T}\right)  r_{z}\nonumber
\end{align}
, i.e. the principal axis $\mathbf{e}_{z,T}$ of the trap is turned by an angle
$\vartheta_{T}$ around the rotation axes $\mathbf{e}_{y,T}$ $\perp\mathbf{B}$
(see Fig.1). \begin{figure}[h]
\centering
\includegraphics[width=0.80\textwidth]{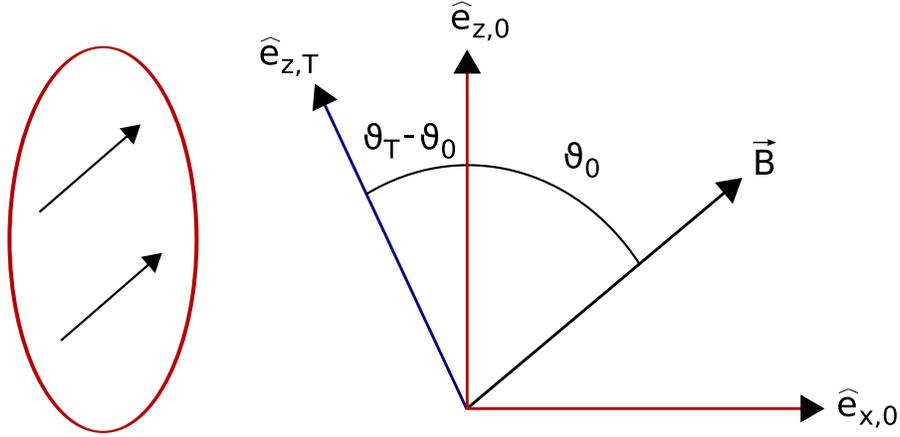}\caption{(Color online) Orientation
of principal axis $\mathbf{e}_{z,T}$ of harmonic trap and orientation of
principal axis $\mathbf{e}_{z,0}$ of Thomas-Fermi ellipsoid $\mathbb{D}_{TF}$
relative to the spin polarizing external magnetic field $\mathbf{B}$. The
inset on the left hand side corresponds to a cut of $\mathbb{D}_{TF}$ with the
symmetry plane $y=0$. Arrows indicate the orientation of the spin polarizing
magnetic field $\mathbf{B}$ relative to the principal axis $\mathbf{e}_{z,0}$
of $\mathbb{D}_{TF}$.}%
\label{Fig. 1}%
\end{figure}

As is indicated in Fig.1, the selfconsistent solution of
(\ref{TF integral equation I}) for the density distribution $n_{TF}\left(
\mathbf{r}\right)  $ reveals, that the principal axis $\mathbf{e}_{z,0}$ of
the Thomas-Fermi cloud $\mathbb{D}_{TF}$ is rotated away from the direction of
the external field by an angle $\vartheta_{0}\neq$ $\vartheta_{T}$.
Accordingly, the density profile associated with the ellipsoidal domain
$\mathbb{D}_{TF}$ has the general form
\begin{equation}
n_{TF}(\mathbf{r})=n_{0}\left(  1-\frac{\widetilde{r}_{x}^{2}}{\lambda_{x}%
^{2}}-\frac{\widetilde{r}_{y}^{2}}{\lambda_{y}^{2}}-\frac{\widetilde{r}%
_{z}^{2}}{\lambda_{z}^{2}}\right)  \label{ansatz density profile}%
\end{equation}
where%
\begin{align}
& \label{coordinates rotated trap}\\
r_{x}\left(  \vartheta_{0}\right)   &  =\widetilde{r}_{x}=\cos\left(
\vartheta_{0}\right)  r_{x}+\sin\left(  \vartheta_{0}\right)  r_{z}\nonumber\\
r_{y}\left(  \vartheta_{0}\right)   &  =\widetilde{r}_{y}=r_{y}\nonumber\\
r_{z}\left(  \vartheta_{0}\right)   &  =\widetilde{r}_{z}=-\sin\left(
\vartheta_{0}\right)  r_{x}+\cos\left(  \vartheta_{0}\right)  r_{z}\nonumber
\end{align}
Only in the highly symmetric case $\vartheta_{T}=0$ the principal axis vector
$\mathbf{e}_{z,0}$ of the ellipsoid $\mathbb{D}_{TF}$ is orientated parallel
to $\mathbf{B}$. For $0<\vartheta_{T}<\frac{\pi}{2}$ it is found from the
selfconsistent solution for the density profile $n_{TF}(\mathbf{r})$, that
$\vartheta_{0}\neq\vartheta_{T}$ , i.e. the principal-axis $\mathbf{e}_{z,0}$
of $\mathbb{D}_{TF}$ is never in alignment with the field $\mathbf{B}$, nor is
it in alignment with the principal axis $\mathbf{e}_{z,T}$ of the trap (see
Fig.1 and Fig.2).

\begin{figure}[h]
\centering
\includegraphics[width=0.80\textwidth]{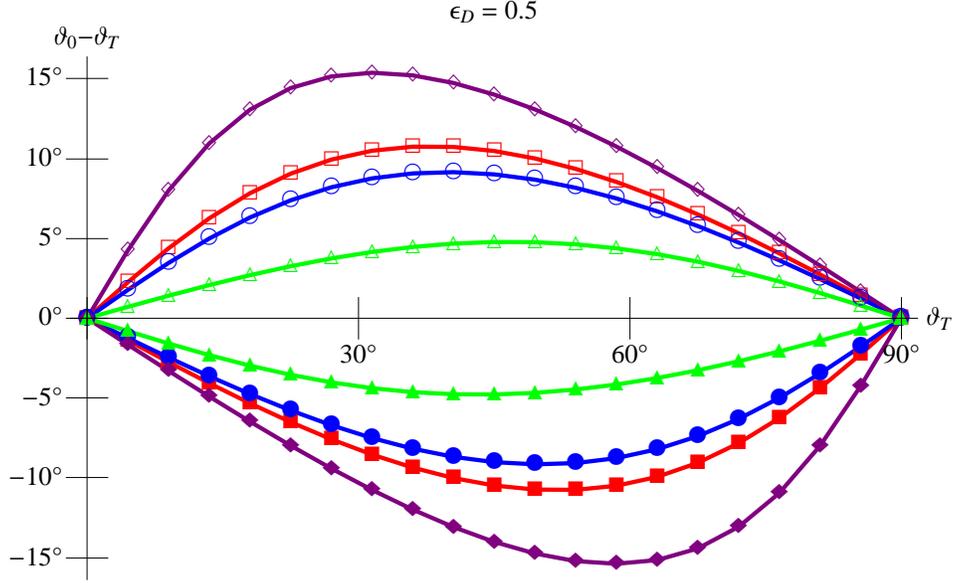} \label{fig:2}%
\caption{(Color online) Angle difference $\vartheta_{0}-\vartheta_{T}$ vs. trap orientation
angle $\vartheta_{T}$ for selfconsistent groundstate density profile
$n_{TF}\left(  \mathbf{r}\right)  $. The curves shown correspond to a dipole
interaction strength $\varepsilon_{D}=0.5$ and to various frequency ratios
$\omega_{x}:\omega_{y}:\omega_{z}$ of the harmonic trap: a) $2:6:3$ purple
empty diamond ; aa) $3:6:2$ purple full diamond; b) $3:2:6$ red empty square;
bb) $6:2:3$ red full square; c) $1:2:2$ blue empty circle; cc) $2:2:1$ blue
full circle; d) $2:3:6$ green empty triangle; dd) $6:3:2$ green full
triangle.}%
\label{Fig. 2}%
\end{figure}

The particle density distribution $n_{TF}(\mathbf{r})$ inside the Thomas-Fermi
ellipsoid $\mathbb{D}_{TF}$ is stratified. Like an onion it consists of a
series of thin homoeoidal shells of constant density
\begin{align}
n_{TF}(\mathbf{r})  &  =n_{0}\left(  1-\nu^{2}\right)
=const\label{strata density I}\\
0  &  \leq\nu\leq1\nonumber
\end{align}
Strata of equal density thus correspond to ellipsoidal shells concentric and
similar to the bounding ellipsoidal shell $\partial\mathbb{D}_{TF}$ , but with
scaled semi-axes $\nu\lambda_{a}$.

For a general tri-axial ellipsoid $\mathbb{D}_{TF}$ the normalization
integral
\begin{equation}
\int_{\mathbb{D}_{TF}}d^{3}r^{\prime}n_{TF}\left(  \mathbf{r}^{\prime}\right)
=N \label{normalisation  Ib}%
\end{equation}
leads to%
\begin{equation}
N=\frac{8\pi}{15}\lambda_{x}\lambda_{y}\lambda_{z}n_{0}
\label{normalisation  Ia}%
\end{equation}
So, the problem is to determine from (\ref{TF integral equation I}) the
three semi axes $\lambda_{x}$ , $\lambda_{y}$ , $\lambda_{z}$ , the
orientational angle $\vartheta_{0}$ and the chemical potential $\mu$.

As has been emphasized by Eberlein et al. in \cite{O'Dell} the central task in
solving the Thomas-Fermi integral equation (\ref{TF integral equation I})
is to calculate the potential function
\begin{equation}
\phi_{TF}\left(  \mathbf{r}\right)  =\frac{1}{4\pi}\int_{\mathbb{D}_{TF}}%
d^{3}r^{\prime}\frac{1}{\left\vert \mathbf{r}-\mathbf{r}^{\prime}\right\vert
}n_{TF}\left(  \mathbf{r}^{\prime}\right)  \label{potential function}%
\end{equation}
for a \emph{heterogeneous} particle density distribution $n_{TF}\left(
\mathbf{r}\right)  $. Then, because%
\begin{equation}
-\nabla^{2}\phi_{TF}\left(  \mathbf{r}\right)  =n_{TF}\left(  \mathbf{r}%
\right)  \label{Poisson  I}%
\end{equation}
, the Thomas-Fermi integral equation (\ref{TF integral equation I})
\begin{equation}
\left(  1-\varepsilon_{D}\right)  n_{TF}\left(  \mathbf{r}\right)
-3\varepsilon_{D}\frac{\partial^{2}}{\partial r_{z}^{2}}\phi_{TF}\left(
\mathbf{r}\right)  =\frac{\mu-V_{T}\left(  \mathbf{r}\right)  }{g_{s}}
\label{TF integral equation II}%
\end{equation}
becomes (in free space) equivalent to a partial differential equation of
potential theory:
\begin{align}
& \label{TF integral equation III}\\
-\left[  \left(  1-\varepsilon_{D}\right)  \left(  \frac{\partial^{2}%
}{\partial r_{x}^{2}}+\frac{\partial^{2}}{\partial r_{y}^{2}}\right)  +\left(
1+2\varepsilon_{D}\right)  \frac{\partial^{2}}{\partial r_{z}^{2}}\right]
\phi_{TF}\left(  \mathbf{r}\right)   &  =\frac{\mu-V_{T}(\mathbf{r})}{g_{s}%
}\nonumber
\end{align}
In order that the differential operator on the left hand side is positive
definite it is required that
\begin{equation}
-\frac{1}{2}<\varepsilon_{D}<1 \label{stability  I}%
\end{equation}
Due to the $d_{z^{2}}$-anisotropy of the dipole-dipole interaction between two
\emph{spin-polarized} atoms at position $\mathbf{r}$ and $\mathbf{r}^{\prime}$
the dipole-dipole interaction part of the potential $U\left(  \mathbf{r}%
,\mathbf{r}^{\prime}\right)  $ is attractive \textbf{or} repulsive, depending
on the orientation of the distance vector $\mathbf{r}$ $-$ $\mathbf{r}%
^{\prime}$ relative to the field vector $\mathbf{B}$. So, if $\varepsilon
_{D}>1$ or $\varepsilon_{D}<-\frac{1}{2}$ , attractive forces prevail and the
system will collapse. Of course, a better criterion for stability is to
calculate the frequencies of the collective modes of the dipolar interacting
BEC. We shall present in the next section new results for quadrupolar like
modes of the density fluctuations around the groundstate density profile
$n_{TF}\left(  \mathbf{r}\right)  $.

The differential equation (\ref{TF integral equation III}) makes it
manifest, that the integral operator with integration domain $\mathbb{D}_{TF}$
and kernel $\frac{1}{\left\vert \mathbf{r}-\mathbf{r}^{\prime}\right\vert }$
in the Thomas-Fermi integral equation (\ref{TF integral equation II} ) for
$\mathbf{r\in}\mathbb{D}_{TF}$ maps a \emph{quadratic} form $n_{TF}\left(
\mathbf{r}\right)  $ spanned by the linearly independent basis functions
$\left\{  1,r_{a}r_{b}\right\}  _{a\leq b\leq\in\left\{  x,y,z\right\}  }$
into a \emph{quartic} form $\phi_{TF}\left(  \mathbf{r}\right)  $ spanned by
linearly independent basis functions $\left\{  1,r_{a}r_{b},r_{a}r_{b}%
r_{c}r_{d}\right\}  _{a\leq b\leq c\leq d\in\left\{  x,y,z\right\}  }$.
Because for a \emph{harmonic} trap potential $\ V_{T}\left(  \mathbf{r}%
\right)  $ the right hand side of (\ref{TF integral equation III}) is (by
definition) a quadratic form, the problem would be exactly solved, provided
the coefficients of the quadratic form presented by the derivatives
$\frac{\partial^{2}}{\partial r_{a}^{2}}\phi_{TF}\left(  \mathbf{r}\right)  $
of the potential function (\ref{potential function}) can be found.

We describe now a very convenient method to determine the coefficients of the
quadratic form $\frac{\partial^{2}}{\partial r_{z}^{2}}\phi_{TF}\left(
\mathbf{r}\right)  $, which is all we need to solve the Thomas-Fermi integral
equation (\ref{TF integral equation II} ). As a matter of fact,
three-dimensional integrals of the type%
\begin{align}
\Phi_{l}\left(  \mathbf{s}\right)   &  =\frac{1}{4\pi}\int_{\mathbb{D}}%
d^{3}s^{\prime}\frac{1}{\left\vert \mathbf{s}-\mathbf{s}^{\prime}\right\vert
}\left(  1-\frac{s_{x}^{\prime2}}{\lambda_{x}^{2}}-\frac{s_{y}^{\prime2}%
}{\lambda_{y}^{2}}-\frac{s_{z}^{\prime2}}{\lambda_{z}^{2}}\right)
^{l}\label{gravitational potential of special mass distribution  I}\\
l  &  =0,1,2,3,...\nonumber\\
\mathbb{D}  &  =\left\{  \ \mathbf{s}^{\prime}\in\mathbb{R}^{3}|\ \frac
{s_{x}^{\prime2}}{\lambda_{x}^{2}}+\frac{s_{y}^{\prime2}}{\lambda_{y}^{2}%
}+\frac{s_{z}^{\prime2}}{\lambda_{z}^{2}}\leq1\right\} \nonumber
\end{align}
have been calculated analytically by S. Chandrasekhar \cite{Chandrasekhar} in
his magisterial treatment of the ellipsoidal figures of equilibrium of
gravitating and rotating gas clouds in astrophysics. He showed, that $\Phi
_{l}\left(  \mathbf{s}\right)  $ can be represented \emph{exactly }in terms of
singularity free fast convergent \emph{one-dimensional} integrals.
Chandrasekhar's result for the three-dimensional integral $\Phi_{l}\left(
\mathbf{s}\right)  $ at an \emph{internal} point of the ellipsoid $\mathbb{D}$
is:%
\begin{align}
\mathbf{s}  &  \in\mathbb{D}_{TF}%
\label{gravitational potential of special mass distribution II}\\
l  &  =0,1,2,3,...\nonumber\\
\Phi_{l}\left(  \mathbf{s}\right)   &  =\frac{\lambda_{x}\lambda_{y}%
\lambda_{z}}{4}\frac{1}{l+1}\int_{0}^{\infty}\frac{du}{\sqrt{\left[  \left(
\lambda_{x}^{2}+u\right)  \left(  \lambda_{y}^{2}+u\right)  \left(
\lambda_{z}^{2}+u\right)  \right]  }}\left(  1-\frac{s_{x}^{2}}{\lambda
_{x}^{2}+u}-\frac{s_{y}^{2}}{\lambda_{y}^{2}+u}-\frac{s_{z}^{2}}{\lambda
_{z}^{2}+u}\right)  ^{l+1}\nonumber
\end{align}

To solve the Thomas-Fermi integral equation we now make use of this result for
the special case $l=1$. Because the Cartesian coordinates $\widetilde{r}_{b}$
of a point presented in the principal axes frame of the Thomas-Fermi ellipsoid
$\mathbb{D}_{TF}$ are connected to the Cartesian coordinates $r_{a}$ of that
same point in the laboratory frame by a rotation,%
\begin{align}
r_{b}\left(  \vartheta_{0}\right)   &  =\widetilde{r}_{b}=\sum_{a\in\left\{
x,y,z\right\}  }\mathcal{R}_{ba}\left(  \vartheta_{0};\mathbf{e}_{y}\right)
r_{a}\label{principal axes frame I}\\
& \nonumber\\
\mathcal{R}_{ba}\left(  \vartheta_{0};\mathbf{e}_{y}\right)   &  =\left[
\begin{array}
[c]{ccc}%
\cos\left(  \vartheta_{0}\right)  & 0 & \sin\left(  \vartheta_{0}\right) \\
0 & 1 & 0\\
-\sin\left(  \vartheta_{0}\right)  & 0 & \cos\left(  \vartheta_{0}\right)
\end{array}
\right]  _{ba}\nonumber
\end{align}
, and taking into account that under such a rotation $\mathcal{R}\left(
\vartheta_{0};\mathbf{e}_{y}\right)  $ we have $\left\vert \mathbf{r}%
-\mathbf{r}^{\prime}\right\vert =\left\vert \widetilde{\mathbf{r}}%
-\widetilde{\mathbf{r}}^{\prime}\right\vert $, we immediately see that the
potential function
\begin{equation}
\phi_{TF}\left(  \mathbf{r}\right)  =n_{0}\Phi_{1}\left[  \widetilde
{\mathbf{r}}\left(  \mathbf{r}\right)  \right]  \label{potential function II}%
\end{equation}
at a position $\mathbf{r}$ $\in\mathbb{D}_{TF}$ is a \emph{quartic} form with
regard to the linearly independent basis functions $\left\{  1,r_{a}%
r_{b},r_{a}r_{b}r_{c}r_{d}\right\}  _{a\leq b\leq c\leq d\in\left\{
x,y,z\right\}  }$. As a second order derivative of a quartic the function
$\frac{\partial^{2}}{\partial r_{z}^{2}}\phi_{TF}\left(  \mathbf{r}\right)  $
is then manifestly a quadratic form, spanned by a linear combination of the
basis functions $\left\{  1,\widetilde{r}_{x}^{2},\widetilde{r}_{y}%
^{2},\widetilde{r}_{z}^{2},\widetilde{r}_{x}\widetilde{r}_{z}\right\}  $ , or
taking into account (\ref{principal axes frame I}), it is spanned by a
linear combination of the basis functions $\left\{  1,r_{x}^{2},r_{y}%
^{2},r_{z}^{2},r_{x}r_{z}\right\}  $. The transformation from one basis system
to the other is accomplished by the orthogonal transformation
(\ref{principal axes frame I}).

The coefficients $c_{ab}$ of the quadratic form
\begin{align}
\mathbf{r}  &  \mathbf{\in}\mathbb{D}_{TF}%
\label{2nd order derivative TF potential}\\
\frac{\partial^{2}}{\partial r_{z}^{2}}\phi_{TF}\left(  \mathbf{r}\right)   &
=\frac{n_{0}}{2}\left(  -c_{00}+c_{xx}r_{x}^{2}+c_{yy}r_{y}^{2}+c_{zz}%
r_{z}^{2}+c_{xz}r_{x}r_{z}\right) \nonumber
\end{align}
depend on the trap orientation angle $\vartheta_{0}$ and the semi-axes
$\lambda_{x}$ , $\lambda_{y}$ , $\lambda_{z}$ of the ellipsoid $\mathbb{D}%
_{TF}$ via the following one-dimensional\emph{\ }integrals:
\begin{align}
a,b  &  \in\left\{  x,y,z\right\} \label{index integrals  I}\\
& \nonumber\\
I_{a}\left(  \lambda_{x},\lambda_{y},\lambda_{z}\right)   &  =\lambda
_{x}\lambda_{y}\lambda_{z}\int_{0}^{\infty}\frac{du}{\sqrt{\left(  \lambda
_{x}^{2}+u\right)  \left(  \lambda_{y}^{2}+u\right)  \left(  \lambda_{z}%
^{2}+u\right)  }}\frac{1}{\left(  \lambda_{a}^{2}+u\right)  }\nonumber\\
& \nonumber\\
I_{ab}\left(  \lambda_{x},\lambda_{y},\lambda_{z}\right)   &  =\lambda
_{x}\lambda_{y}\lambda_{z}\int_{0}^{\infty}\frac{du}{\sqrt{\left(  \lambda
_{x}^{2}+u\right)  \left(  \lambda_{y}^{2}+u\right)  \left(  \lambda_{z}%
^{2}+u\right)  }}\frac{1}{\left(  \lambda_{a}^{2}+u\right)  \left(
\lambda_{b}^{2}+u\right)  }\nonumber
\end{align}
In the appendix \ref{appendixA} some of the properties of these so called \emph{\ index
integrals} are listed. We find
\begin{align}
& \label{coefficients selfconsistent quadratic form}\\
c_{00}  &  =\sin^{2}\left(  \vartheta_{0}\right)  I_{x}+\cos^{2}\left(
\vartheta_{0}\right)  I_{z}\nonumber\\
& \nonumber\\
c_{xx}  &  =\left[
\begin{array}
[c]{c}%
\cos^{2}\left(  \vartheta_{0}\right)  \sin^{2}\left(  \vartheta_{0}\right)
\left(  I_{xx}+I_{zz}\right)  +\left[  \cos^{4}\left(  \vartheta_{0}\right)
+\sin^{4}\left(  \vartheta_{0}\right)  \right]  I_{zx}\\
+2\sin^{2}\left(  \vartheta_{0}\right)  \cos^{2}\left(  \vartheta_{0}\right)
\left(  I_{xx}-2I_{zx}+I_{zz}\right)
\end{array}
\right] \nonumber\\
& \nonumber\\
c_{yy}  &  =\sin^{2}\left(  \vartheta_{0}\right)  I_{xy}+\cos^{2}\left(
\vartheta_{0}\right)  I_{zy}\nonumber\\
& \nonumber\\
c_{zz}  &  =3\cdot\left[  \sin^{4}\left(  \vartheta_{0}\right)  I_{xx}%
+\cos^{4}\left(  \vartheta_{0}\right)  I_{zz}+2\sin^{2}\left(  \vartheta
_{0}\right)  \cos^{2}\left(  \vartheta_{0}\right)  I_{zx}\right] \nonumber\\
& \nonumber\\
c_{xz}  &  =6\sin\left(  \vartheta_{0}\right)  \cos\left(  \vartheta
_{0}\right)  \left[  \sin^{2}\left(  \vartheta_{0}\right)  \left(
I_{xx}-I_{zx}\right)  +\cos^{2}\left(  \vartheta_{0}\right)  \left(
I_{zx}-I_{zz}\right)  \right] \nonumber
\end{align}

It is advantageous to work in the geometry under consideration not with the
basis functions $\left\{  1,r_{x}^{2},r_{y}^{2},r_{z}^{2},r_{x}r_{z}\right\}
$, but with the basis functions $\left\{  1,\widetilde{r}_{x}^{2}%
,\widetilde{r}_{y}^{2},\widetilde{r}_{z}^{2},\widetilde{r}_{x}\widetilde
{r}_{z}\right\}  $ obtained by a rotation of the coordinate system around the
axis $\mathbf{e}_{y}$ by the trap orientation angle $\vartheta_{0}$ as defined
in (\ref{principal axes frame I}). The exact solution of the Thomas-Fermi
integral equation (\ref{TF integral equation II}) is then obtained
inserting the corresponding explicit expressions for the quadratic form
$\frac{\partial^{2}}{\partial r_{z}^{2}}\phi_{TF}\left(  \mathbf{r}\right)  $
and the trap potential $V_{T}(\mathbf{r})$. From the condition, that the
prefactors of the linearly independent basis functions $\left\{
1,\widetilde{r}_{x}^{2},\widetilde{r}_{y}^{2},\widetilde{r}_{z}^{2}%
,\widetilde{r}_{x}\widetilde{r}_{z}\right\}  $ in
(\ref{TF integral equation II}) should vanish identically, the following
set of coupled selfconsistency equations is found:
\begin{align}
\left\{  1-\varepsilon_{D}+\frac{3}{2}\varepsilon_{D}\left[  \sin^{2}\left(
\vartheta_{0}\right)  I_{x}+\cos^{2}\left(  \vartheta_{0}\right)
I_{z}\right]  \right\}  n_{0}  &  =\frac{\mu}{g_{s}}\label{selfconsistent  Ia}%
\\
& \nonumber
\end{align}%
\begin{align}
& \label{selfconsistent  Ib}\\
\left\{
\begin{array}
[c]{c}%
\frac{1-\varepsilon_{D}}{\lambda_{x}^{2}}\\
\\
+\frac{3\varepsilon_{D}}{2}\left[  \cos^{2}\left(  \vartheta_{0}\right)
I_{zx}+3\sin^{2}\left(  \vartheta_{0}\right)  I_{xx}\right]
\end{array}
\right\}  n_{0}  &  =\frac{m^{\star}}{2g_{s}}\left[  \omega_{x}^{2}\cos
^{2}\left(  \vartheta_{T}-\vartheta_{0}\right)  +\omega_{z}^{2}\sin^{2}\left(
\vartheta_{T}-\vartheta_{0}\right)  \right] \nonumber
\end{align}%
\begin{align}
& \label{selfconsistent  Ic}\\
\left\{
\begin{array}
[c]{c}%
\frac{1-\varepsilon_{D}}{\lambda_{y}^{2}}\\
\\
+\frac{3\varepsilon_{D}}{2}\left[  \cos^{2}\left(  \vartheta_{0}\right)
I_{zy}+\sin^{2}\left(  \vartheta_{0}\right)  I_{xy}\right]
\end{array}
\right\}  n_{0}  &  =\frac{m^{\star}}{2g_{s}}\omega_{y}^{2}\nonumber
\end{align}%
\begin{align}
& \label{selfconsistent  Id}\\
\left\{
\begin{array}
[c]{c}%
\frac{1-\varepsilon_{D}}{\lambda_{z}^{2}}\\
\\
+\frac{3\varepsilon_{D}}{2}\left[  3\cos^{2}\left(  \vartheta_{0}\right)
I_{zz}+\sin^{2}\left(  \vartheta_{0}\right)  I_{xz}\right]
\end{array}
\right\}  n_{0}  &  =\frac{m^{\star}}{2g_{s}}\left[  \omega_{x}^{2}\sin
^{2}\left(  \vartheta_{T}-\vartheta_{0}\right)  +\omega_{z}^{2}\cos^{2}\left(
\vartheta_{T}-\vartheta_{0}\right)  \right] \nonumber
\end{align}%
\begin{align}
& \label{selfconsistent  Ie}\\
\frac{3\varepsilon_{D}}{2}\sin\left(  2\vartheta_{0}\right)  I_{xz}\ n_{0}  &
=\frac{m^{\star}}{2g_{s}}\frac{\omega_{x}^{2}-\omega_{z}^{2}}{2}\cdot
\sin\left(  2\vartheta_{T}-2\vartheta_{0}\right) \nonumber
\end{align}
The normalization condition connects the density $n_{0}$ at the center of the
Thomas-Fermi domain $\mathbb{D}_{TF}$ to the product of the semi-axes:
\begin{equation}
n_{0}=\frac{15}{8\pi}\frac{N}{\lambda_{x}\lambda_{y}\lambda_{z}}
\label{normalisation  III}%
\end{equation}

Let us first write the selfconsistency equations without dipole interaction
setting $\varepsilon_{D}=0$. There follows%
\[
n_{0}^{\left(  0\right)  }=\frac{\mu^{\left(  0\right)  }}{g_{s}}%
\]%
\begin{equation}
\frac{1}{\left[  \lambda_{a}^{\left(  0\right)  }\right]  ^{2}}=\frac
{m^{\star}}{2\mu^{\left(  0\right)  }}\omega_{a}^{2}
\label{semi axes epsilon-D=0}%
\end{equation}
Using the normalization%
\begin{equation}
n_{0}^{\left(  0\right)  }=\frac{15}{8\pi}\frac{N}{\lambda_{x}^{\left(
0\right)  }\lambda_{y}^{\left(  0\right)  }\lambda_{z}^{\left(  0\right)  }}
\label{normalisation  IV}%
\end{equation}
and introducing the definitions%
\begin{align}
\omega &  =\left(  \omega_{x}\omega_{y}\omega_{z}\right)  ^{\frac{1}{3}%
}\label{geometric mean frequency}\\
a_{\omega}  &  =\left(  \frac{\hbar}{m^{\star}\omega}\right)  ^{\frac{1}{2}%
}\nonumber
\end{align}
we obtain for the chemical potential $\mu^{\left(  0\right)  }$ and the
semi-axes $\lambda_{a}^{\left(  0\right)  }$ of a BEC inside an anisotropic
harmonic trap in the Thomas-Fermi regime well known results:
\begin{equation}
\mu^{\left(  0\right)  }\left(  N\right)  =\left(  \frac{15}{4\pi}\frac{4\pi
Na_{s}}{a_{\omega}}\right)  ^{\frac{2}{5}}\frac{\hslash\omega}{2}
\label{chemical potential epsilon-D=0}%
\end{equation}%
\begin{align}
\lambda_{a}^{\left(  0\right)  }\left(  N\right)   &  =\left(  \frac
{2\mu^{\left(  0\right)  }}{m^{\star}\omega_{a}^{2}}\right)  ^{\frac{1}{2}%
}=\frac{\omega}{\omega_{a}}\Lambda\label{semi axes epsilon-D=0  II}\\
\Lambda &  =a_{\omega}\left(  \frac{15}{4\pi}\frac{4\pi Na_{s}}{a_{\omega}%
}\right)  ^{\frac{1}{5}}\nonumber
\end{align}
The exponent $\frac{2}{5}$ is characteristic for the large $N$ scaling of the
chemical potential $\mu^{\left(  0\right)  }\left(  N\right)  $ of a BEC
confined inside a \emph{harmonic} trap \cite{Pethick and Smith}.

The ensuing calculations simplify making use of elementary scaling relations
that hold for single index and double index integrals:%
\begin{align}
I_{a}\left(  \lambda_{x},\lambda_{y},\lambda_{z}\right)   &  =I_{a}\left(
\frac{\lambda_{x}}{\lambda_{z}},\frac{\lambda_{y}}{\lambda_{z}},1\right)
\equiv\overline{I}_{a}\label{index integrals  II}\\
& \nonumber\\
\lambda_{c}^{2}\ I_{ab}\left(  \lambda_{x},\lambda_{y},\lambda_{z}\right)   &
=\frac{\lambda_{c}^{2}}{\lambda_{z}^{2}}\ I_{ab}\left(  \frac{\lambda_{x}%
}{\lambda_{z}},\frac{\lambda_{y}}{\lambda_{z}},1\right)  \equiv\frac
{\lambda_{c}^{2}}{\lambda_{z}^{2}}\ \overline{I}_{ab}\nonumber
\end{align}
Using the obvious relation%
\begin{equation}
\frac{n_{0}}{n_{0}^{\left(  0\right)  }}=\frac{\lambda_{x}^{\left(  0\right)
}\lambda_{y}^{\left(  0\right)  }\lambda_{z}^{\left(  0\right)  }}{\lambda
_{x}\lambda_{y}\lambda_{z}} \label{normalisation   V}%
\end{equation}
the selfconsistency problem posed by (\ref{selfconsistent Ia}%
)-(\ref{selfconsistent Ie}) may then be reduced to three coupled equations for
the ratios $\frac{\lambda_{x}}{\lambda_{z}}$ , $\frac{\lambda_{y}}{\lambda
_{z}}$ and the equilibrium orientation angle $\vartheta_{0}$ as functions of
the trap orientation angle $\vartheta_{T}$ , the trap frequencies $\omega_{a}$
and the dipole interaction strength parameter $\varepsilon_{D}$ :
\begin{align}
& \label{selfconsistent  IV a}\\
\frac{\lambda_{x}^{2}}{\lambda_{z}^{2}}  &  =\frac{\omega_{x}^{2}\sin
^{2}\left(  \vartheta_{T}-\vartheta_{0}\right)  +\omega_{z}^{2}\cos^{2}\left(
\vartheta_{T}-\vartheta_{0}\right)  }{\omega_{x}^{2}\cos^{2}\left(
\vartheta_{T}-\vartheta_{0}\right)  +\omega_{z}^{2}\sin^{2}\left(
\vartheta_{T}-\vartheta_{0}\right)  }\cdot\frac{1-\varepsilon_{D}%
+\frac{3\varepsilon_{D}}{2}\frac{\lambda_{x}^{2}}{\lambda_{z}^{2}}\left[
\cos^{2}\left(  \vartheta_{0}\right)  \overline{I}_{zx}+3\sin^{2}\left(
\vartheta_{0}\right)  \overline{I}_{xx}\right]  }{1-\varepsilon_{D}%
+\frac{3\varepsilon_{D}}{2}\left[  3\cos^{2}\left(  \vartheta_{0}\right)
\overline{I}_{zz}+\sin^{2}\left(  \vartheta_{0}\right)  \overline{I}%
_{xz}\right]  }\nonumber
\end{align}%
\begin{align}
& \label{selfconsistent  IV b}\\
\frac{\lambda_{y}^{2}}{\lambda_{z}^{2}}  &  =\frac{\omega_{x}^{2}\sin
^{2}\left(  \vartheta_{T}-\vartheta_{0}\right)  +\omega_{z}^{2}\cos^{2}\left(
\vartheta_{T}-\vartheta_{0}\right)  }{\omega_{y}^{2}}\cdot\frac{1-\varepsilon
_{D}+\frac{3\varepsilon_{D}}{2}\frac{\lambda_{y}^{2}}{\lambda_{z}^{2}}\left[
\cos^{2}\left(  \vartheta_{0}\right)  \overline{I}_{yz}+\sin^{2}\left(
\vartheta_{0}\right)  \overline{I}_{xy}\right]  }{1-\varepsilon_{D}%
+\frac{3\varepsilon_{D}}{2}\left[  3\cos^{2}\left(  \vartheta_{0}\right)
\overline{I}_{zz}+\sin^{2}\left(  \vartheta_{0}\right)  \overline{I}%
_{xz}\right]  }\nonumber
\end{align}%
\begin{align}
& \label{selfconsistent  IV c}\\
\tan\left(  2\vartheta_{0}\right)   &  =\frac{\left(  \omega_{x}^{2}%
-\omega_{z}^{2}\right)  \cdot\sin\left(  2\vartheta_{T}\right)  }{\left(
\omega_{x}^{2}-\omega_{z}^{2}\right)  \cos\left(  2\vartheta_{T}\right)
+3\varepsilon_{D}\frac{\lambda_{y}^{2}}{\lambda_{z}^{2}}\overline{I}_{xz}%
\frac{\omega_{y}^{2}}{1-\varepsilon_{D}+\frac{3\varepsilon_{D}}{2}%
\frac{\lambda_{y}^{2}}{\lambda_{z}^{2}}\left[  \cos^{2}\left(  \vartheta
_{0}\right)  \overline{I}_{zy}+\sin^{2}\left(  \vartheta_{0}\right)
\overline{I}_{xy}\right]  }}\nonumber
\end{align}
We have found, that the set of selfconsistency equations
(\ref{selfconsistent IV a}), (\ref{selfconsistent IV b}) and
(\ref{selfconsistent IV c}) may be conveniently solved numerically by the
method of fixed point iteration. Using identities like%
\[
\cos^{2}\left(  \vartheta_{0}\right)  =\frac{1}{2}+\frac{1}{2}\frac{1}%
{\sqrt{1+\tan^{2}\left(  2\vartheta_{0}\right)  }}%
\]
the evaluation of trigonometric functions in the iteration process can be
completely avoided.

Once the ratios $\frac{\lambda_{x}}{\lambda_{z}}$ , $\frac{\lambda_{y}%
}{\lambda_{z}}$ and the orientation angle $\vartheta_{0}$ are known, it
follows directly from (\ref{selfconsistent Id}) and
(\ref{normalization V}):
\begin{align}
& \label{semi axes at finite value of epislon-D}\\
\lambda_{z}  &  =\left[  \frac{\frac{\omega_{z}^{2}}{\omega_{x}\omega_{y}}%
}{\frac{\lambda_{x}}{\lambda_{z}}\frac{\lambda_{y}}{\lambda_{z}}}%
\frac{1-\varepsilon_{D}+\frac{3\varepsilon_{D}}{2}\left[  3\cos^{2}\left(
\vartheta_{0}\right)  \overline{I}_{zz}+\sin^{2}\left(  \vartheta_{0}\right)
\overline{I}_{xz}\right]  }{\frac{\omega_{x}^{2}}{\omega_{z}^{2}}\sin
^{2}\left(  \vartheta_{T}-\vartheta_{0}\right)  +\cos^{2}\left(  \vartheta
_{T}-\vartheta_{0}\right)  }\right]  ^{\frac{1}{5}}\lambda_{z}^{\left(
0\right)  }\nonumber\\
& \nonumber\\
\lambda_{x}  &  =\frac{\lambda_{x}}{\lambda_{z}}\lambda_{z}\;\;;\;\;\lambda
_{y}=\frac{\lambda_{y}}{\lambda_{z}}\lambda_{z}\nonumber
\end{align}

For the special case of an \emph{isotropic} harmonic trap \cite{Pethick and
Smith} the principal effect of the dipole-dipole interaction on the
groundstate density profile $n_{TF}\left(  \mathbf{r}\right)  $ of a dipolar
interacting BEC is the well known elongation of the semi-axis $\lambda_{z}$
parallel to $\mathbf{B}$, and the distortion of the semi-axes $\lambda
_{x}=\lambda_{y}$ perpendicular to $\mathbf{B}$ towards smaller values:
\begin{equation}
\frac{\lambda_{x}}{\lambda_{z}}=\frac{1-\frac{1}{5}\varepsilon_{D}%
+...}{1+\frac{2}{5}\varepsilon_{D}+...} \label{elongation I}%
\end{equation}
In Fig.2 and Fig.3 selfconsistent solutions of the coupled equations
(\ref{selfconsistent IV a}), (\ref{selfconsistent IV b}) and
(\ref{selfconsistent IV c}) for the equilibrium angle $\vartheta_{0}%
-\vartheta_{T}$ and the semi-axes $\lambda_{a}$ are plotted as functions of
the trap orientation angle $\vartheta_{T}$ for various magnetic dipole
interaction strength parameters $\varepsilon_{D}$ assuming a tri-axial trap
anisotropy ratio $\omega_{x}:\omega_{y}:\omega_{z}=6:3:2$ .

\begin{figure}[h]
\centering\includegraphics[width=0.80\textwidth]{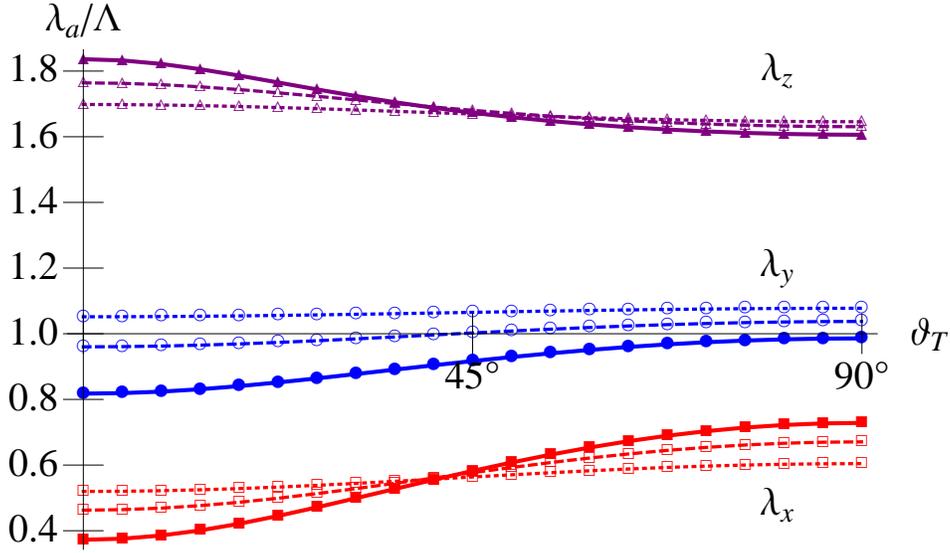}
\label{fig:3-1}\caption{(Color online) Plot of semi-axes $\lambda_{x}$ , $\lambda_{y}$ and
$\lambda_{z}$ of selfconsistent groundstate density profile $n_{TF}%
(\mathbf{r})$ vs. trap orientation angle $\vartheta_{T}$ for various dipolar
interaction strength $\varepsilon_{D}$: a) $\varepsilon_{D}=0.2$ dotted line ;
b) $\varepsilon_{D}=0.5$ dashed line; c) $\varepsilon_{D}=0.8$ solid line. The
ratio of trap frequencies is $\omega_{x}:\omega_{y}:\omega_{z}=6:3:2$ .}%
\label{Fig. 3}%
\end{figure}

Once the semi-axes $\lambda_{a}$ of the particle density $n_{TF}\left(
\mathbf{r}\right)  $ (\ref{ansatz density profile}) are determined, then
(\ref{normalisation V}) gives us the value $n_{0}$ of the density of the
ellipsoidal shaped BEC at its center. In Fig.4 the ratio $\frac{n_{0}}%
{n_{0}^{\left(  0\right)  }}$ is plotted vs. the trap orientation angle
$\vartheta_{T}$ , the inset showing for selected trap orientation angles
$\vartheta_{T}$ $\in\left\{  0,\frac{\pi}{4},\frac{\pi}{2}\right\}  $ cuts of
the corresponding selfconsistently determined Thomas-Fermi ellipsoid
$\mathbb{D}_{TF}$ with the symmetry plane $y=0$. For a prolate trap the
density is largest for $\vartheta_{T}=0$ , because then the net mutual dipole
force between atom pairs inside the domain $\mathbb{D}_{TF}$ is attractive. As
$\vartheta_{T}$ increases the density $n_{0}$ becomes smaller and assumes a
minimum at $\vartheta_{T}=90^{\circ}$ , because for a parallel alignment the
net mutual dipole force between atom pairs inside the domain $\mathbb{D}_{TF}$
is repulsive.

\begin{figure}[h]
\centering
\includegraphics[width=0.80\textwidth]{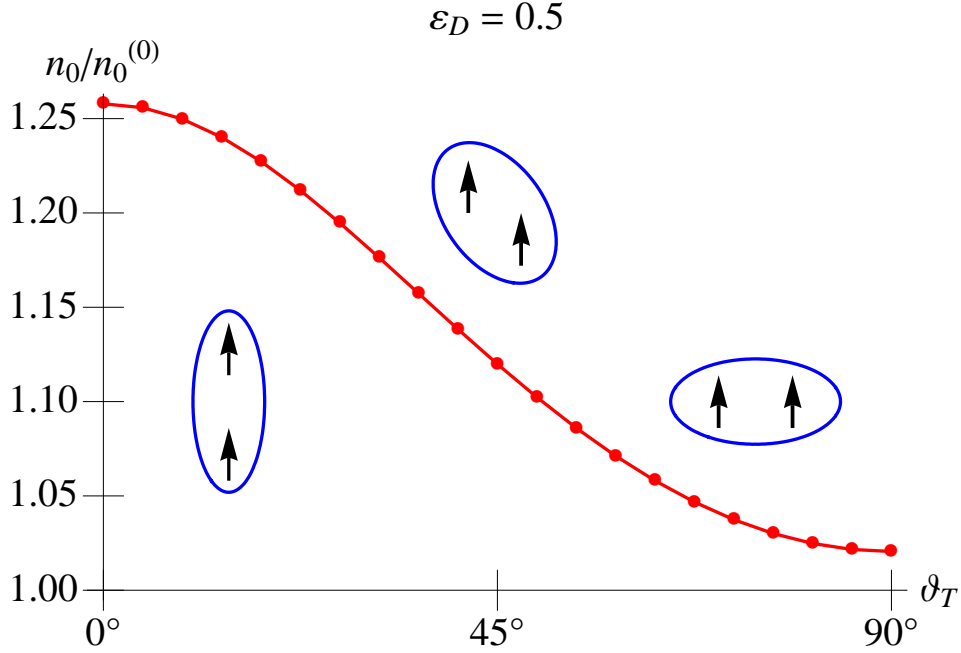} \label{fig:5}%
\caption{(Color online) Selfconsistent particle density $n_{0}$ at center position of
anisotropic harmonic trap plotted vs. trap orientation angle $\vartheta_{T}$.
The inset displays cuts of the Thomas-Fermi ellipsoid $\mathbb{D}_{TF}$ with
the symmetry plane $y=0$ for trap orientation angles $\vartheta_{T}=0^{\circ}%
$, $\vartheta_{T}=45^{\circ}$ and $\vartheta_{T}=90^{\circ}$. The ratio of
trap frequencies is $\omega_{x}:\omega_{y}:\omega_{z}=2:2:1$ , the dipole
interaction strength is $\varepsilon_{D}=0.5$ . }%
\label{Fig. 4}%
\end{figure}

Finally, there follows from (\ref{selfconsistent Ia}) an explicit formula
for the chemical potential:%
\begin{equation}
\mu=\left\{  1-\varepsilon_{D}+\frac{3}{2}\varepsilon_{D}\left[  \sin
^{2}\left(  \vartheta_{0}\right)  \overline{I}_{x}+\cos^{2}\left(
\vartheta_{0}\right)  \overline{I}_{z}\right]  \right\}  \frac{n_{0}}%
{n_{0}^{\left(  0\right)  }}\cdot\mu^{\left(  0\right)  }\left(  N\right)
\label{result chemical potential finite epsilon}%
\end{equation}
The dependence of $\mu$ on particle number $N$ is solely described by the
factor $\mu^{\left(  0\right)  }\left(  N\right)  $ , i.e. the ratio
$\frac{\mu}{\mu^{\left(  0\right)  }}$ is independent on particle number $N$.
In Fig. 5 the chemical potential $\mu$ is plotted vs. the trap orientation
angle $\vartheta_{T}$ for different values of the dipole interaction strength
$\varepsilon_{D}$.While for an \emph{isotropic} harmonic trap the chemical
potential $\mu$ doesn't change to first order in $\varepsilon_{D}$ , one finds
for an \emph{anisotropic} harmonic trap, making a straightforward expansion to
the first order in the dipole interaction strength $\varepsilon_{D}$ , for the
case of an oblate (pancake shaped) trap that $\left[  \frac{d\mu}%
{d\varepsilon_{D}}\right]  _{\varepsilon_{D}=0}>0$, and for a prolate (cigar
shaped) trap that $\left[  \frac{d\mu}{d\varepsilon_{D}}\right]
_{\varepsilon_{D}=0}<0$ , respectively.

\begin{figure}[h]
\centering
\includegraphics[width=0.80\textwidth]{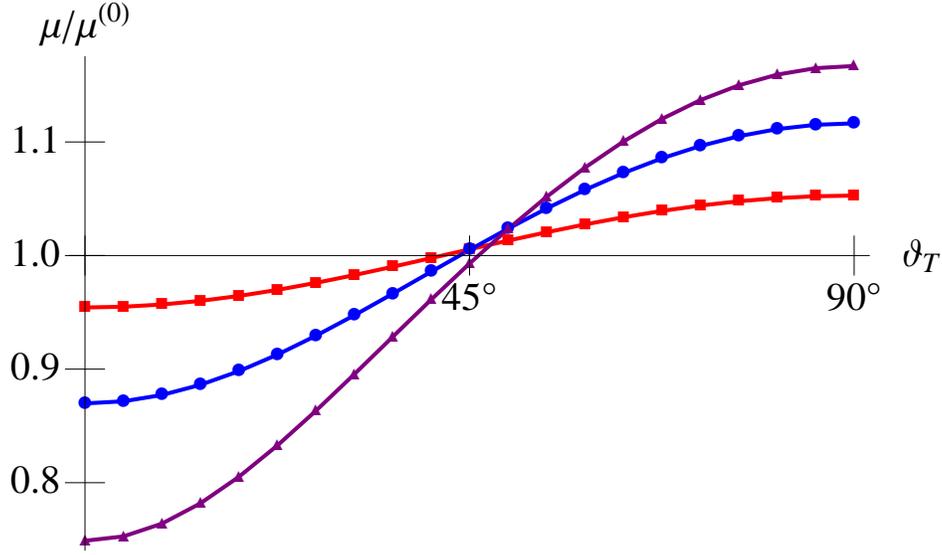} \label{fig:6}%
\caption{(Color online) Chemical potential $\mu$ vs. trap orientation angle $\vartheta_{T}$
for different values of dipole interaction strength. a) $\varepsilon_{D}=0.2$
red square ; b) $\varepsilon_{D}=0.5$ blue circle ; c) $\varepsilon_{D}=0.8$
purple triangle. The ratio of trap frequencies is $\omega_{x}:\omega
_{y}:\omega_{z}=6:3:2$ .}%
\label{Fig. 5}%
\end{figure}

The displayed characteristic dependence of chemical potential $\mu$ on the
trap orientation angle $\vartheta_{T}$ should be observable as the release
energy $E_{r}$ of a spin-polarized dipolar interacting BEC confined in a
harmonic trap, when the trap potential is suddenly switched off to zero, and
subsequently the dilute atom gas cloud undergoes a \emph{ballistic} expansion
\cite{Lahaye et al}. We find within the range of validity of the Thomas-Fermi
approximation that there holds also in the presence of long ranged
dipole-dipole interactions
\begin{equation}
E_{r}=E_{int}=\frac{2}{7}\mu N \label{release energy of BEC in harmonic trap}%
\end{equation}

As a matter of fact, the total energy
\begin{equation}
E=\left\langle \widehat{H}\right\rangle _{\Psi}=\left\langle \widehat{H}%
_{kin}+\widehat{H}_{pot}+\widehat{H}_{int}\right\rangle _{\Psi}
\label{total energy of groundstate}%
\end{equation}
, and the interaction energy $\ E_{int}=\left\langle \widehat{H}%
_{int}\right\rangle _{\Psi}$ are connected in the groundstate $\Psi$ of the
BEC, see (\ref{Hartree Groundstate}), by the general relation%

\begin{equation}
E=\mu N-E_{int} \label{Energy in the groundstate II}%
\end{equation}
This applies, because the \emph{optimal} one-particle wave function
$\psi\left(  \mathbf{r}\right)  $ building the $N$ particle groundstate $\Psi$
solves the GP equation, so that the expectation value $\left\langle
\widehat{H}_{kin}\right\rangle _{\Psi}$of the kinetic energy can be
re-expressed via the GP equation in terms of the chemical potential $\mu$ and
the interaction energy $E_{int}$ . On the other hand, the total energy $E$ of
the BEC is connected to the chemical potential $\mu$ by the general relation%
\begin{equation}
\mu=\frac{\partial E}{\partial N} \label{chemical potential}%
\end{equation}
It follows from (\ref{result chemical potential finite epsilon}), and the
established scaling (\ref{chemical potential epsilon-D=0}) of $\mu^{\left(  0\right)  }\left(  N\right)  \propto
N^{\frac{2}{5}}$ for a large particle number $N\gg1$, that up to a constant
that is independent on $N$ there also holds for a spin-polarized dipolar BEC confined inside a
\emph{harmonic} trap the well known relation \cite{Pethick and Smith}
\begin{equation}
E=\frac{5}{7}\mu N \label{total energy  of BEC in harmonic trap II}%
\end{equation}
, and therefore:%
\begin{equation}
E_{int}=\mu N-E=\frac{2}{7}\mu N
\label{interaction energy of BEC in terms of chemical potential for harmonic trap}%
\end{equation}

\section{Collective Modes of Small Amplitude Density Oscillations}

\subsection{Parametrization of Low-Lying Excitations in Tri-Axial Harmonic Trap}

An important test of the macroscopic quantum physics of a BEC is the study of
elementary excitations above the groundstate. One technique to excite low
energy collective modes of a BEC is to suddenly modify the trap potential. For
example, shifting the center of the trap excites the dipole modes, that is the
motion of the center of mass of a BEC cloud around its equilibrium position in
a harmonic trap. Changing the curvature of the trap by switching the trap
frequencies may excite the breather mode. In an anisotropic harmonic trap
there also exist the so called sissors modes \cite{Stringari I}, which can be
excited by rotating a principal axis of the trap, thus pushing the atom cloud
in the trap away from equilibrium. A recently reported elegant new
experimental technique excites a BEC by modulating the field dependence of the
atomic scattering length $a_{s}$ near to a magnetic Feshbach resonance
\cite{Lahaye II}, \cite{Pollack II}.

We calculate in this section for the case of a harmonic anisotropic trap with
arbitrary trap orientation angle $\vartheta_{T}$ the small amplitude
collective modes of a dipolar interacting spin-polarized BEC at very low
energy, so that the wavelength of the excitations becomes comparable to the
size of the Thomas-Fermi length $\Lambda$. For a large number $N>>1$ of
particles in the BEC the particle density $n\left(  \mathbf{r},t\right)  $ and
the macroscopic Josephson phase $S\left(  \mathbf{r},t\right)  $ are
\emph{conjugate }variables, so that the collective dynamics of the system
(ignoring a small quantum pressure) is governed by the standard canonical
equations of motion of macroscopic quantum physics \cite{Pethick and Smith}:%
\begin{align}
& \label{Josephson equation at zero quantum pressure}\\
\hslash\frac{\partial}{\partial t}S\left(  \mathbf{r},t\right)   &
=-\frac{\hbar^{2}}{2m^{\star}}\sum_{a\in\left\{  x,y,z\right\}  }\left(
\frac{\partial S\left(  \mathbf{r},t\right)  }{\partial r_{a}}\right)
^{2}-V_{T}\left(  \mathbf{r}\right)  -\int_{\mathbb{D}\left(  t\right)  }%
d^{3}r^{\prime}\ U\left(  \mathbf{r},\mathbf{r}^{\prime}\right)  n\left(
\mathbf{r}^{\prime},t\right) \nonumber
\end{align}%
\begin{align}
& \label{continuity equation for particle density}\\
\frac{\partial}{\partial t}n\left(  \mathbf{r},t\right)  +\frac{\hslash
}{m^{\star}}\sum_{a\in\left\{  x,y,z\right\}  }\frac{\partial}{\partial r_{a}%
}\left[  n\left(  \mathbf{r},t\right)  \frac{\partial S\left(  \mathbf{r}%
,t\right)  }{\partial r_{a}}\right]   &  =0\nonumber
\end{align}
The gradient of the phase, the velocity field $\mathbf{v}\left(
\mathbf{r},t\right)  =\frac{\hslash}{m^{\star}}\nabla\delta s\left(
\mathbf{r},t\right)  $, is directly connected to the density of particle
current, $\mathbf{j\left(  \mathbf{r},t\right)  }=n\left(  \mathbf{r}%
,t\right)  \mathbf{v}\left(  \mathbf{r},t\right)  $ \cite{Pethick and Smith}.
The quasiclassical Josephson equation
(\ref{Josephson equation at zero quantum pressure}) and the continuity
equation (\ref{continuity equation for particle density}) need to be solved
subject to the normalization condition%
\begin{equation}
\int_{\mathbb{D}(t)}d^{3}r\;n\left(  \mathbf{r},t\right)  =N
\label{normalization V}%
\end{equation}
, where for a fluctuating particle density $n\left(  \mathbf{r},t\right)  $
the positivity domain
\begin{equation}
\mathbb{D}(t)=\left\{  \mathbf{r}\in\mathbb{R}^{3}|\;n\left(  \mathbf{r}%
,t\right)  >0\right\}  \label{time dependet integration domain}%
\end{equation}
depends ( in principle) on time too!

Let us assume a perturbation expansion of phase and density of the form%
\begin{align}
S\left(  \mathbf{r},t\right)   &  =-\frac{\mu}{\hbar}\cdot t+\delta s\left(
\mathbf{r},t\right) \label{perturbation expansion density and phase}\\
& \nonumber\\
n\left(  \mathbf{r},t\right)   &  =n_{TF}\left(  \mathbf{r}\right)  +\delta
n\left(  \mathbf{r},t\right) \nonumber
\end{align}
, with $\delta s\left(  \mathbf{r},t\right)  $ and $\delta n\left(
\mathbf{r},t\right)  $ denoting small fluctuations of phase and particle
density around the groundstate of the BEC.

For convenience we work from now on, unless otherwise explicitely stated, in
the principal axes frame of the Thomas-Fermi ellipsoid $\mathbb{D}_{TF}$. As
is indicated in Fig. 1, for a trap orientation angle $\vartheta_{T}\neq0$ ,
the magnetic field $\mathbf{B}\ $ is not aligned parallel to the principal
axis vector $\mathbf{e}_{z,0}$ of the ellipsoid $\mathbb{D}_{TF}$. In this
case the interaction energy $U\left(  \mathbf{r},\mathbf{r}^{\prime}\right)  $
between two spin-polarized atoms at position $\mathbf{r}$ and $\mathbf{r}%
^{\prime}$ , both carrying a magnetic dipole moment $\left\langle
\mathbf{M}\right\rangle =\left(  2\mu_{B}S\right)  \mathbf{m}$ orientated
(anti-)parallel to $\mathbf{B}$ is then given by%
\begin{equation}
U\left(  \mathbf{r},\mathbf{r}^{\prime}\right)  =g_{s}\left[  \left(
1-\varepsilon_{D}\right)  \delta^{\left(  3\right)  }\left(  \mathbf{r}%
-\mathbf{r}^{\prime}\right)  \ -3\varepsilon_{D}\left(  \mathbf{m}%
\cdot\mathbf{\nabla}_{\mathbf{r}}\right)  ^{2}\frac{1}{4\pi}\ \frac
{1}{\left\vert \mathbf{r}-\mathbf{r}^{\prime}\right\vert }\right]
\label{interaction  III}%
\end{equation}
In the geometry under consideration the axis $\mathbf{e}_{z,0}$ of the
ellipsoid $\mathbb{D}_{TF}$ is rotated around the axis $\mathbf{e}_{y,T}$ of
the trap by an angle $\vartheta_{0}$ , and we assume $\mathbf{m\perp e}_{y}$
:
\begin{equation}
\left(  \mathbf{m}\cdot\mathbf{\nabla}_{\mathbf{r}}\right)  ^{2}=\sin
^{2}\left(  \vartheta_{0}\right)  \frac{\partial^{2}}{\partial r_{x}^{2}}%
+\cos^{2}\left(  \vartheta_{0}\right)  \frac{\partial^{2}}{\partial r_{z}^{2}%
}+\sin\left(  2\vartheta_{0}\right)  \frac{\partial^{2}}{\partial
r_{x}\partial r_{z}} \label{interaction  IIIb}%
\end{equation}
Upon linearization of (\ref{Josephson equation at zero quantum pressure})
and (\ref{continuity equation for particle density}) we reproduce to order
zero in the expansion the Thomas-Fermi integral equation
(\ref{TF integral equation I}) determining the equilibrium density profile
$n_{TF}\left(  \mathbf{r}\right)  $ and the chemical potential $\mu$ , as
discussed already in the previous section.

To derive the equations of motion for the fluctuations of the phase $\delta
s\left(  \mathbf{r},t\right)  $ and the density $\delta n\left(
\mathbf{r},t\right)  $ let us first consider for a kernel $K\left(
\mathbf{r},\mathbf{r}^{\prime}\right)  $ that couples to particle density
$n\left(  \mathbf{r}^{\prime},t\right)  $ the associated fluctuation%
\begin{equation}
\delta K\left(  t\right)  =\int_{\mathbb{D}\left(  t\right)  }d^{3}r^{\prime
}\ K\left(  \mathbf{r},\mathbf{r}^{\prime}\right)  n\left(  \mathbf{r}%
^{\prime},t\right)  -\int_{\mathbb{D}_{TF}}d^{3}r^{\prime}\ K\left(
\mathbf{r},\mathbf{r}^{\prime}\right)  n_{TF}\left(  \mathbf{r}^{\prime
}\right)  \label{particle difference I}%
\end{equation}
In principle there are two contributions to $\delta K\left(  t\right)  $. One
is generated by the time dependence of the density distribution, the other
results from a change of the integration domain $\mathbb{D}\left(  t\right)
$. Introducing the Heaviside distribution%
\begin{equation}
\Theta_{H}\left(  x\right)  =\frac{1+\mbox{sign}(x)}{2} \label{Heaviside}%
\end{equation}
we rewrite (\ref{particle difference I}) as%
\begin{equation}
\delta K\left(  t\right)  =\int d^{3}r^{\prime}K\left(  \mathbf{r}%
,\mathbf{r}^{\prime}\right)  \left\{
\begin{array}
[c]{c}%
\Theta_{H}\left[  n_{TF}\left(  \mathbf{r}^{\prime}\right)  +\delta n\left(
\mathbf{r}^{\prime},t\right)  \right]  \left[  n_{TF}\left(  \mathbf{r}%
^{\prime}\right)  +\delta n\left(  \mathbf{r}^{\prime},t\right)  \right] \\
\\
-\Theta_{H}\left[  n_{TF}\left(  \mathbf{r}^{\prime}\right)  \right]
n_{TF}\left(  \mathbf{r}^{\prime}\right)
\end{array}
\right\}  \label{fluctuation dK}%
\end{equation}
An expansion to the first order in the small quantitiy $\delta n$ leads to%
\begin{equation}
\delta K\left(  t\right)  =\int d^{3}r^{\prime}K\left(  \mathbf{r}%
,\mathbf{r}^{\prime}\right)  \delta n\left(  \mathbf{r}^{\prime},t\right)
\left\{  \Theta\left[  n_{TF}\left(  \mathbf{r}^{\prime}\right)  \right]
+n_{TF}\left(  \mathbf{r}^{\prime}\right)  \delta_{D}\left[  n_{TF}\left(
\mathbf{r}^{\prime}\right)  \right]  \right\}  +o\left(  \left\vert \delta
n\right\vert ^{2}\right)  \label{particle difference II}%
\end{equation}
, where $\delta_{D}\left(  x\right)  $ denotes the Dirac delta-distribution:
\begin{equation}
\delta_{D}\left(  x\right)  =\frac{d}{dx}\Theta_{H}\left(  x\right)
\label{Dirac delta}%
\end{equation}
Here, the term proportional to $n_{TF}\left(  \mathbf{r}^{\prime}\right)
\delta_{D}\left[  n_{TF}\left(  \mathbf{r}^{\prime}\right)  \right]  $
corresponds to a surface integral over the boundary $\partial\mathbb{D}_{TF}$
of the Tomas-Fermi domain $\mathbb{D}_{TF}$. However, because $n_{TF}\left(
\mathbf{r}^{\prime}\right)  \equiv0$ for $\mathbf{r}^{\prime}$ $\in$
$\partial\mathbb{D}_{TF}$ the value of this surface integral is zero. This
means the fluctuation of the integration domain $\mathbb{D}\left(  t\right)  $
around the shape of the \emph{equilibrium} cloud $\mathbb{D}_{TF}$ as caused
by a density fluctuation $\delta n=n\left(  \mathbf{r},t\right)
-n_{TF}\left(  \mathbf{r}\right)  $ represents only a small correction to
$\delta K\left(  t\right)  $ beyond first order accuracy:
\begin{equation}
\delta K\left(  t\right)  =\int_{\mathbb{D}_{TF}}d^{3}r^{\prime}K\left(
\mathbf{r},\mathbf{r}^{\prime}\right)  \delta n\left(  \mathbf{r}^{\prime
},t\right)  +o\left(  \left\vert \delta n\right\vert ^{2}\right)
\label{particle difference III}%
\end{equation}

It follows now directly from
(\ref{Josephson equation at zero quantum pressure}) for the time derivative
of the phase fluctuation $\delta s\left(  \mathbf{r},t\right)  $ ,
substituting into (\ref{particle difference III}) the special case
$K\left(  \mathbf{r},\mathbf{r}^{\prime}\right)  =U\left(  \mathbf{r}%
,\mathbf{r}^{\prime}\right)  $, that
\begin{equation}
\hslash\frac{\partial}{\partial t}\delta s\left(  \mathbf{r},t\right)
=-\int_{\mathbb{D}_{TF}}d^{3}r^{\prime}\ U\left(  \mathbf{r},\mathbf{r}%
^{\prime}\right)  \delta n\left(  \mathbf{r}^{\prime},t\right)
\label{fluctuation phase}%
\end{equation}
, and further from (\ref{continuity equation for particle density})
\begin{align}
a  &  \in\left\{  x,y,z\right\} \label{wave equation}\\
0  &  =\frac{\partial^{2}}{\partial t^{2}}\delta n\left(  \mathbf{r},t\right)
+\frac{1}{m^{\star}}\sum_{a}\frac{\partial}{\partial r_{a}}\left[
n_{TF}\left(  \mathbf{r}\right)  \frac{\partial}{\partial r_{a}}\hslash
\frac{\partial}{\partial t}\delta s\left(  \mathbf{r},t\right)  \right]
\nonumber
\end{align}
This is the well known wave equation describing the collective density
excitations of a BEC in the quantum hydrodynamic limit\cite{Pethick and Smith}
for the case of a non local interaction potential $U\left(  \mathbf{r}%
,\mathbf{r}^{\prime}\right)  $.

In the ensuing discussion we consider the collective modes of small amplitude
oscillations around the equilibrium density distribution $n_{TF}\left(
\mathbf{r}\right)  $ in the trap. The corresponding density fluctuations
$\delta n\left(  \mathbf{r},t\right)  $ may be expanded with respect to a set
of linearly independent multinomials $\left\{  1,r_{a},r_{a}r_{b},r_{a}%
r_{b}r_{c},...\right\}  _{a\leq b\leq c...}$ :%
\begin{align}
a,b,c  &  \in\left\{  x,y,z\right\}
\label{expansion of density fluctuation in a polynomial basis}\\
\delta n\left(  \mathbf{r},t\right)   &  =\delta n^{\left(  0\right)  }\left(
t\right)  +\sum_{a}\delta n_{a}^{\left(  1\right)  }\left(  t\right)
r_{a}+\sum_{a,b}\delta n_{ab}^{\left(  2\right)  }\left(  t\right)  r_{a}%
r_{b}+\sum_{a,b,c}\delta n_{abc}^{\left(  3\right)  }\left(  t\right)
r_{a}r_{b}r_{c}...\nonumber
\end{align}
We restrict here to the terms of zero order , first order and second order of
the density fluctuation $\delta n\left(  \mathbf{r},t\right)  $ as described
by amplitudes $\delta n^{\left(  0\right)  }\left(  t\right)  $ , $\delta
n_{a}^{\left(  1\right)  }\left(  t\right)  $ and $\delta n_{ab}^{\left(
2\right)  }\left(  t\right)  $ , respectively. Third and higher order terms
proportional to $\delta n_{abc}^{\left(  3\right)  }\left(  t\right)  $ etc.
we shall consider elsewhere.

A very convenient approach to the determination of the small amplitude
oscillations $\delta n\left(  \mathbf{r},t\right)  $ around the equilibrium
density $n_{TF}\left(  \mathbf{r}\right)  \equiv n_{TF}\left(
\mathbf{r;\lambda}\right)  $ is to parametrize the density fluctuations
$\delta n\left(  \mathbf{r},t\right)  $ in terms of a \emph{displacement
}vectorfield $\mathbf{\eta}\left(  \mathbf{r},t\right)  $ and in terms of a
\emph{dilatation }amplitude vector $\mathbf{\zeta}\left(  t\right)  $:%
\begin{equation}
\delta n\left(  \mathbf{r},t\right)  =n_{TF}\left[  \mathbf{r}+\mathbf{\eta
}\left(  \mathbf{r},t\right)  ;\mathbf{\lambda+\zeta}\left(  t\right)
\right]  -n_{TF}\left(  \mathbf{r;\lambda}\right)
\label{parametrization of density fluctuation 0}%
\end{equation}
Neglecting small higher order terms regarding the size of the amplitudes
$\left\vert \mathbf{\eta}\right\vert $ and $\left\vert \mathbf{\zeta
}\right\vert $ , a general density fluctuation $\delta n\left(  \mathbf{r}%
,t\right)  $ around the groundstate of the BEC cloud is then
\begin{align}
b  &  \in\left\{  x,y,z\right\}
\label{parametrization of density fluctuation I}\\
\delta n\left(  \mathbf{r},t\right)   &  =\sum_{b}\left[  \frac{\partial
n_{TF}\left(  \mathbf{r;\lambda}\right)  }{\partial r_{b}}\eta_{b}\left(
\mathbf{r},t\right)  +\frac{\partial n_{TF}\left(  \mathbf{r;\lambda}\right)
}{\partial\lambda_{b}}\zeta_{b}\left(  t\right)  \right]  +o\left(  \left\vert
\mathbf{\eta}\right\vert ^{2}+\left\vert \mathbf{\zeta}\right\vert ^{2}\right)
\nonumber
\end{align}
It follows directly from atom number conservation, and substituting into
(\ref{particle difference III}) the special case $K\left(  \mathbf{r}%
,\mathbf{r}^{\prime}\right)  \equiv1$ , that
\begin{equation}
\int_{\mathbb{D}_{TF}}d^{3}r^{\prime}\delta n\left(  \mathbf{r}^{\prime
},t\right)  =0 \label{constraint density fluctuation}%
\end{equation}
Upon insertion of (\ref{parametrization of density fluctuation I}) into
(\ref{constraint density fluctuation}) , and using the theorem of
Gau\ss ,\ we see that the displacement vectorfield $\mathbf{\eta}\left(
\mathbf{r},t\right)  $ is necessarily a \emph{solenoidal} vectorfield:%
\begin{equation}
\mbox{div}\mathbf{\eta}\left(  \mathbf{r},t\right)  =0
\label{solenoidal constraint  Ia}%
\end{equation}
Making a partial integration this property of $\mathbf{\eta}\left(
\mathbf{r},t\right)  $ implies for the phase fluctuation
(\ref{fluctuation phase}):
\begin{align}
b  &  \in\left\{  x,y,z\right\} \label{fluctuation phase  I}\\
\hslash\frac{\partial}{\partial t}\delta s\left(  \mathbf{r},t\right)   &
=\sum_{b}\left[
\begin{array}
[c]{c}%
\int_{\mathbb{D}_{TF}}d^{3}r^{\prime}\ \eta_{b}\left(  \mathbf{r}^{\prime
},t\right)  \frac{\partial U\left(  \mathbf{r},\mathbf{r}^{\prime}\right)
}{\partial r_{b}^{\prime}}n_{TF}\left(  \mathbf{r}^{\prime}\mathbf{;\lambda
}\right) \\
\\
-\zeta_{b}\left(  t\right)  \frac{\partial}{\partial\lambda_{b}}%
\int_{\mathbb{D}_{TF}}d^{3}r^{\prime}\ U\left(  \mathbf{r},\mathbf{r}^{\prime
}\right)  n_{TF}\left(  \mathbf{r}^{\prime}\mathbf{;\lambda}\right)
\end{array}
\right] \nonumber
\end{align}
A simplification results for standard two body interaction forces among two
atoms, say at position $\mathbf{r}$ and $\mathbf{r}^{\prime}$ , that obey
Newton's law "actio = reactio":%
\begin{equation}
\frac{\partial U\left(  \mathbf{r},\mathbf{r}^{\prime}\right)  }{\partial
r_{b}^{\prime}}=-\frac{\partial U\left(  \mathbf{r},\mathbf{r}^{\prime
}\right)  }{\partial r_{b}} \label{actio=reactio}%
\end{equation}
Then a general density fluctuation $\delta n\left(  \mathbf{r},t\right)  $
around equilibrium, as parametrized by
(\ref{parametrization of density fluctuation I}) in terms of a solenoidal
displacement $\mathbf{\eta}\left(  \mathbf{r},t\right)  $ and a
\emph{dilatation }amplitude $\mathbf{\zeta}\left(  t\right)  $, is connected
to the time derivative of the fluctuation $\delta s\left(  \mathbf{r}%
,t\right)  $ of the Josephson phase by
\begin{align}
b  &  \in\left\{  x,y,z\right\} \label{fluctuation phase II}\\
\hslash\frac{\partial}{\partial t}\delta s\left(  \mathbf{r},t\right)   &
=-\sum_{b}\left[
\begin{array}
[c]{c}%
\frac{\partial}{\partial r_{b}}\int_{\mathbb{D}_{TF}}d^{3}r^{\prime}U\left(
\mathbf{r},\mathbf{r}^{\prime}\right)  n_{TF}\left(  \mathbf{r}^{\prime
}\mathbf{;\lambda}\right)  \ \eta_{b}\left(  \mathbf{r}^{\prime},t\right) \\
\\
+\zeta_{b}\left(  t\right)  \frac{\partial}{\partial\lambda_{b}}%
\int_{\mathbb{D}_{TF}}d^{3}r^{\prime}\ U\left(  \mathbf{r},\mathbf{r}^{\prime
}\right)  n_{TF}\left(  \mathbf{r}^{\prime}\mathbf{;\lambda}\right)
\end{array}
\right] \nonumber
\end{align}
Indeed, this exact representation of the quasiclassical hydrodynamic phase
fluctuations of a BEC for interactions $U\left(  \mathbf{r},\mathbf{r}%
^{\prime}\right)  $ that obey (\ref{actio=reactio}) represents a convenient
starting point for our analytic calculation of the collective density
oscillations of a spin-polarized dipolar interacting BEC in a trap.

Like the density fluctuations in
(\ref{expansion of density fluctuation in a polynomial basis}) we may also
expand the Cartesian components $\eta_{a}\left(  \mathbf{r},t\right)  $ of the
displacement vectorfield:%
\begin{align}
a,b,c  &  \in\left\{  x,y,z\right\}
\label{expansion displacement vectorfield  I}\\
\eta_{a}\left(  \mathbf{r},t\right)   &  =\eta_{a}^{\left(  0\right)  }\left(
t\right)  +\sum_{b}\eta_{ab}^{\left(  1\right)  }\left(  t\right)
r_{b}\ +\sum_{b,c}\eta_{abc}^{\left(  2\right)  }\left(  t\right)  r_{b}%
r_{c}....\nonumber
\end{align}
The zero order term $\eta_{a}^{\left(  0\right)  }\left(  t\right)  $
describes a \emph{homogeneous} displacement of the center of mass of a BEC
cloud in a trap. Actually, any system of particles that interact via two-body
forces obeying to (\ref{actio=reactio}), has the property that the motion
of the center of mass separates from the equations of motion of the other
degrees of freedom of the system. As a result the frequency of the dipole
modes of a trapped atom gas cloud is independent on any such interactions,
because the center of mass of the cloud moves like a single particle of mass
$Nm^{\star}$ in the external trap potential $V_{T}(\mathbf{r})$. For a
harmonic trap the frequencies of the dipole modes coincide therefore with the
bare frequencies $\omega_{a}$ of the trap. In experiments this feature is useful to measure and calibrate the trap frequencies.\\
The first order terms $\eta_{ab}^{\left(  1\right)  }\left(  t\right)  $ together with the displacement
amplitudes $\zeta_{b}\left(  t\right)  $ are connected to density oscillations
with $s$-wave and $d$-wave symmetry. Second order displacement amplitudes like
$\eta_{abc}^{\left(  2\right)  }\left(  t\right)  $ are connected to the
octupolar collective density excitations $\delta n_{abc}^{\left(  3\right)
}\left(  t\right)  $. These and even higher order modes we shall consider elsewhere.

It follows directly from (\ref{parametrization of density fluctuation I})
that only certain linear combinations of the first order displacement
amplitudes $\eta_{ab}^{\left(  1\right)  }\left(  t\right)  $ together with
the dilatation amplitudes $\zeta_{b}\left(  t\right)  $ couple to $s$-wave and
$d$-wave symmetry density oscillation amplitudes $\delta n^{\left(  0\right)
}\left(  t\right)  $ and $\delta n_{ab}^{\left(  2\right)  }\left(  t\right)
$, while the homogeneous zero order displacement amplitudes $\eta_{a}^{\left(
0\right)  }\left(  t\right)  $ couple to the dipole modes:
\begin{align}
& \label{density fluctuation II}\\
\delta n\left(  \mathbf{r},t\right)   &  =2n_{0}\cdot\left[
\begin{array}
[c]{c}%
\ \frac{1}{\lambda_{x}^{2}}\rho_{xx}(t)r_{x}^{2}\ +\frac{1}{\lambda_{y}^{2}%
}\rho_{yy}(t)r_{y}^{2}+\frac{1}{\lambda_{z}^{2}}\rho_{zz}(t)r_{z}^{2}\\
\\
-\frac{1}{2}\left(  1-\frac{r_{x}^{2}}{\lambda_{x}^{2}}-\frac{r_{y}^{2}%
}{\lambda_{y}^{2}}-\frac{r_{z}^{2}}{\lambda_{z}^{2}}\right)  \rho_{00}(t)\\
\\
+\sum_{a<b}\ \frac{1}{\lambda_{a}\lambda_{b}}\rho_{ab}(t)r_{a}r_{b}\\
\\
+\frac{1}{\lambda_{x}^{2}}\rho_{x}(t)r_{x}+\frac{1}{\lambda_{y}^{2}}\rho
_{y}(t)r_{y}+\frac{1}{\lambda_{z}^{2}}\rho_{z}(t)r_{z}%
\end{array}
\right] \nonumber
\end{align}
, where
\begin{align}
a,b  &  \in\left\{  x,y,z\right\} \label{quadrupolar density fluctuations}\\
\rho_{ab}(t)  &  =\delta_{ab}\left[  \frac{\zeta_{a}\left(  t\right)
}{\lambda_{a}}-\eta_{aa}^{\left(  1\right)  }(t)\right]  -\left(
1-\delta_{ab}\right)  \lambda_{a}\lambda_{b}\left[  \frac{1}{\lambda_{a}^{2}%
}\eta_{ab}^{\left(  1\right)  }(t)+\frac{1}{\lambda_{b}^{2}}\eta_{ba}^{\left(
1\right)  }(t)\right] \nonumber\\
\rho_{00}(t)  &  =\sum_{a}\rho_{aa}(t)\nonumber\\
\rho_{a}(t)  &  =-\eta_{a}^{\left(  0\right)  }(t)\nonumber
\end{align}
Particle number conservation implies the solenoidal constraint%
\begin{equation}
\eta_{xx}^{\left(  1\right)  }\left(  t\right)  +\eta_{yy}^{\left(  1\right)
}\left(  t\right)  +\eta_{zz}^{\left(  1\right)  }\left(  t\right)  =0
\label{solenoidal constraint  Ic}%
\end{equation}
, so that there follows immediately%
\begin{equation}
\rho_{00}(t)=\sum_{a}\rho_{aa}(t)=\frac{\zeta_{x}\left(  t\right)  }%
{\lambda_{x}}+\frac{\zeta_{y}\left(  t\right)  }{\lambda_{y}}+\frac{\zeta
_{z}\left(  t\right)  }{\lambda_{z}} \label{solenoidal constraint  Id}%
\end{equation}

In general, a BEC cloud may get excited by a combination of actions, involving
translations of the trap minimum, rotations of the trap axes, or changes of
the curvature of the trap. The homogeneous displacement amplitudes $\eta
_{a}^{\left(  0\right)  }\left(  t\right)  $ correspond to infinitesimal
translations of the position of the center of the BEC cloud. Thus, a density
oscillation with a dipolar $p$-wave symmetry proportional to $\rho_{a}(t)$ can
be excited by a translation of the minimum of the trap. Being mainly
interested in the effect of interactions on the collective modes, however, we
set in the following without loss of generality $\eta_{a}^{\left(  0\right)
}\left(  t\right)  =0$. With regard to the first order off diagonal
displacement amplitudes $\eta_{ab}^{\left(  1\right)  }\left(  t\right)  $ we
easily identify anti-symmetric displacement amplitudes $\eta_{ab}^{\left(
1\right)  }\left(  t\right)  =-\eta_{ba}^{\left(  1\right)  }\left(  t\right)
$ as \emph{infinitesimal rotations} around a rotation axis perpendicular to
the $r_{a}r_{b}$-plane, while symmetric off diagonal amplitudes $\eta
_{ab}^{\left(  1\right)  }\left(  t\right)  =\eta_{ba}^{\left(  1\right)
}\left(  t\right)  $ correspond to\emph{\ transverse shear}. So, in the
geometry under consideration density oscillations with a quadrupolar $d_{ab}%
$-symmetry proportional to $\rho_{ab}(t)$ can be excited by rotations or by
transversal shear of the trap. Density oscillations displaying an isotropic
$s-$wave symmetry proportional to $\rho_{00}(t)$, and also showing a
quadrupolar $d_{z}^{2}$-wave or $d_{x^{2}-y^{2}}$-wave symmetry proportional
to certain linear combinations of the diagonal amplitudes$\rho_{aa}(t)$, can
be excited by a sudden change of the curvature of the trap.

According to (\ref{fluctuation phase II}) the time derivative $\hslash
\frac{\partial}{\partial t}\delta s\left(  \mathbf{r},t\right)  $ of a phase
fluctuation associated with such a density fluctuation $\delta n\left(
\mathbf{r},t\right)  $ is given by%
\begin{align}
a,b,c  &  \in\left\{  x,y,z\right\} \label{fluctuation phase III}\\
\hslash\frac{\partial}{\partial t}\delta s\left(  \mathbf{r},t\right)   &
=-g_{s}\left\{
\begin{array}
[c]{c}%
\sum_{b,c}\eta_{bc}^{\left(  1\right)  }(t)\frac{\partial}{\partial r_{b}%
}\left[
\begin{array}
[c]{c}%
\left(  1-\varepsilon_{D}\right)  n_{TF}\left(  \mathbf{r}\right)  r_{c}\\
\\
-3\varepsilon_{D}\ \left(  \mathbf{m}\cdot\mathbf{\nabla}_{\mathbf{r}}\right)
^{2}\frac{1}{4\pi}\int_{\mathbb{D}_{TF}}d^{3}r^{\prime}\frac{1}{\left\vert
\mathbf{r}-\mathbf{r}^{\prime}\right\vert }n_{TF}\left(  \mathbf{r}^{\prime
}\right)  \ r_{c}^{\prime}%
\end{array}
\right] \\
\\
+\sum_{a}\zeta_{a}\left(  t\right)  \frac{\partial}{\partial\lambda_{a}%
}\left[
\begin{array}
[c]{c}%
\left(  1-\varepsilon_{D}\right)  n_{TF}\left(  \mathbf{r}\right) \\
\\
-3\varepsilon_{D}\ \left(  \mathbf{m}\cdot\mathbf{\nabla}_{\mathbf{r}}\right)
^{2}\frac{1}{4\pi}\int_{\mathbb{D}_{TF}}d^{3}r^{\prime}\frac{1}{\left\vert
\mathbf{r}-\mathbf{r}^{\prime}\right\vert }n_{TF}\left(  \mathbf{r}^{\prime
}\right)
\end{array}
\right]
\end{array}
\right\} \nonumber
\end{align}
The three-dimensional integrals over the Thomas-Fermi ellipsoid $\mathbb{D}%
_{TF}$ we express now as one-dimensional integrals using Chandrasekhar's
integrals (\ref{gravitational potential of special mass distribution I})
and (\ref{gravitational potential of special mass distribution II}):
\begin{equation}
\frac{1}{4\pi}\int_{\mathbb{D}_{TF}}d^{3}r^{\prime}\frac{1}{\left\vert
\mathbf{r}-\mathbf{r}^{\prime}\right\vert }n_{TF}\left(  \mathbf{r}^{\prime
}\right)  =n_{0}\Phi_{1}\left(  \mathbf{r}\right)
\label{potential function I}%
\end{equation}
, and also%
\begin{align}
& \label{first moment of TF density}\\
\frac{1}{4\pi}\int_{\mathbb{D}_{TF}}d^{3}r^{\prime}\frac{1}{\left\vert
\mathbf{r}-\mathbf{r}^{\prime}\right\vert }n_{TF}\left(  \mathbf{r}^{\prime
}\right)  \ r_{c}^{\prime}  &  =\mathbf{-}\;n_{0}\frac{\lambda_{c}^{2}}%
{4}\frac{\partial}{\partial r_{c}}\Phi_{2}\left(  \mathbf{r}\right) \nonumber
\end{align}
The crucial trick to proof this representation for the first moment of the
Thomas-Fermi density profile $n_{TF}\left(  \mathbf{r}\right)  $ is to use the
identity
\begin{equation}
n_{TF}\left(  \mathbf{r}\right)  r_{c}=\left(  -n_{0}\frac{\lambda_{c}^{2}}%
{4}\frac{\partial}{\partial r_{c}}\right)  \left(  1-\frac{r_{x}^{2}}%
{\lambda_{x}^{2}}-\frac{r_{y}^{2}}{\lambda_{y}^{2}}-\frac{r_{z}^{2}}%
{\lambda_{z}^{2}}\right)  ^{2} \label{first moment identity}%
\end{equation}
One finds then upon partial integration a surface integral over the boundary
$\partial\mathbb{D}_{TF}$ , and a volume integral over the Thomas-Fermi domain
$\mathbb{D}_{TF}$. However, the surface integral vanishes identically taking
into account that $n_{TF}\left(  \mathbf{r}^{\prime}\right)  \equiv0$ for
$\mathbf{r}^{\prime}\in\partial\mathbb{D}_{TF}$. So only the volume integral
contributes, confirming the result (\ref{first moment of TF density}).

The wave equation (\ref{wave equation}) for the density fluctuations we
rewrite now%
\begin{equation}
0=\frac{\partial^{2}}{\partial t^{2}}\delta n\left(  \mathbf{r},t\right)
+\frac{1}{m^{\star}}\left\{
\begin{array}
[c]{c}%
n_{TF}\left(  \mathbf{r}\right)  \nabla_{r}^{2}\hslash\frac{\partial}{\partial
t}\delta s\left(  \mathbf{r},t\right) \\
\\
+\sum_{a}\frac{\partial n_{TF}\left(  \mathbf{r}\right)  }{\partial r_{a}%
}\frac{\partial}{\partial r_{a}}\hslash\frac{\partial}{\partial t}\delta
s\left(  \mathbf{r},t\right)
\end{array}
\right\}  \label{wave equation II}%
\end{equation}
To calculate the term $\nabla_{r}^{2}\hslash\frac{\partial}{\partial t}\delta
s\left(  \mathbf{r},t\right)  $ we need
\begin{align}
l=1,2  &  ,...\label{Laplace identity of potential functions}\\
\nabla_{r}^{2}\Phi_{l}\left(  \mathbf{r}\right)   &  =-\left(  1-\frac
{r_{x}^{2}}{\lambda_{x}^{2}}-\frac{r_{y}^{2}}{\lambda_{y}^{2}}-\frac{r_{z}%
^{2}}{\lambda_{z}^{2}}\right)  ^{l}\nonumber
\end{align}
Straightforward calculations lead to%
\begin{align}
&  \nabla_{r}^{2}\hslash\frac{\partial}{\partial t}\delta s\left(
\mathbf{r},t\right) \label{Laplace phase fluctuation}\\
&  =-4g_{s}n_{0}\left\{
\begin{array}
[c]{c}%
\left(  1-\varepsilon_{D}\right)  \left[  \frac{1}{\lambda_{x}^{2}}\rho
_{xx}\left(  t\right)  +\frac{1}{\lambda_{y}^{2}}\rho_{yy}\left(  t\right)
+\frac{1}{\lambda_{z}^{2}}\rho_{zz}\left(  t\right)  \right] \\
\\
+3\varepsilon_{D}\left[  \frac{\cos^{2}\left(  \vartheta_{0}\right)  }%
{\lambda_{z}^{2}}\rho_{zz}\left(  t\right)  +\frac{\sin^{2}\left(
\vartheta_{0}\right)  }{\lambda_{x}^{2}}\rho_{xx}\left(  t\right)  +\frac
{\sin\left(  2\vartheta_{0}\right)  }{2\lambda_{x}\lambda_{z}}\rho_{xz}\left(
t\right)  \right] \\
\\
\mathbf{+}\left[
\begin{array}
[c]{c}%
\frac{1-\varepsilon_{D}}{2}\left(  \frac{1}{\lambda_{x}^{2}}+\frac{1}%
{\lambda_{y}^{2}}+\frac{1}{\lambda_{z}^{2}}\right) \\
+\frac{3\varepsilon_{D}}{2}\left(  \frac{\cos^{2}\left(  \vartheta_{0}\right)
}{\lambda_{z}^{2}}+\frac{\sin^{2}\left(  \vartheta_{0}\right)  }{\lambda
_{x}^{2}}\right)
\end{array}
\right]  \rho_{00}\left(  t\right)
\end{array}
\right\} \nonumber
\end{align}
Likewise we obtain%
\begin{align}
a,b,c  &  \in\left\{  x,y,z\right\} \label{div (n grad ds/dt))}\\
&  \sum_{a}\frac{\partial n_{TF}\left(  \mathbf{r}\right)  }{\partial r_{a}%
}\frac{\partial}{\partial r_{a}}\hslash\frac{\partial}{\partial t}\delta
s\left(  \mathbf{r},t\right) \nonumber\\
&  =-4n_{0}^{2}g_{s}\cdot\left[
\begin{array}
[c]{c}%
\left(  1-\varepsilon_{D}\right)  \sum_{a,c}\frac{r_{a}}{\lambda_{a}^{2}%
}\left[  \frac{1}{\lambda_{a}^{2}}\eta_{ac}^{\left(  1\right)  }\left(
t\right)  +\frac{1}{\lambda_{c}^{2}}\eta_{ca}^{\left(  1\right)  }\left(
t\right)  \right]  r_{c}\\
\\
-\frac{3}{2}\varepsilon_{D}\ \sum_{a,b,c}\frac{r_{a}}{4\lambda_{a}^{2}}%
\frac{\partial}{\partial r_{a}}\left[  \eta_{bc}^{\left(  1\right)  }\left(
t\right)  \lambda_{c}^{2}\frac{\partial^{2}}{\partial r_{b}\partial r_{c}%
}\left(  \mathbf{m}\cdot\mathbf{\nabla}_{\mathbf{r}}\right)  ^{2}\Phi
_{2}\left(  \mathbf{r}\right)  \right] \\
\\
-\left(  1-\varepsilon_{D}\right)  \sum_{a,b}\frac{\zeta_{b}\left(  t\right)
}{\lambda_{b}}\left(  1+2\delta_{ab}\right)  \frac{r_{a}^{2}}{\lambda_{a}^{4}%
}\\
\\
+\frac{3}{2}\varepsilon_{D}\sum_{a,b}\frac{r_{a}}{\lambda_{a}^{2}}%
\frac{\partial}{\partial r_{a}}\left[  \zeta_{b}\left(  t\right)  \left(
-\frac{1}{\lambda_{b}}+\frac{\partial}{\partial\lambda_{b}}\right)  \left(
\mathbf{m}\cdot\mathbf{\nabla}_{\mathbf{r}}\right)  ^{2}\Phi_{1}\left(
\mathbf{r}\right)  \right]
\end{array}
\right] \nonumber
\end{align}
A glance at Chandrasekhar's representation
(\ref{gravitational potential of special mass distribution II}) for the
potential functions $\Phi_{l}\left(  \mathbf{r}\right)  $ of inhomogenous
ellipsoids reveals, that for a point $\mathbf{r}$ inside the ellipsoid
$\mathbb{D}_{TF}$ the potential function for $l=1$ is a $4$-th order
multinomial in the variables $\left\{  1,r_{x}^{2},r_{y}^{2},r_{z}%
^{2}\right\}  $ with coefficients proportional to the index
integrals $I_{a}$ and $I_{ab}$ , and for $l=2$ it is a $6$-th order
multinomial with coefficients proportional to the index
integrals $I_{a}$ , $I_{ab}$ and $I_{abc}$ (see appendix \ref{appendixA}). The task to
calculate the term (\ref{div (n grad ds/dt))}) is therefore reduced to
calculate linear combinations of derivatives of certain multinomials in the
variables $\left\{  1,r_{x}^{2},r_{y}^{2},r_{z}^{2}\right\}  $ :
\begin{align}
a,b,c  &  \in\left\{  x,y,z\right\}  \label{first row  of  div (n grad ds/dt)}%
\\
&  \sum_{a}\frac{r_{a}}{4\lambda_{a}^{2}}\frac{\partial}{\partial r_{a}}%
\sum_{b,c}\eta_{bc}^{\left(  1\right)  }\left(  t\right)  \lambda_{c}%
^{2}\ \frac{\partial^{2}}{\partial r_{b}\partial r_{c}}\left(  \mathbf{m}%
\cdot\mathbf{\nabla}_{\mathbf{r}}\right)  ^{2}\Phi_{2}\left(  \mathbf{r}%
\right) \nonumber\\
& \nonumber\\
&  =-\left[
\begin{array}
[c]{c}%
F_{xx}\left(  t\right)  r_{x}^{2}+F_{yy}\left(  t\right)  r_{y}^{2}%
+F_{zz}\left(  t\right)  r_{z}^{2}\\
+F_{xy}\left(  t\right)  r_{x}r_{y}+F_{yz}\left(  t\right)  r_{y}r_{z}%
+F_{xz}\left(  t\right)  r_{x}r_{z}%
\end{array}
\right] \nonumber
\end{align}
and
\begin{align}
a,b,c  &  \in\left\{  x,y,z\right\} \label{second row  of  div( n grad ds/dt)}%
\\
&  \sum_{a}\frac{r_{a}}{\lambda_{a}^{2}}\frac{\partial}{\partial r_{a}}%
\sum_{b}\zeta_{b}\left(  t\right)  \left(  -\frac{1}{\lambda_{b}}%
+\frac{\partial}{\partial\lambda_{b}}\right)  \left(  \mathbf{m}%
\cdot\mathbf{\nabla}_{\mathbf{r}}\right)  ^{2}\Phi_{1}\left(  \mathbf{r}%
\right) \nonumber\\
& \nonumber\\
&  =-\left[
\begin{array}
[c]{c}%
G_{xx}\left(  t\right)  r_{x}^{2}+G_{yy}\left(  t\right)  r_{y}^{2}%
+G_{zz}\left(  t\right)  r_{z}^{2}\\
+G_{xy}\left(  t\right)  r_{x}r_{y}+G_{yz}\left(  t\right)  r_{y}r_{z}%
+G_{xz}\left(  t\right)  r_{x}r_{z}%
\end{array}
\right] \nonumber\\
& \nonumber\\
G_{xy}  &  \equiv0\equiv G_{yz}\nonumber
\end{align}
In terms of the triple index integrals (see appendix \ref{appendixA})
\begin{align}
& \label{triple index integrals}\\
I_{abc}\left(  \lambda_{x},\lambda_{y},\lambda_{z}\right)   &  =\lambda
_{x}\lambda_{y}\lambda_{z}\int_{0}^{\infty}\frac{du}{\sqrt{\left(  \lambda
_{x}^{2}+u\right)  \left(  \lambda_{y}^{2}+u\right)  \left(  \lambda_{z}%
^{2}+u\right)  }}\frac{1}{\left(  \lambda_{a}^{2}+u\right)  \left(
\lambda_{b}^{2}+u\right)  \left(  \lambda_{c}^{2}+u\right)  }\nonumber\\
a,b,c  &  \in\left\{  x,y,z\right\} \nonumber
\end{align}
the coefficients $F_{ab}(t)$ are determined as linear combinations of the
displacement fluctuation amplitudes $\eta_{ab}^{\left(  1\right)  }(t)$ , and
the coefficients $G_{ab}(t)$ are determined as linear combination of the
dilatation fluctuation amplitudes $\zeta_{a}(t)$. To evaluate the gradient
terms in the wave equation (\ref{wave equation II}), however, only the
differences $G_{ab}(t)-F_{ab}(t)$ are needed, which can be represented as
linear combinations of the fluctuation amplitudes $\rho_{ab}\left(  t\right)
$ defined in (\ref{quadrupolar density fluctuations}). Explicit expressions
for $G_{ab}(t)-F_{ab}(t)$ in terms of the triple index integrals $I_{abc}$ are
presented in the appendix \ref{appendixC}.

Altogether we find for the gradient part of (\ref{wave equation II}) the
following second order multinomial in the variables $\left\{  1,\frac{r_{x}%
}{\lambda_{x}},\frac{r_{y}}{\lambda_{y}},\frac{r_{z}}{\lambda_{z}}\right\}  $
:
\begin{align}
a  &  \in\left\{  x,y,z\right\}
\label{spatial derivative terms of wave equation}\\
&  \frac{1}{m^{\star}}\sum_{a}\left[  \frac{\partial n_{TF}\left(
\mathbf{r}\right)  }{\partial r_{a}}\frac{\partial}{\partial r_{a}}%
\hslash\frac{\partial}{\partial t}\delta s\left(  \mathbf{r},t\right)
+n_{TF}\left(  \mathbf{r}\right)  \frac{\partial^{2}}{\partial r_{a}^{2}%
}\hslash\frac{\partial}{\partial t}\delta s\left(  \mathbf{r},t\right)
\right] \nonumber\\
& \nonumber\\
&  =\frac{4n_{0}^{2}g_{s}}{m^{\star}}\left[  -w_{00}\left(  t\right)
\cdot1+\sum_{a}w_{aa}\left(  t\right)  \cdot\frac{r_{a}^{2}}{\lambda_{a}^{2}%
}+\sum_{a<b}w_{ab}\left(  t\right)  \cdot\frac{r_{a}r_{b}}{\lambda_{a}%
\lambda_{b}}\right] \nonumber
\end{align}
For the coefficient proportional to unity we find
\begin{align}
& \label{coefficient of unity derivative term of wave equation}\\
w_{00}\left(  t\right)   &  =\left\{
\begin{array}
[c]{c}%
\left(  1-\varepsilon_{D}\right)  \left[  \frac{1}{\lambda_{x}^{2}}\rho
_{xx}\left(  t\right)  +\frac{1}{\lambda_{y}^{2}}\rho_{yy}\left(  t\right)
+\frac{1}{\lambda_{z}^{2}}\rho_{zz}\left(  t\right)  \right] \\
\\
+3\varepsilon_{D}\left[  \frac{\cos^{2}\left(  \vartheta_{0}\right)  }%
{\lambda_{z}^{2}}\rho_{zz}\left(  t\right)  +\frac{\sin^{2}\left(
\vartheta_{0}\right)  }{\lambda_{x}^{2}}\rho_{xx}\left(  t\right)  +\frac
{\sin\left(  2\vartheta_{0}\right)  }{2\lambda_{x}\lambda_{z}}\rho_{xz}\left(
t\right)  \right] \\
\\
\mathbf{+}\left[  \frac{1-\varepsilon_{D}}{2}\left(  \frac{1}{\lambda_{x}^{2}%
}+\frac{1}{\lambda_{y}^{2}}+\frac{1}{\lambda_{z}^{2}}\right)  +\frac
{3\varepsilon_{D}}{2}\left(  \frac{\cos^{2}\left(  \vartheta_{0}\right)
}{\lambda_{z}^{2}}+\frac{\sin^{2}\left(  \vartheta_{0}\right)  }{\lambda
_{x}^{2}}\right)  \right]  \rho_{00}\left(  t\right)
\end{array}
\right\} \nonumber
\end{align}
The coefficients of the diagonal terms $\frac{r_{a}^{2}}{\lambda_{a}^{2}}$ for
$a\in\left\{  x,y,z\right\}  $ in
(\ref{spatial derivative terms of wave equation}) are
\begin{align}
& \label{diagonal coefficients spatial derivative term of wave equation}\\
w_{aa}\left(  t\right)   &  =w_{00}\left(  t\right)  +\left(  1-\varepsilon
_{D}\right)  \frac{1}{\lambda_{a}^{2}}\ \left[  2\rho_{aa}\left(  t\right)
+\rho_{00}\left(  t\right)  \right]  +\frac{3}{2}\varepsilon_{D}\lambda
_{a}^{2}\left[  G_{aa}(t)-F_{aa}(t)\right] \nonumber
\end{align}
, while the coefficients of the off diagonal terms $\frac{r_{a}r_{b}}%
{\lambda_{a}\lambda_{b}}$ for $a,b\in\left\{  x,y,z\right\}  $ and $a<b$ are
\begin{align}
& \label{off diagonal coefficients of derivative term of wave equation}\\
w_{ab}\left(  t\right)   &  =\left(  1-\varepsilon_{D}\right)  \left(
\frac{1}{\lambda_{a}^{2}}+\frac{1}{\lambda_{b}^{2}}\right)  \rho_{ab}\left(
t\right)  +\frac{3}{2}\varepsilon_{D}\lambda_{a}\lambda_{b}\left[
G_{ab}(t)-F_{ab}\left(  t\right)  \right] \nonumber
\end{align}

It follows directly from what has been said that the right hand side of
(\ref{wave equation II}) represents a second order quadratic form:
\begin{equation}
0\overset{!}{=}\left\{
\begin{array}
[c]{c}%
-\left[  \frac{1}{2}\frac{\partial^{2}}{\partial t^{2}}\rho_{00}%
(t)+\frac{2n_{0}g_{s}}{m^{\star}}w_{00}\left(  t\right)  \right]  \cdot1\\
+\sum_{a}\left[  \frac{\partial^{2}}{\partial t^{2}}\rho_{aa}(t)+\frac{1}%
{2}\frac{\partial^{2}}{\partial t^{2}}\rho_{00}(t)+\frac{2n_{0}g_{s}}%
{m^{\star}}w_{aa}\left(  t\right)  \right]  \cdot\frac{r_{a}^{2}}{\lambda
_{a}^{2}}\\
+\sum_{a<b}\left[  \frac{\partial^{2}}{\partial t^{2}}\rho_{ab}(t)+\frac
{2n_{0}g_{s}}{m^{\star}}w_{ab}\left(  t\right)  \right]  \cdot\frac{r_{a}%
r_{b}}{\lambda_{a}\lambda_{b}}%
\end{array}
\right\}
\label{system of ordinary differential equations for quadrupolar density fluctuation}%
\end{equation}
Equating the coefficients of the linearly independent basis functions
$1,\frac{r_{a}^{2}}{\lambda_{a}^{2}},\frac{r_{a}r_{b}}{\lambda_{a}\lambda_{b}%
}$ for $a,b\in\left\{  x,y,z\right\}  $ to zero leads to a set of seven
coupled ordinary differential equations for the sought fluctuation amplitudes
$\rho_{ab}(t)$.

As a matter of fact, the equation for the variable $\rho_{00}(t)$ is obsolete,
because the solenoidal constraint (\ref{solenoidal constraint Id}) implies
\begin{equation}
\rho_{00}(t)=\rho_{xx}(t)+\rho_{yy}(t)+\rho_{zz}(t) \label{Q00  II}%
\end{equation}
This is consistent because certain identities obeyed by the triple index
integrals $I_{abc}$ imply the following sum rule (see appendix \ref{appendixB}):%
\begin{equation}
\sum_{a}w_{aa}\left(  t\right)  =5w_{00}\left(  t\right)  \label{sum rule}%
\end{equation}
Indeed, adding the differential equations for the diagonal fluctuation
amplitudes proportional to $\frac{r_{a}^{2}}{\lambda_{a}^{2}}$ leads
immediately to
\begin{equation}
\frac{1}{2}\frac{\partial^{2}}{\partial t^{2}}\rho_{00}(t)+\frac{2n_{0}g_{s}%
}{m^{\star}}w_{00}\left(  t\right)  =0 \label{prefactor of unity}%
\end{equation}
Consequently the derivative term $\frac{\partial^{2}}{\partial t^{2}}\rho
_{00}(t)$ and the term $w_{00}(t)$ in the differential equations
(\ref{system of ordinary differential equations for quadrupolar density fluctuation}%
) for the diagonal density fluctuation amplitudes $\rho_{aa}(t)$ cancel each
other. We obtain finally the following six differential equations for six
fluctuation amplitudes $\rho_{ab}(t)$:
\begin{align}
a,b  &  \in\left\{  x,y,z\right\}
\label{differential equations second harmonic fluctuation amplitudes}\\
0  &  =\frac{\partial^{2}}{\partial t^{2}}\rho_{aa}(t)+\frac{2n_{0}g^{\left(
s\right)  }}{m^{\star}}\left[
\begin{array}
[c]{c}%
\left(  1-\varepsilon_{D}\right)  \frac{1}{\lambda_{a}^{2}}\ \left[
3\rho_{aa}\left(  t\right)  +\sum_{b\neq a}\rho_{bb}\left(  t\right)  \right]
\\
\\
+\frac{3}{2}\varepsilon_{D}\lambda_{a}^{2}\left[  G_{aa}(t)-F_{aa}(t)\right]
\end{array}
\right] \nonumber\\
& \nonumber\\
0  &  =\frac{\partial^{2}}{\partial t^{2}}\rho_{ab}(t)+\frac{2n_{0}g^{\left(
s\right)  }}{m^{\star}}\left[
\begin{array}
[c]{c}%
\left(  1-\varepsilon_{D}\right)  \left(  \frac{1}{\lambda_{a}^{2}}+\frac
{1}{\lambda_{b}^{2}}\right)  \rho_{ab}\left(  t\right) \\
\\
+\frac{3}{2}\varepsilon_{D}\lambda_{a}\lambda_{b}\left[  G_{ab}(t)-F_{ab}%
\left(  t\right)  \right]
\end{array}
\right] \nonumber
\end{align}
To determine the eigenmodes of oscillation we look for a solution of the form%
\begin{equation}
\rho_{ab}(t)=\widehat{\rho}_{ab}\left(  \Omega\right)  \cos\left(  \Omega
t+\delta_{\Omega}\right)  \label{ansatz eigen modes}%
\end{equation}
, where $\Omega$ is the eigenfrequency of the mode, and $\widehat{\rho}%
_{ab}\left(  \Omega\right)  $ denotes a component of the associated
eigenvector:
\begin{equation}
\frac{2n_{0}g^{\left(  s\right)  }}{m^{\star}}\left[
\begin{array}
[c]{cccccc}%
C_{xx,xx} & C_{xx,yy} & C_{xx,zz} & C_{xx,xz} & 0 & 0\\
C_{yy,xx} & C_{yy,yy} & C_{yy,zz} & C_{yy,xz} & 0 & 0\\
C_{zz,xx} & C_{zz,yy} & C_{zz,zz} & C_{zz,xz} & 0 & 0\\
C_{xz,xx} & C_{xz,yy} & C_{xz,zz} & C_{xz,xz} & 0 & 0\\
0 & 0 & 0 & 0 & C_{yz,yz} & C_{yz,xy}\\
0 & 0 & 0 & 0 & C_{xy,yz} & C_{xy,xy}%
\end{array}
\right]  \left[
\begin{array}
[c]{c}%
\widehat{\rho}_{xx}\left(  \Omega\right) \\
\widehat{\rho}_{yy}\left(  \Omega\right) \\
\widehat{\rho}_{zz}\left(  \Omega\right) \\
\widehat{\rho}_{xz}\left(  \Omega\right) \\
\widehat{\rho}_{yz}\left(  \Omega\right) \\
\widehat{\rho}_{xy}\left(  \Omega\right)
\end{array}
\right]  =\Omega^{2}\left[
\begin{array}
[c]{c}%
\widehat{\rho}_{xx}\left(  \Omega\right) \\
\widehat{\rho}_{yy}\left(  \Omega\right) \\
\widehat{\rho}_{zz}\left(  \Omega\right) \\
\widehat{\rho}_{xz}\left(  \Omega\right) \\
\widehat{\rho}_{yz}\left(  \Omega\right) \\
\widehat{\rho}_{xy}\left(  \Omega\right)
\end{array}
\right]  \label{collective mode eigenvalue problem}%
\end{equation}

We find it convenient to eliminate the interaction constant using
(\ref{selfconsistent Ic}):
\begin{equation}
2n_{0}\frac{g^{\left(  s\right)  }}{m^{\star}}=\frac{\omega_{y}^{2}\lambda
_{y}^{2}}{1-\varepsilon_{D}+\frac{3\varepsilon_{D}}{2}\lambda_{y}^{2}\left[
\cos^{2}\left(  \vartheta_{0}\right)  I_{zy}+\sin^{2}\left(  \vartheta
_{0}\right)  I_{xy}\right]  }
\label{elimination interaction constant using selfconsistency}%
\end{equation}
It follows upon inspection of the coupled differential equations
(\ref{differential equations second harmonic fluctuation amplitudes}) for the
coefficients $C_{ab,cd}$ explicit expressions, which are listed in the
appendix \ref{appendixC}.

The collective modes associated with the $4\times4$-sub matrix in
(\ref{collective mode eigenvalue problem}) describe small amplitude
oscillations of the density, which are linear combinations of $s$-wave and
quadrupolar $d_{x^{2}-y^{2}}$ , $d_{z^{2}}$ and $d_{xz}$-waves, while the
modes associated with the $2\times2$-sub matrix describe small amplitude
oscillations of the density consisting solely of combinations of quadrupolar
$d_{yz}$-and $d_{xy}$-waves:%
\begin{equation}
\delta n_{\Omega}\left(  \mathbf{r},t\right)  =2n_{0}\cdot\left[
\begin{array}
[c]{c}%
\ \frac{1}{\lambda_{x}^{2}}\widehat{\rho}_{xx}(\Omega)r_{x}^{2}\ +\frac
{1}{\lambda_{y}^{2}}\widehat{\rho}_{yy}(\Omega)r_{y}^{2}+\frac{1}{\lambda
_{z}^{2}}\widehat{\rho}_{zz}(\Omega)r_{z}^{2}\\
\\
-\frac{1}{2}\left(  1-\frac{r_{x}^{2}}{\lambda_{x}^{2}}-\frac{r_{y}^{2}%
}{\lambda_{y}^{2}}-\frac{r_{z}^{2}}{\lambda_{z}^{2}}\right)  \widehat{\rho
}_{00}(\Omega)\\
\\
+\sum_{a<b}\ \frac{1}{\lambda_{a}\lambda_{b}}\widehat{\rho}_{ab}(\Omega
)r_{a}r_{b}%
\end{array}
\right]  \cos\left(  \Omega t+\delta_{\Omega}\right)
\label{small amplitude eigen modes of density}%
\end{equation}
By construction there holds%
\[
\int_{\mathbb{D}_{TF}}d^{3}r\ \delta n_{\Omega}\left(  \mathbf{r},t\right)
=0
\]
It is instructive to visualize the spatial dependence of the eigenmodes of
small amplitude oscillations of the density by plotting the
\emph{instantaneous} boundary of the BEC cloud when only a single mode with
eigenfrequency $\Omega$ is excited. This instantaneous boundary is implicitely
defined as the surface
\begin{equation}
n_{\Omega}(\mathbf{r},t)=n_{TF}\left(  \mathbf{r}\right)  +\delta n_{\Omega
}\left(  \mathbf{r},t\right)  \overset{!}{=}0 \label{instantaneous surface}%
\end{equation}

Finally, let us discuss which collective modes can be excited by changing the
trap potential, always keeping the trap strictly harmonic while changing it.
It follows directly from (\ref{quadrupolar density fluctuations}):
\begin{align}
a,b  &  \in\left\{  x,y,z\right\}
\label{coupling of modes to rotation, shear and curvature of the trap}\\
\widehat{\rho}_{ab}(\Omega)  &  =\delta_{ab}\left[  \frac{\widehat{\zeta}%
_{a}\left(  \Omega\right)  }{\lambda_{a}}-\widehat{\eta}_{aa}^{\left(
1\right)  }(\Omega)\right]  -\left(  1-\delta_{ab}\right)  \lambda_{a}%
\lambda_{b}\left[  \frac{1}{\lambda_{a}^{2}}\widehat{\eta}_{ab}^{\left(
1\right)  }(\Omega_{ab})+\frac{1}{\lambda_{b}^{2}}\widehat{\eta}_{ba}^{\left(
1\right)  }(\Omega_{ab})\right] \nonumber
\end{align}
Sudden changes of the trap potential may excite collective density
oscillations around the quantum degenerate groundstate. For example, a
rotation around a trap axis perpendicular to the $ab$-plane, as represented by
the anti-symmetric components of the tensor $\widehat{\eta}_{ab}^{\left(
1\right)  }$ , or changes of the curvature of the trap, as represented by
dilatation amplitudes $\widehat{\zeta}_{a}\left(  \Omega\right)  $, but also
transversal or longitudinal shear movements of the trap, as represented by the
symmetric components of the tensor $\widehat{\eta}_{ab}^{\left(  1\right)  }$,
can be used to excite the collective modes (\ref{density fluctuation II}) of
the particle density of a trapped BEC cloud. A sudden translation of the
origin of a \emph{harmonic} trap, on the other hand, only excites the dipole
modes with eigenfrequency $\Omega_{a}\equiv\omega_{a}$. It should be noted,
that during these collective oscillations of a spin polarized dipolar BEC
cloud, as described by the density fluctuation
(\ref{small amplitude eigen modes of density}), the atoms always keep the
orientation of their magnetic moments strictly along the external polarizing
field $\mathbf{B}$.

\subsection{Pure Scissors Modes and Mixed Monopole- Quadrupole Excitations.}

Consider a harmonic trap where the principal axis $\mathbf{e}%
_{z,T}$ of the trap is aligned parallel to the polarizing external field
$\mathbf{B}$, i.e. $\vartheta_{T}=0$. In this case the off diagonal matrix
elements $C_{xy,yz}$ ,$C_{yz,xy}$ , $C_{xz,aa}$ and $C_{aa,xz}$ vanish
identically for arbitrary strength $\varepsilon_{D}$ of the dipole interaction
parameter. There follows then a simpler eigenvalue problem determining the
eigenmodes of the small amplitude density oscillations:
\begin{align}
\vartheta_{T}  &  =0\label{collective mode eigenvalue problem II}\\
& \nonumber\\
\frac{2n_{0}g^{\left(  s\right)  }}{m^{\star}}\left[
\begin{array}
[c]{cccccc}%
C_{xx,xx} & C_{xx,yy} & C_{xx,zz} & 0 & 0 & 0\\
C_{yy,xx} & C_{yy,yy} & C_{yy,zz} & 0 & 0 & 0\\
C_{zz,xx} & C_{zz,yy} & C_{zz,zz} & 0 & 0 & 0\\
0 & 0 & 0 & C_{xz,xz} & 0 & 0\\
0 & 0 & 0 & 0 & C_{yz,yz} & 0\\
0 & 0 & 0 & 0 & 0 & C_{xy,xy}%
\end{array}
\right]  \left[
\begin{array}
[c]{c}%
\widehat{\rho}_{xx}\left(  \Omega\right) \\
\widehat{\rho}_{yy}\left(  \Omega\right) \\
\widehat{\rho}_{zz}\left(  \Omega\right) \\
\widehat{\rho}_{xz}\left(  \Omega\right) \\
\widehat{\rho}_{yz}\left(  \Omega\right) \\
\widehat{\rho}_{xy}\left(  \Omega\right)
\end{array}
\right]   &  =\Omega^{2}\left[
\begin{array}
[c]{c}%
\widehat{\rho}_{xx}\left(  \Omega\right) \\
\widehat{\rho}_{yy}\left(  \Omega\right) \\
\widehat{\rho}_{zz}\left(  \Omega\right) \\
\widehat{\rho}_{xz}\left(  \Omega\right) \\
\widehat{\rho}_{yz}\left(  \Omega\right) \\
\widehat{\rho}_{xy}\left(  \Omega\right)
\end{array}
\right] \nonumber
\end{align}

Three modes with indices $a\neq b$ display a pure quadrupolar $d_{xz}$-,
$d_{yz}$- and $d_{xy}$-symmetry. Also there exists a mixed symmetry coupling
between two basis functions with $d$-wave symmetry and one basis function with
$s$-wave symmetry. This is reminiscent of the symmetry of the discrete group
$D_{4h}$ lifting the $5$-fold degeneracy of the $l=2$ spherical harmonics into
three one-dimensional manifolds, namely $A_{1g}$ , $B_{1g}$ and $B_{2g}$- ,
and a two-dimensional $E_{g}$-manifold. The one-dimensional (trivial)
representation of the isotropic basis function with $s$-wave symmetry we refer
to as $a_{1g}$.

In the geometry under consideration the $E_{g}$-manifold is spanned by basis
functions with $d_{yz}$- and $d_{xy}$-symmetry, while $B_{2g}$ is spanned by a
single basis function with $d_{xz}$-symmetry, and $B_{1g}$ is spanned by a
single basis function with $d_{x^{2}-y^{2}}$-symmetry. The one-dimensional
manifold $A_{1g}$ represents a fixed linear combination of basis elements with
$d_{z^{2}}$- and $s$-wave symmetry. So, the upper $3\times3$ block in
(\ref{collective mode eigenvalue problem II}) describes a coupling between
members of the$\ a_{1g}$ , $A_{1g}$-and $B_{1g}$- manifolds. For $\omega
_{x}=\omega_{y}\neq\omega_{z}$ there exists a pure $B_{1g}$-mode, and two
coupled modes with mixed $a_{1g}$- and $A_{1g}$-symmetry.

The eigenfrequencies of the $B_{2g}$-and $E_{g}$-modes are obtained from the
diagonal matrix elements $C_{xz,xz}$ , $C_{yz,yz}$ and $C_{xy,xy}$ taking the
limit $\vartheta_{0}\rightarrow0$:
\begin{align}
& \nonumber\\
\Omega_{xz}^{2}  &  =\omega_{y}^{2}\left(  \frac{\lambda_{y}^{2}}{\lambda
_{x}^{2}}+\frac{\lambda_{y}^{2}}{\lambda_{z}^{2}}\right)  \frac{\left(
1-\varepsilon_{D}\right)  +\frac{9}{2}\varepsilon_{D}\cdot\frac{\lambda
_{x}^{2}}{\lambda_{z}^{2}}\overline{I}_{xzz}}{1-\varepsilon_{D}+\frac
{3\varepsilon_{D}}{2}\frac{\lambda_{y}^{2}}{\lambda_{z}^{2}}\overline{I}_{zy}%
}\nonumber\\
& \label{collective modes scissors modes II}\\
\Omega_{yz}^{2}  &  =\omega_{y}^{2}\left(  1+\frac{\lambda_{y}^{2}}%
{\lambda_{z}^{2}}\right)  \frac{\left(  1-\varepsilon_{D}\right)  +\frac{9}%
{2}\varepsilon_{D}\frac{\lambda_{y}^{2}}{\lambda_{z}^{2}}\overline{I}_{yzz}%
}{1-\varepsilon_{D}+\frac{3\varepsilon_{D}}{2}\frac{\lambda_{y}^{2}}%
{\lambda_{z}^{2}}\overline{I}_{zy}}\nonumber\\
& \nonumber\\
\Omega_{xy}^{2}  &  =\omega_{y}^{2}\left(  1+\frac{\lambda_{y}^{2}}%
{\lambda_{x}^{2}}\right)  \frac{\left(  1-\varepsilon_{D}\right)  +\frac{3}%
{2}\varepsilon_{D}\frac{\lambda_{x}^{2}}{\lambda_{z}^{2}}\frac{\lambda_{y}%
^{2}}{\lambda_{z}^{2}}\overline{I}_{xyz}}{1-\varepsilon_{D}+\frac
{3\varepsilon_{D}}{2}\frac{\lambda_{y}^{2}}{\lambda_{z}^{2}}\overline{I}_{zy}%
}\nonumber
\end{align}
The spatial variation of the associated density fluctuation of these modes is
purely two-dimensional
\begin{align}
a,b  &  \in\left\{  x,y,z\right\} \label{scissors modes III}\\
a  &  \neq b\nonumber\\
& \nonumber\\
\widehat{\rho}_{a^{\prime}b^{\prime}}\left(  \Omega_{ab}\right)   &
=\delta_{aa^{\prime}}\delta_{bb^{\prime}}\nonumber\\
& \nonumber\\
\delta n_{\Omega_{ab}}\left(  \mathbf{r},t\right)   &  =2n_{0}\widehat{\rho
}_{ab}(\Omega_{ab})\frac{r_{a}r_{b}}{\lambda_{a}\lambda_{b}}\cos\left(
\Omega_{ab}t+\delta_{\Omega_{ab}}\right) \nonumber
\end{align}
In the limit $\varepsilon_{D}\rightarrow0$ it is found that $\frac{\lambda
_{a}}{\lambda_{b}}\rightarrow\frac{\omega_{b}}{\omega_{a}}$ . Then one obtains
for a BEC without dipole-dipole interactions confined inside a harmonic trap:
\begin{align}
a  &  \neq b\label{scissors modes I}\\
\lim\limits_{\varepsilon_{D}\rightarrow0}\Omega_{ab}  &  =\sqrt{\omega_{a}%
^{2}+\omega_{b}^{2}}\nonumber
\end{align}
These are the so called "scissors" modes first predicted by Gu\'{e}ry-Odelin
and Stringari \cite{Stringari I}, and then observed in experiment
\cite{Cozzini}, \cite{Marago }.

In order to specify conditions that enable excitation of the scissors modes
(\ref{scissors modes III}) for a dipolar BEC cloud confined in a harmonic trap
we point out, that the components $\widehat{\rho}_{a^{\prime}b^{\prime}%
}\left(  \Omega_{ab}\right)  =\delta_{aa^{\prime}}\delta_{bb^{\prime}}\ $of
the eigenvectors of the respective modes are connected to the off diagonal
displacement amplitudes $\widehat{\eta}_{ab}^{\left(  1\right)  }$, see
(\ref{quadrupolar density fluctuations}), by
\begin{align}
a  &  \neq b\label{scissors modes II}\\
\widehat{\rho}_{ab}\left(  \Omega_{ab}\right)   &  =-\lambda_{a}\lambda
_{b}\left[  \frac{1}{\lambda_{a}^{2}}\widehat{\eta}_{ab}^{\left(  1\right)
}(\Omega_{ab})+\frac{1}{\lambda_{b}^{2}}\widehat{\eta}_{ba}^{\left(  1\right)
}(\Omega_{ab})\right] \nonumber
\end{align}
For an infinitesimal rotation of the BEC cloud around one of its symmetry
axes, say $\mathbf{e}_{c,0}=\mathbf{e}_{a,0}\wedge$ $\mathbf{e}_{b,0}$ , the
associated displacement amplitude is anti-symmetric, $\widehat{\eta}%
_{ab}^{\left(  1\right)  }=-\widehat{\eta}_{ba}^{\left(  1\right)  }$. So one
recognizes immediately that in the highly symmetric case $\vartheta_{T\ }=0$ a
scissors mode with amplitude $\widehat{\rho}_{ab}\left(  \Omega_{ab}\right)  $
cannot be excited by a rotation around a principal axis of the BEC cloud
perpendicular to the $ab$-plane, if the semi-axes $\lambda_{a}$ and
$\lambda_{b}$ of the BEC cloud in that plane are equal, i.e. $\lambda
_{a}=\lambda_{b}$. However, even then a scissors mode may get excited by a
sudden \emph{tranverse shear} movement of the trap as described by a symmetric
displacement amplitude $\widehat{\eta}_{ab}^{\left(  1\right)  }=\widehat
{\eta}_{ba}^{\left(  1\right)  }$. If the BEC is confined inside a harmonic
trap with tri-axial symmetry, one may always excite the scissors modes
$\Omega_{ab}$ by a sudden infinitesimal rotation of the trap potential around
a symmetry axis perpendicular to the respective $ab$-plane.

Let us now discuss the coupled modes corresponding to the $3\times3$-sub block
in (\ref{collective mode eigenvalue problem II}). These are small amplitude
oscillations of the density that are linear combinations of the three diagonal
amplitudes $\widehat{\rho}_{aa}(\Omega)$. In the limit $\varepsilon
_{D}\rightarrow0$ the corresponding eigenfrequencies and eigenvectors of the
triplet of coupled modes can be obtained solving a cubic equation for the
frequencies $\Omega^{\left(  0\right)  }$:
\begin{align}
& \label{collective modes for  eps_D=0}\\
\left[
\begin{array}
[c]{ccc}%
3\omega_{x}^{2} & \omega_{x}^{2} & \omega_{x}^{2}\\
\omega_{y}^{2} & 3\omega_{y}^{2} & \omega_{y}^{2}\\
\omega_{z}^{2} & \omega_{z}^{2} & 3\omega_{z}^{2}%
\end{array}
\right]  \left[
\begin{array}
[c]{c}%
\widehat{\rho}_{xx}\left(  \Omega^{\left(  0\right)  }\right) \\
\widehat{\rho}_{yy}\left(  \Omega^{\left(  0\right)  }\right) \\
\widehat{\rho}_{zz}\left(  \Omega^{\left(  0\right)  }\right)
\end{array}
\right]   &  =\left[  \Omega^{\left(  0\right)  }\right]  ^{2}\left[
\begin{array}
[c]{c}%
\widehat{\rho}_{xx}\left(  \Omega^{\left(  0\right)  }\right) \\
\widehat{\rho}_{yy}\left(  \Omega^{\left(  0\right)  }\right) \\
\widehat{\rho}_{zz}\left(  \Omega^{\left(  0\right)  }\right)
\end{array}
\right] \nonumber
\end{align}
One easily sees, that for a tri-axial trap the eigenmodes of this triplet are
mixtures of basis functions with isotropic $s$-wave and quadrupolar $d_{z^{2}%
}$-wave and $d_{x^{2}-y^{2}}$-wave symmetry, respectively.

When the harmonic trap has a uniaxial (cylindrical) symmetry, $\omega_{z}%
\neq\ \omega_{y}=\omega_{x}=\omega_{\perp}$ , simple analytic formulas for the
eigenfrequencies and eigenmodes of the density oscillations of a BEC cloud can
be derived from (\ref{collective modes for eps_D=0}) that apply for
$\varepsilon_{D}=0$. One easily obtains for the three eigenfrequencies
$\Omega_{x^{2}-y^{2}}^{\left(  0\right)  }$ , $\Omega_{s\ \ }^{\left(
0\right)  }$ , $\Omega_{z^{2}\ \ }^{\left(  0\right)  }$ well known results
first derived by Stringari \cite{Stringari II}. In the appendix \ref{appendixD} we present a
detailed discussion of these modes as a function of the anisotropy ratio%
\begin{equation}
\nu=\frac{\omega_{z}}{\omega_{\perp}}
\label{cylindrical trap anisotropy ratio}%
\end{equation}

\subsection{Spherical Harmonic Trap}

For the special case of a \emph{spherical} harmonic trap, say with trap frequency
$\omega_{a}\equiv\ \omega$, setting $\lambda_{a}^{\left(
0\right)  }\equiv\Lambda$, we immediately find from (\ref{collective modes for  eps_D=0}) for a BEC without dipole-dipole interaction (see appendix \ref{appendixD}):
\begin{align}
\Omega &  =\Omega_{x^{2}-y^{2}}^{\left(  0\right)  }=\sqrt{2}\omega
_{\ }\label{spherical trap eps_D=0  collective mode d_x^2-y^2-wave}\\
\delta n_{\Omega}\left(  \mathbf{r},t\right)   &  =2n_{0}\cos\left(  \Omega
t+\delta_{\Omega}\right)  \frac{r_{x}^{2}-r_{y}^{2}}{\Lambda^{2}}\nonumber
\end{align}%
\begin{align}
\Omega &  \equiv\Omega_{+\ \ }^{\left(  0\right)  }=\sqrt{5}\omega
\label{spherical trap eps_D=0  collective mode s-wave}\\
\delta n_{\Omega}\left(  \mathbf{r},t\right)   &  =n_{0}\cos\left(  \Omega
t+\delta_{\Omega}\right)  \left(  5\frac{r_{x}^{2}+r_{y}^{2}+r_{z}^{2}%
}{\Lambda^{2}}-3\right) \nonumber
\end{align}%
\begin{align}
\Omega &  =\Omega_{-\ \ }^{\left(  0\right)  }=\sqrt{2}\omega
\label{spherical trap eps_D=0  collective mode d_z^2 -wave}\\
\delta n_{\Omega}\left(  \mathbf{r},t\right)   &  =n_{0}\cos\left(  \Omega
t+\delta_{\Omega}\right)  \frac{2r_{z}^{2}-r_{x}^{2}-r_{y}^{2}}{\Lambda^{2}%
}\nonumber
\end{align}
So for $\varepsilon_{D}=0$, a BEC
cloud confined inside a harmonic \emph{spherical} trap may get excited as an
$s$-wave breather mode with frequency $\Omega_{+\ \ }^{\left(  0\right)
}=\sqrt{5}\omega$ , or as a quintuplet of degenerate modes with quadrupolar
symmetry and frequency $\Omega_{x^{2}-y^{2}}^{\left(  0\right)  }%
=\Omega_{-\ \ }^{\left(  0\right)  }=\Omega_{xz}^{\left(  0\right)  }%
=\Omega_{yz}^{\left(  0\right)  }=\Omega_{xy}^{\left(  0\right)  }=$ $\sqrt
{2}\omega$, namely three scissors modes with $d_{xz}$-, $d_{yz}$-, $d_{xy}%
$-symmetry, and two modes with $d_{x^{2}-y^{2}}$- and $d_{z^{2}}$-symmetry.

Next we take into account the effect of the dipole-dipole interaction.
According to (\ref{elongation I}) for $\varepsilon_{D}>0$ the groundstate
of a spin-polarized dipolar BEC cloud confined in a \emph{spherical} trap with
trap frequency $\omega_{a}\equiv\ \omega$ displays uni-axial symmetry along
the direction of the magnetic field $\mathbf{B}$, so that $\lambda_{x}%
=\lambda_{y}<\lambda_{z}$. Let us check if for $\varepsilon_{D}\neq0$ the
modes of a dipolar BEC cloud confined in a \emph{spherical} trap are
qualitatively similar to the aforementioned collective modes of a BEC cloud
without dipole-dipole interaction, $\varepsilon_{D}=0$, for the case of a
prolate trap with \emph{cylindrical }symmetry: $\omega_{z}<\ \omega_{y}%
=\omega_{x}$.

Indeed, for a spherical trap with trap frequency $\omega$ we have
$\vartheta_{0}=0$ , so that all matrix elements in
(\ref{collective mode eigenvalue problem II}) can be expressed in terms of the
following expressions:
\begin{align}
A  &  =\frac{1-\varepsilon_{D}+\frac{3}{2}\varepsilon_{D}\cdot\frac
{\lambda_{y}^{4}}{\lambda_{z}^{4}}\overline{I}_{yyz}}{1-\varepsilon_{D}%
+\frac{3\varepsilon_{D}}{2}\frac{\lambda_{y}^{2}}{\lambda_{z}^{2}}\overline
{I}_{zy}}\label{matrix elements spherical trap eps_D>0}\\
B  &  =\frac{1-\varepsilon_{D}+\frac{9}{2}\varepsilon_{D}\cdot\frac
{\lambda_{y}^{2}}{\lambda_{z}^{2}}\overline{I}_{yzz}}{1-\varepsilon_{D}%
+\frac{3\varepsilon_{D}}{2}\frac{\lambda_{y}^{2}}{\lambda_{z}^{2}}\overline
{I}_{zy}}\nonumber\\
C  &  =\frac{1-\varepsilon_{D}+\frac{15}{2}\varepsilon_{D}\overline{I}_{zzz}%
}{1-\varepsilon_{D}+\frac{3\varepsilon_{D}}{2}\frac{\lambda_{y}^{2}}%
{\lambda_{z}^{2}}\overline{I}_{zy}}\nonumber
\end{align}
It follows then from (\ref{collective mode eigenvalue problem II}):
\begin{align}
\Omega_{xz}^{2}  &  =\Omega_{yz}^{2}=\left(  1+\frac{\lambda_{y}^{2}}%
{\lambda_{z}^{2}}\right)  \omega^{2}%
B\label{scissors modes vs. eps_D  for spherical trap}\\
& \nonumber\\
\Omega_{xy}^{2}  &  =2\omega^{2}A\nonumber\\
& \nonumber\\
\widehat{\rho}_{a^{\prime}b^{\prime}}\left(  \Omega_{ab}\right)   &
=\delta_{aa^{\prime}}\delta_{bb^{\prime}}\nonumber
\end{align}
So, the scissors modes with $d_{xz}$-and $d_{yz}$-symmetry remain degenerate.

For $\varepsilon_{D}\neq0$ the $3\times3$ sub block in
(\ref{collective mode eigenvalue problem II}) represents a triplet of
coupled modes. For the case of a dipolar BEC confined in a \emph{spherical}
trap there follows
\begin{equation}
\omega^{2}\left[
\begin{array}
[c]{ccc}%
3A & A & B\\
A & 3A & B\\
\frac{\lambda_{y}^{2}}{\lambda_{z}^{2}}B & \frac{\lambda_{y}^{2}}{\lambda
_{z}^{2}}B & 3\frac{\lambda_{y}^{2}}{\lambda_{z}^{2}}C
\end{array}
\right]  \left[
\begin{array}
[c]{c}%
\widehat{\rho}_{xx}\left(  \Omega\right) \\
\widehat{\rho}_{yy}\left(  \Omega\right) \\
\widehat{\rho}_{zz}\left(  \Omega\right)
\end{array}
\right]  =\Omega^{2}\left[
\begin{array}
[c]{c}%
\widehat{\rho}_{xx}\left(  \Omega\right) \\
\widehat{\rho}_{yy}\left(  \Omega\right) \\
\widehat{\rho}_{zz}\left(  \Omega\right)
\end{array}
\right]  \label{s-wave & d_z^2&d_x^2-y^2 modes vs. eps_D  spherical trap}%
\end{equation}
It is easy to see that the mode with quadrupolar $d_{x^{2}-y^{2}}$ - symmetry
remains an exact eigenstate for $\varepsilon_{D}\neq0$:
\begin{align}
\Omega_{x^{2}-y^{2}}^{2}  &  =2\omega^{2}A=\Omega_{xy}^{2}
\label{eigenfrequency and eigen vector d_x^2-y^2  for arbitrary eps_D for spherical trap}%
\\
& \nonumber\\
\left[
\begin{array}
[c]{c}%
\widehat{\rho}_{xx}\left(  \Omega_{x^{2}-y^{2}}\right) \\
\widehat{\rho}_{yy}\left(  \Omega_{x^{2}-y^{2}}\right) \\
\widehat{\rho}_{zz}\left(  \Omega_{x^{2}-y^{2}}\right)
\end{array}
\right]   &  =\left[
\begin{array}
[c]{c}%
1\\
-1\\
0
\end{array}
\right] \nonumber
\end{align}
So, for $\varepsilon_{D}\neq0$ the quadrupolar modes with $d_{x^{2}-y^{2}}%
$-and $d_{xy}$-symmetry remain degenerate for the case of a spherical harmonic trap.

Next we show, that the isotropic breather mode of a dipolar BEC inside a
spherical trap with frequency $\omega$ is an eigenstate of the small amplitude
density oscillations of the BEC cloud, displaying an exact (!) $s$-wave
symmetry for any value of the dipole interaction strength $\varepsilon_{D}%
\neq0$ :
\begin{align}
\Omega_{s}^{2}  &  =5\omega^{2}
\label{eigenfrequency and eigen vector s-wave breather mode  for arbitrary eps_D for spherical trap}%
\\
& \nonumber\\
\left[
\begin{array}
[c]{c}%
\widehat{\rho}_{xx}\left(  \Omega_{s}\right) \\
\widehat{\rho}_{yy}\left(  \Omega_{s}\right) \\
\widehat{\rho}_{zz}\left(  \Omega_{s}\right)
\end{array}
\right]   &  =\left[
\begin{array}
[c]{c}%
1\\
1\\
1
\end{array}
\right] \nonumber
\end{align}
If this claim was correct then it should be true that
\begin{align}
4A+B  &  =5 \label{identities for breather mode in spherical trap for eps_D>0}%
\\
\frac{\lambda_{y}^{2}}{\lambda_{z}^{2}}\left(  2B+3C\right)   &  =5\nonumber
\end{align}
Indeed, making use of identities (\ref{identity index integrals III}%
),(\ref{identity index integrals IV}) obeyed by the triple index integrals
$I_{abc}$ we see that%
\begin{align}
4\frac{\lambda_{y}^{2}}{\lambda_{z}^{2}}\overline{I}_{yyz}+3\overline
{I}_{yzz}  &  =5\overline{I}_{zy}%
\label{identities for breather mode in spherical trap for eps_D>0    Ia}\\
& \nonumber\\
4A+B  &  =4\frac{1-\varepsilon_{D}+\frac{3}{2}\varepsilon_{D}\cdot
\frac{\lambda_{y}^{4}}{\lambda_{z}^{4}}\overline{I}_{yyz}}{1-\varepsilon
_{D}+\frac{3\varepsilon_{D}}{2}\frac{\lambda_{y}^{2}}{\lambda_{z}^{2}%
}\overline{I}_{zy}}+\frac{1-\varepsilon_{D}+\frac{9}{2}\varepsilon_{D}%
\cdot\frac{\lambda_{y}^{2}}{\lambda_{z}^{2}}\overline{I}_{yzz}}{1-\varepsilon
_{D}+\frac{3\varepsilon_{D}}{2}\frac{\lambda_{y}^{2}}{\lambda_{z}^{2}%
}\overline{I}_{zy}}\nonumber\\
&  =\frac{5\left(  1-\varepsilon_{D}\right)  +\frac{3}{2}\varepsilon_{D}%
\cdot\frac{\lambda_{y}^{2}}{\lambda_{z}^{2}}\left(  4\frac{\lambda_{y}^{2}%
}{\lambda_{z}^{2}}\overline{I}_{yyz}+3\overline{I}_{yzz}\right)
}{1-\varepsilon_{D}+\frac{3\varepsilon_{D}}{2}\frac{\lambda_{y}^{2}}%
{\lambda_{z}^{2}}\overline{I}_{zy}}\nonumber\\
&  =5\nonumber
\end{align}
, and also%
\begin{align}
5\overline{I}_{zzz}+2\frac{\lambda_{y}^{2}}{\lambda_{z}^{2}}\overline
{I}_{yzz}  &  =5\overline{I}_{zz}%
\label{identities for breather mode in spherical trap for eps_D>0    Ib}\\
& \nonumber\\
\frac{\lambda_{y}^{2}}{\lambda_{z}^{2}}\left(  2B+3C\right)   &
=2\frac{\lambda_{y}^{2}}{\lambda_{z}^{2}}\frac{1-\varepsilon_{D}+\frac{9}%
{2}\varepsilon_{D}\cdot\frac{\lambda_{y}^{2}}{\lambda_{z}^{2}}\overline
{I}_{yzz}}{1-\varepsilon_{D}+\frac{3\varepsilon_{D}}{2}\frac{\lambda_{y}^{2}%
}{\lambda_{z}^{2}}\overline{I}_{zy}}+3\frac{\lambda_{y}^{2}}{\lambda_{z}^{2}%
}\frac{1-\varepsilon_{D}+\frac{15}{2}\varepsilon_{D}\overline{I}_{zzz}%
}{1-\varepsilon_{D}+\frac{3\varepsilon_{D}}{2}\frac{\lambda_{y}^{2}}%
{\lambda_{z}^{2}}\overline{I}_{zy}}\nonumber\\
&  =\frac{\lambda_{y}^{2}}{\lambda_{z}^{2}}\frac{5\left(  1-\varepsilon
_{D}\right)  +\frac{9}{2}\varepsilon_{D}\left(  2\frac{\lambda_{y}^{2}%
}{\lambda_{z}^{2}}\overline{I}_{yzz}+5\overline{I}_{zzz}\right)
}{1-\varepsilon_{D}+\frac{3\varepsilon_{D}}{2}\frac{\lambda_{y}^{2}}%
{\lambda_{z}^{2}}\overline{I}_{zy}}\nonumber\\
&  =5\frac{\lambda_{y}^{2}}{\lambda_{z}^{2}}\frac{1-\varepsilon_{D}+\frac
{9}{2}\varepsilon_{D}\overline{I}_{zz}}{1-\varepsilon_{D}+\frac{3\varepsilon
_{D}}{2}\frac{\lambda_{y}^{2}}{\lambda_{z}^{2}}\overline{I}_{zy}}\nonumber\\
&  =5\nonumber
\end{align}
The last line follows because the selfconsistency equation
(\ref{selfconsistent IV b}) implies for the case of a \emph{spherical} trap:%
\begin{equation}
\frac{\lambda_{y}^{2}}{\lambda_{z}^{2}}=\frac{1-\varepsilon_{D}+\frac
{3\varepsilon_{D}}{2}\frac{\lambda_{y}^{2}}{\lambda_{z}^{2}}\overline{I}_{yz}%
}{1-\varepsilon_{D}+\frac{9\varepsilon_{D}}{2}\overline{I}_{zz}}
\label{identities for breather mode in spherical trap for eps_D>0    II}%
\end{equation}
Because for a finite value $\varepsilon_{D}>0$ a spin-polarized dipolar BEC
cloud confined in a spherical harmonic trap with frequency $\omega$ has the
shape of an uniaxial (prolate) ellipsoid orientated parallel to $\mathbf{B}$ ,
so that $\lambda_{x}=\lambda_{y}<\lambda_{z}$ , we find it remarkable that the
isotropic breather mode
(\ref{eigenfrequency and eigen vector s-wave breather mode for arbitrary eps_D for spherical trap}%
) remains an \emph{exact} eigenmode of the small amplitude density
fluctuations with $s$-wave symmetry, oscilllating at a \emph{constant}
frequency $\Omega_{s}=\sqrt{5}\omega$ that is \emph{independent} on the value
of the dipole interaction strength for $-\frac{1}{2}<\varepsilon_{D}<1$.

Knowledge of two eigenvalues is sufficient to determine the third one from the
trace of the coefficient matrix in
(\ref{s-wave & d_z^2&d_x^2-y^2 modes vs. eps_D spherical trap}):%
\begin{equation}
\Omega_{s}^{2}+\Omega_{x^{2}-y^{2}}^{2}+\Omega_{z^{2}}^{2}=6\omega^{2}\left(
A+\frac{\lambda_{y}^{2}}{2\lambda_{z}^{2}}C\right)
\label{trace identity 3x3 block matrix}%
\end{equation}
This leads for the eigenfrequency and the eigenvector of the density
oscillations with a predominant $d_{z^{2}}$-symmetry to the result:
\begin{align}
\Omega_{z^{2}}^{2}  &  =\omega^{2}\left(  4A+3\frac{\lambda_{y}^{2}}%
{\lambda_{z}^{2}}C-5\right)
\label{eigenfrequency and eigen vector d_z^ 2  for arbitrary eps_D for spherical trap}%
\\
& \nonumber\\
\left[
\begin{array}
[c]{c}%
\widehat{\rho}_{xx}\left(  \Omega_{z^{2}\ }\right) \\
\widehat{\rho}_{yy}\left(  \Omega_{z^{2}\ }\right) \\
\widehat{\rho}_{zz}\left(  \Omega_{z^{2}\ }\right)
\end{array}
\right]   &  =\left[
\begin{array}
[c]{c}%
-\frac{\lambda_{z}^{2}}{\lambda_{y}^{2}}\frac{5-4A}{2B}\\
\\
-\frac{\lambda_{z}^{2}}{\lambda_{y}^{2}}\frac{5-4A}{2B}\\
\\
1
\end{array}
\right] \nonumber
\end{align}

\begin{figure}[h]
\centering
\includegraphics[width=0.90\textwidth]{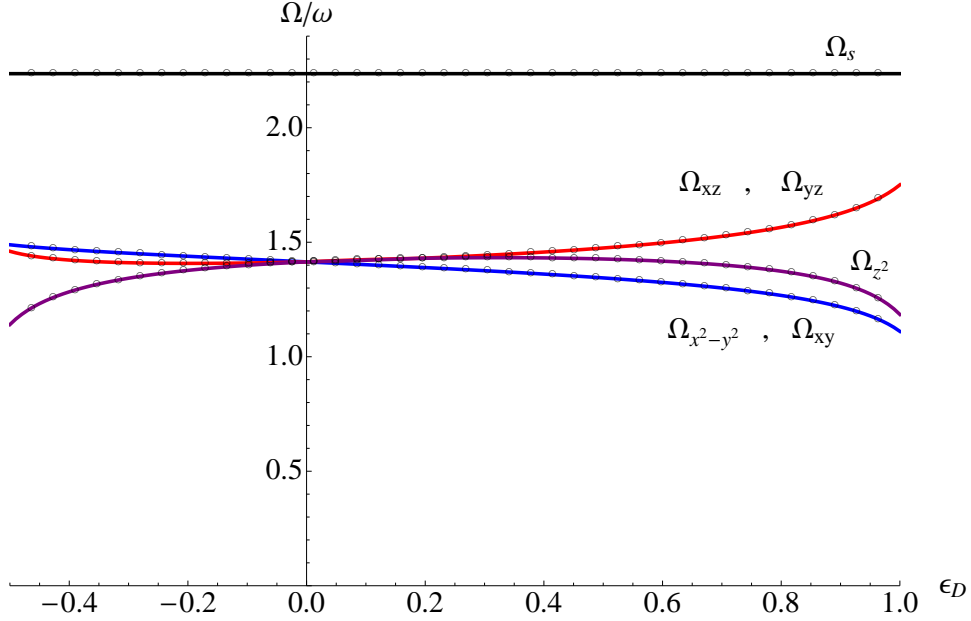}\caption{(Color online)
Eigenfrequencies $\Omega_{s}$ , $\Omega_{z^{2}}$ , $\Omega_{xz}=\Omega_{yz}$
and $\Omega_{xy}$ $=\Omega_{x^{2}-y^{2}}$ of small amplitude density
oscillations of BEC cloud vs. dipole interaction strength $\varepsilon_{D}$
for a \emph{spherical }harmonic trap with trap frequeny $\omega$.}%
\label{Fig. 6}%
\end{figure}

It follows from what has been said that the degeneracy of the small amplitude
collective modes of a dipolar BEC cloud confined in a spherical harmonic trap
is only partially lifted for $\varepsilon_{D}\neq0$. For a spherical trap the
modes with $d_{x^{2}-y^{2}}$-and with $d_{xy}$-symmetry, and also the modes
with $d_{yz}$- and $d_{xz}$-symmetry remain degenerate, irrespective of the
value of the dipole interaction $\varepsilon_{D}$. In Fig.\ref{Fig. 6} we plot
the collective mode frequencies $\Omega_{s}$ ,$\Omega_{z^{2}\ \ }$,
$\Omega_{xy}$ and $\Omega_{xz}$ vs. the interaction strength parameter
$\varepsilon_{D}$. For small $\left\vert \varepsilon_{D}\right\vert \ $the
splitting of the quadrupolar modes $\Omega_{z^{2}\ \ }$, $\Omega_{xy}$ and
$\Omega_{xz}$ is weak. Most remarkably, the breather mode $\Omega_{s}$
displays for $-\frac{1}{2}<\varepsilon_{D}<1$ an exact $s$-wave symmetry, the
eigenfrequency assuming a constant value $\Omega_{s}=\sqrt{5}\omega$, even though for $\varepsilon_D\neq 0$ the shape of the  groundstate is not isotropic.

The following reason can be given for the breather mode frequency of a dipolar
BEC being independent on the dipole interaction strength $\varepsilon_{D}$ for
an \emph{isotropic} harmonic trap. The microscopic Hamiltonian of a dipolar interacting gas cloud
consisting of $N$ atoms is
\begin{align}
\widehat{H}  &  =\widehat{H}_{kin}+\widehat{H}_{pot}+\widehat{H}%
_{int}\label{Hamiltonian of spin-polarized dipolar BEC in free space}\\
\widehat{H}_{kin}  &  =\sum_{n=1}^{N}\frac{1}{2m^{\star}}\sum_{a\in\left\{
x,y,z\right\}  }p_{a}^{\left(  n\right)  }p_{a}^{\left(  n\right)
}\nonumber\\
\widehat{H}_{pot}  &  =\sum_{n=1}^{N}\frac{m^{\star}\omega^{2}}{2}\sum
_{a\in\left\{  x,y,z\right\}  }r_{a}^{\left(  n\right)  }r_{a}^{\left(
n\right)  }\nonumber\\
\widehat{H}_{int}  &  =\frac{1}{2}\sum_{\substack{n,n^{\prime}=1\\n^{\prime
}\neq n}}^{N}U\left(  \mathbf{r}^{\left(  n\right)  },\mathbf{r}^{\left(
n^{\prime}\right)  }\right) \nonumber
\end{align}
The breather mode (or monopole mode) of small amplitude collective density
oscillations of such an atom cloud may get excited by a sudden change of the
curvature of the trap potential, say by changing the trap frequency
$\omega\rightarrow\omega+\delta\omega$. The associated excitation operator is
\begin{equation}
\delta\widehat{V}=m^{\star}\omega\delta\omega\sum_{n=1}^{N}\sum_{a\in\left\{
x,y,z\right\}  }r_{a}^{\left(  n\right)  }r_{a}^{\left(  n\right)  }
\label{excitation operator monopole mode}%
\end{equation}
It is important to realize, that the interaction potential $U\left(  \mathbf{r}^{\left(
n\right)  },\mathbf{r}^{\left(  n^{\prime}\right)}\right)$ for the spin-polarized dipolar BEC in (\ref{interaction I}) transforms under a
scaling transformation $\mathbf{r}\rightarrow\Lambda\mathbf{r}$ like a
homogeneous function with scaling degree $-3$:%
\begin{equation}
U\left(  \Lambda\mathbf{r}^{\left(  n\right)  },\Lambda\mathbf{r}^{\left(
n^{\prime}\right)  }\right)  =\Lambda^{-3}U\left(  \mathbf{r}^{\left(
n\right)  },\mathbf{r}^{\left(  n^{\prime}\right)  }\right)
\label{homogeneous function of degree -3}%
\end{equation}
Together with Newton's law of action and reaction (\ref{actio=reactio}) this implies
\begin{equation}
\left[  \left[  \delta\widehat{V},\widehat{H}\right]  ,\widehat{H}\right]
=2\hbar^{2}\omega\delta\omega\ \left(  2\widehat{H}_{pot}-2\widehat{H}%
_{kin}-3\widehat{H}_{int}\right)  \label{commutator I}%
\end{equation}
If $\Psi_{0}$ denotes the groundstate and $E_0$ the groundstate energy of the system under consideration, there
holds%
\begin{align}
0  &  =\left\langle \Psi_{0},\left(  \left[  \delta\widehat{V},\widehat
{H}\right]  E_{0}-E_{0}\left[  \delta\widehat{V},\widehat{H}\right]  \right)
\Psi_{0}\right\rangle \label{virial I}\\
&  =\left\langle \Psi_{0},\left[  \left[  \delta\widehat{V},\widehat
{H}\right]  ,\widehat{H}\right]  \ \Psi_{0}\right\rangle \nonumber
\end{align}
Inserting the double commutator (\ref{commutator I}) it is found, that the
full interaction energy of a dipolar interacting BEC in the groundstate is
proportional to a difference of kinetic and potential energy only:
\begin{equation}
\left\langle \widehat{H}_{int}\right\rangle _{\Psi_{0}}=\frac{2}%
{3}\left\langle \widehat{H}_{pot}\right\rangle _{\Psi_{0}}-\frac{2}%
{3}\left\langle \widehat{H}_{kin}\right\rangle _{\Psi_{0}} \label{virial II}%
\end{equation}
It should be emphasized, that if in (\ref{interaction I}) the scaling degree of the long ranged interaction (\ref{interaction  Imd}) under $\mathbf{r\rightarrow}%
\Lambda\,\mathbf{r}$ was different from the scaling degree $-3$ of the short ranged $s$-wave contact interaction (\ref{interaction  Is}), the derived virial identity (\ref{virial II}) would not apply!

Next we employ a well known sum rule \cite{Bohigas},\cite{Stringari II}
providing an upper bound for the low-lying excitation energies $E_{1}-E_{0}$
that can be excited by a hermitean perturbation operator $\delta\widehat{V}$
:
\begin{equation}
\left(  E_{1}-E_{0\ }\right)  ^{2}\leq\frac
{\left\langle \Psi_{0},\left[  \left[  \delta\widehat{V},\widehat{H}\right]
,\left[  \left[  \delta\widehat{V},\widehat{H}\right]  ,\widehat{H}\right]
\right]  \Psi_{0}\right\rangle }{\left\langle \Psi_{0},\left[  \delta
\widehat{V},\left[  \widehat{H},\delta\widehat{V}\right]  \right]  \Psi
_{0}\right\rangle } \label{sum rule upper bound I}%
\end{equation}
For the operator $\delta\widehat{V}$ exciting the breather mode, see (\ref{excitation operator monopole mode}), it is found%
\begin{equation}
\left[  \delta\widehat{V},\left[  \widehat{H},\delta\widehat{V}\right]
\right]  =8\hbar^{2}\left(  \delta\omega\right)  ^{2}\widehat{H}_{pot}
\label{commutator II}%
\end{equation}
and also%
\begin{align}
& \label{commutator III}\\
\left[  \left[  \delta\widehat{V},\widehat{H}\right]  ,\left[  \left[
\delta\widehat{V},\widehat{H}\right]  ,\widehat{H}\right]  \right]   &
=4\hbar^{4}\left(  \omega\delta\omega\right)  ^{2}\ \left(  4\widehat{H}%
_{pot}+4\widehat{H}_{kin}+9\widehat{H}_{int}\right) \nonumber
\end{align}

From what has been said there follows now for the frequency $\Omega_{s}$ of
the breather mode an upper bound:
\begin{align}
\left(  \hbar\Omega_{s\ }\right)  ^{2} &  \leq\left(
\hbar\omega\right)  ^{2}\frac{2\left\langle \widehat{H}_{pot}\right\rangle
_{\Psi_{0}}+2\left\langle \widehat{H}_{kin}\right\rangle _{\Psi_{0}}+\frac
{9}{2}\left\langle \widehat{H}_{int}\right\rangle _{\Psi_{0}}}{\left\langle
\widehat{H}_{pot}\right\rangle _{\Psi_{0}}}\label{sum rule upper bound II}\\
&  =\left(  \hbar\omega\right)  ^{2}\left(  5-\frac{\left\langle \widehat
{H}_{kin}\right\rangle _{\Psi_{0}}}{\left\langle \widehat{H}_{pot}%
\right\rangle _{\Psi_{0}}}\right)  \nonumber
\end{align}
For the optimized groundstate (\ref{Hartree Groundstate}) of a BEC, as
constructed from a solution to the Gross-Pitaevskii equation (\ref{GP}), the
ratio of kinetic to potential energy scales like
\begin{equation}
\frac{\left\langle \widehat{H}_{kin}\right\rangle _{\Psi_{0}}}{\left\langle
\widehat{H}_{pot}\right\rangle _{\Psi_{0}}}=o\left(  N^{-\frac{4}{5}}\right)
\label{ratio kinetic/potential  large N}%
\end{equation}
So, in the Thomas-Fermi approximation the derived upper bound for the breather
mode frequency is indeed independent on the strength of the dipole-dipole
interaction parameter $\varepsilon_{D}$. The fact, that this upper bound
actually coincides with the previously derived result $\Omega_{s}=\sqrt
{5}\omega$ , which was obtained solving the eigenvalue problem
(\ref{s-wave & d_z^2&d_x^2-y^2 modes vs. eps_D spherical trap}) for the small
amplitude collective modes of density oscillations, suggests that the spectral
weight of the mode is indeed exhausted by the specified excitation operator
$\delta\widehat{V}$ (\ref{excitation operator monopole mode}) of the monopole mode.

It is instructive to visualize the spatial variation of the associated density
eigenmodes $\delta n_{\Omega}\left(  \mathbf{r},t\right)  $ by plotting the
instantaneous surface of the BEC cloud as defined by
(\ref{instantaneous surface}). In Fig.7 and in Fig.8 these eigenmodes are
plotted at stroboscopic times $t=0$ , $t=\frac{\pi}{2\Omega}$ and $t=\frac
{\pi}{\Omega}$ , corresponding to maximal, zero and minimal deviation from the
boundary $\partial\mathbb{D}_{TF}$ of the groundstate cloud $\mathbb{D}_{TF}$
, respectively. The plots shown are based on a selfconsistent calculation of
the groundstate cloud for a dipole interaction strength parameter
$\varepsilon_{D}=0.7$ , assuming that the BEC cloud is confined inside a
\emph{spherical} harmonic trap with trap frequency $\omega$. The amplitudes of
the respective eigenmodes $\delta n_{\Omega}\left(  \mathbf{r},t\right)  $ of
the density fluctuation have been scaled by a suitable factor for each mode
separately to make the typical shapes better visible. The $s$-wave breather
mode is clearly distinguished in its appearance from the three characteristic
scissors modes with their $d_{xz}$ , $d_{yz}$ and $d_{xy}$-wave symmetry, and
also the $d_{x^{2}-y^{2}}$-wave and $d_{z}^{2}$-wave quadrupolar modes.

\clearpage

\begin{figure}[h]
\centering
\includegraphics[width=0.80\textwidth]{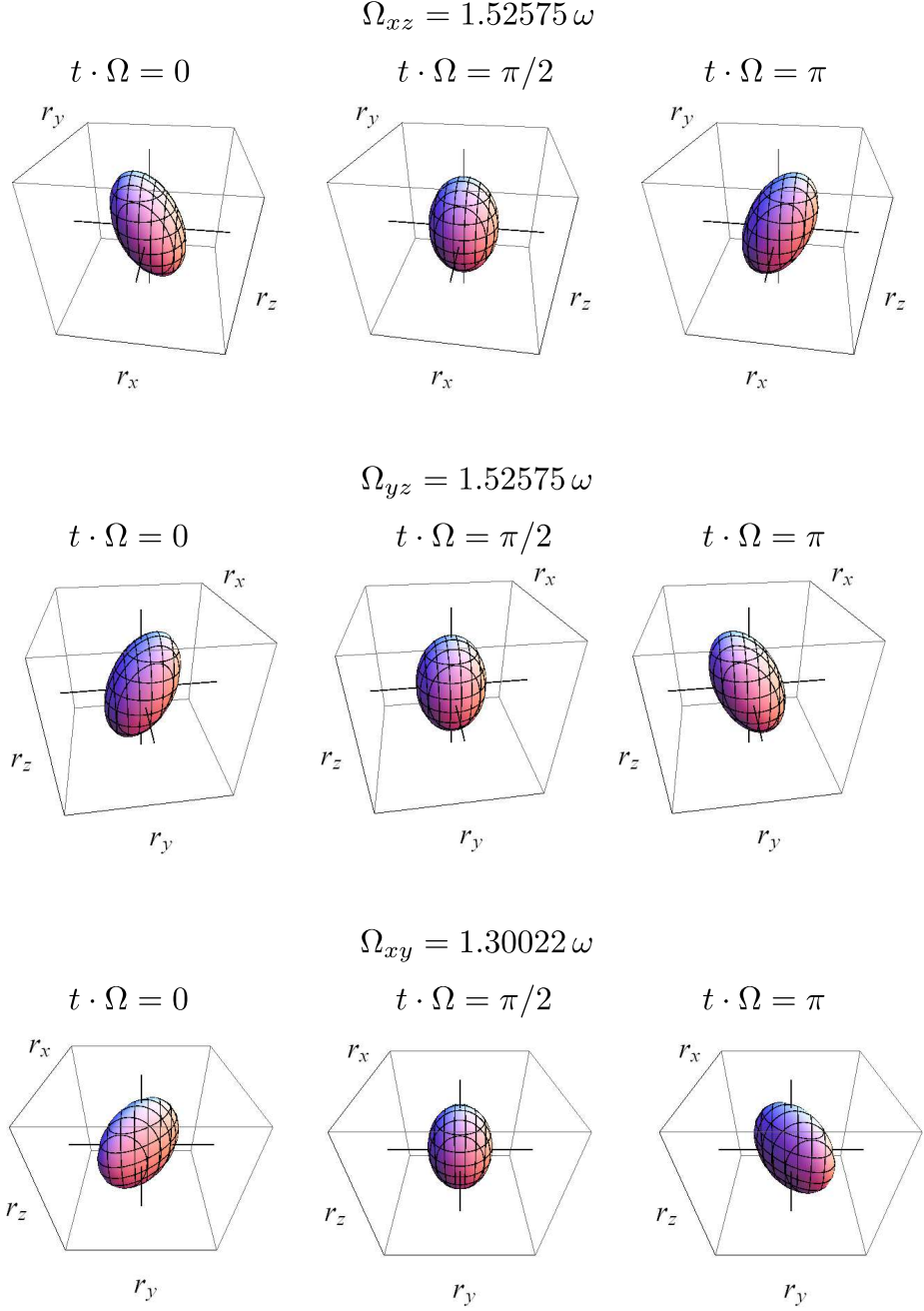}
\caption{(Color online) Visualization of density fluctuations $n_{\Omega}(\mathbf{r}%
,t)=n_{TF}(\mathbf{r})+\delta n_{\Omega}(\mathbf{r},t)$ of scissors modes for
dipolar BEC cloud confined inside a spherical trap for a dipole interaction
strength $\varepsilon_{D}=0.7$. }%
\label{visualization of scissors modes}%
\end{figure}

\clearpage\begin{figure}[h]
\centering
\includegraphics[width=0.80\textwidth]{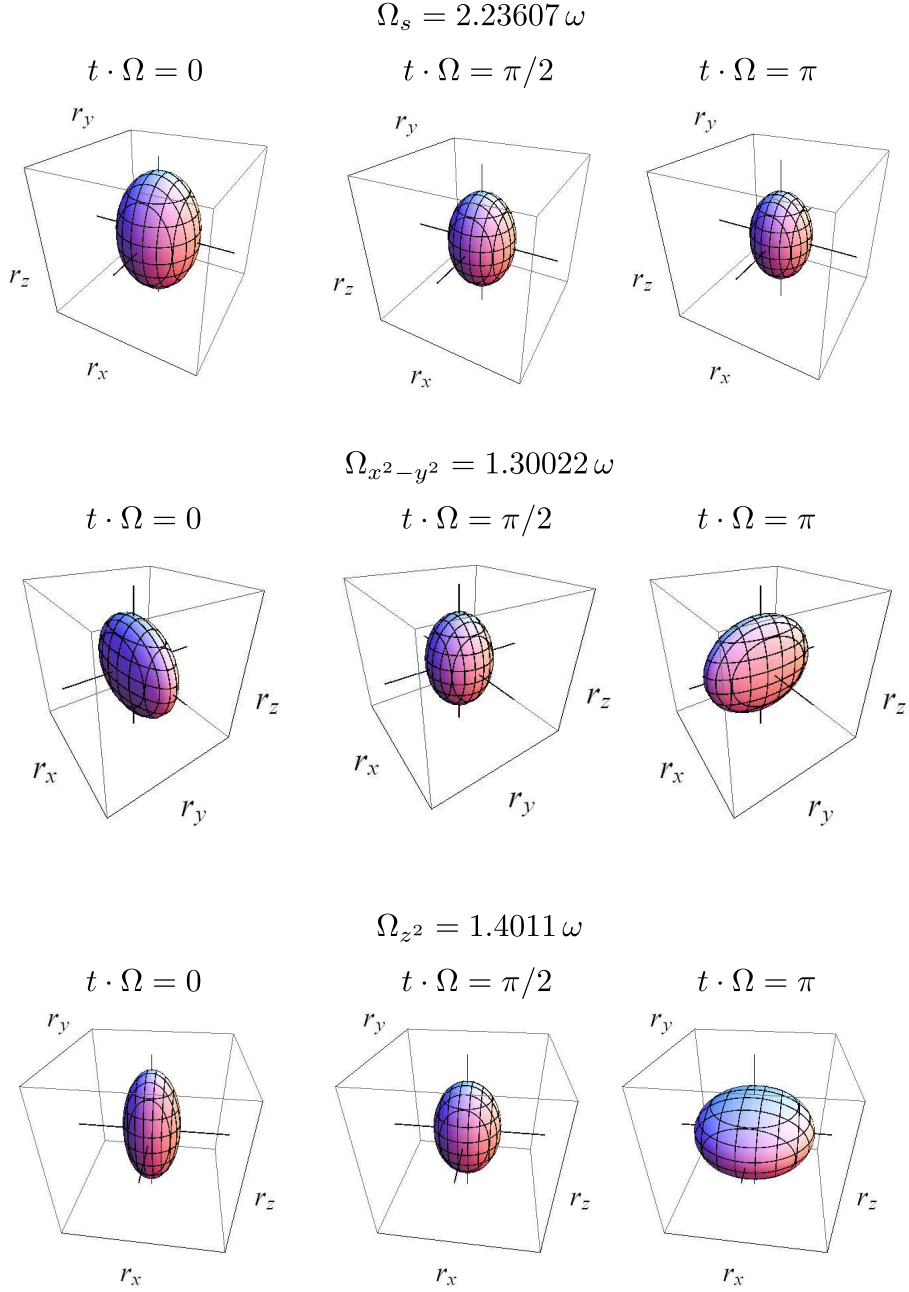}
\caption{(Color online) Visualization of density fluctuations $n_{\Omega}(\mathbf{r}%
,t)=n_{TF}(\mathbf{r})+\delta n_{\Omega}(\mathbf{r},t)$ of dipolar BEC cloud
confined inside a spherical trap for a dipole interaction strength
$\varepsilon_{D}=0.7$. First row isotropic breather mode $\Omega_{s}$ ; second
row quadrupolar mode $\Omega_{x^{2}-y^{2}}$ ; third row quadrupolar mode
$\Omega_{z^{2}}$. }%
\label{Viusualization Breather and Quadrupolar modes}%
\end{figure}

\subsection{Spectrum of Low-Lying Excitations for the Case $\vartheta_{T}=0$.}

We now discuss the collective density oscillations of a
dipolar BEC cloud confined in a tri-axial harmonic trap in the highly
symmetric case, when the principal axis $\mathbf{e}_{z,T}$ of the trap is
orientated colinear to the spin polarizing magnetic field $\mathbf{B}$, so
that $\vartheta_{T}=0$. In Fig.9 and Fig.10 the collective mode frequencies
$\Omega$ corresponding to the solution of the eigenvalue problem
(\ref{collective mode eigenvalue problem II}) are plotted vs. the dipole
interaction strength $\varepsilon_{D}$. Shown are three scissors modes with
$d_{xz}$ , $d_{yz}$ and $d_{xy}$-wave symmetry, and three hybridized modes
combined from basis elements with $s$-wave, $d_{z^{2}}$-wave and
$d_{x^{2}-y^{2}}$-wave symmetry. The anisotropy ratio chosen is $\omega
_{x}:\omega_{y}:\omega_{z}=$ $712:128:942$ in Fig.9, and in reversed order
$\omega_{x}:\omega_{y}:\omega_{z}=942:128:712$ in Fig.10,
respectively.\begin{figure}[h]
\centering
\includegraphics[width=0.90\textwidth]{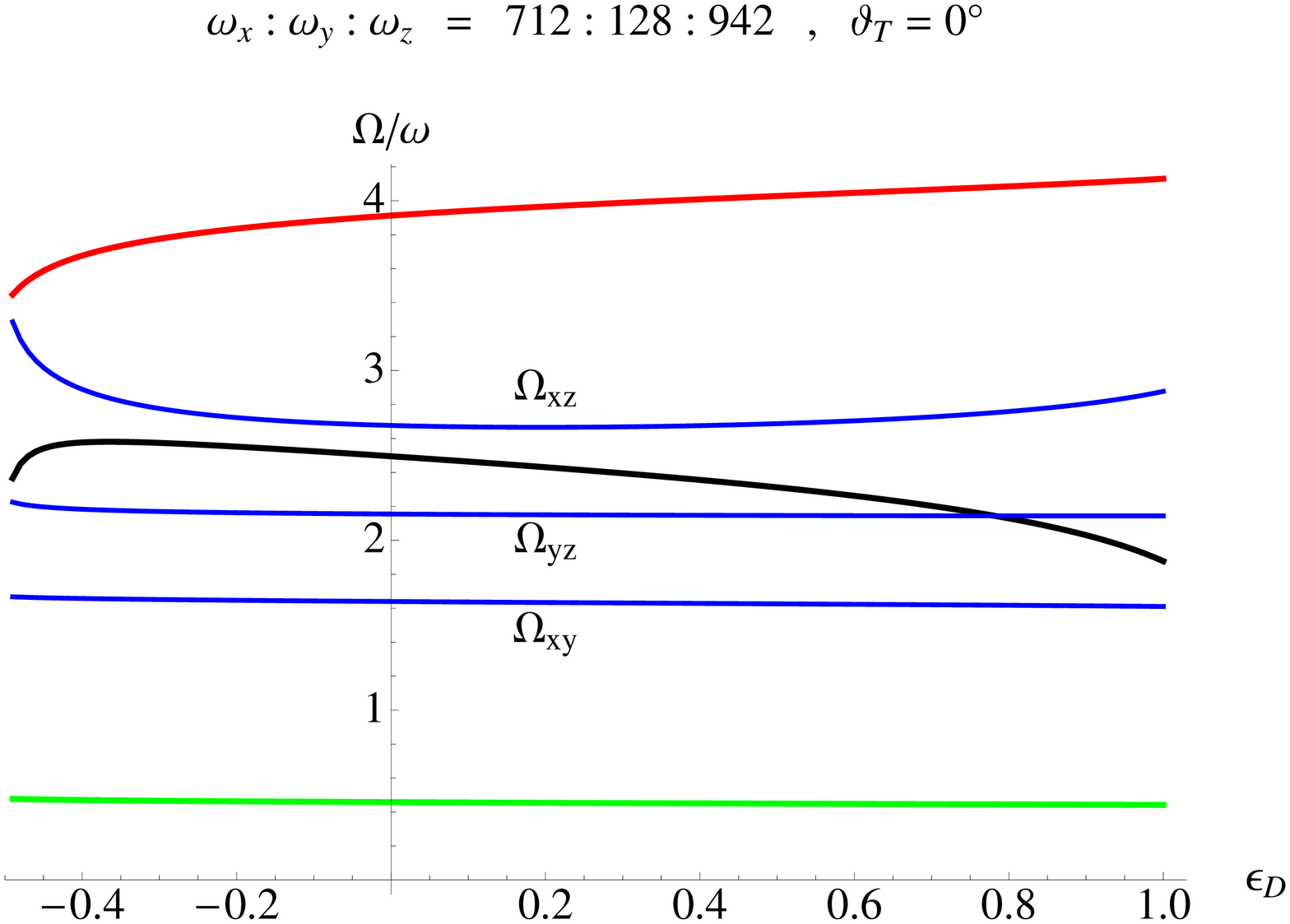}\caption{(Color online)
Eigenfrequencies of small amplitude collective modes combining isotropic
$s$-wave and quadrupolar $d$-wave basis elements vs. dipole interaction
strength $\varepsilon_{D}$ for dipolar BEC cloud confined in a harmonic trap
with anisotropy ratio $\omega_{x}:\omega_{y}:\omega_{z}=712:128:942$ in the
highly symmetric case $\vartheta_{T}=0$. Displayed are three scissors modes
(blue lines) and three hybridized modes that are combinations of $s$-wave,
$d_{x^{2}-y^{2}}$-and $d_{z^{2}}$-basis elements (red line, black line and
green line).}%
\label{Fig.13}%
\end{figure}

\begin{figure}[h]
\centering
\includegraphics[width=0.90\textwidth]{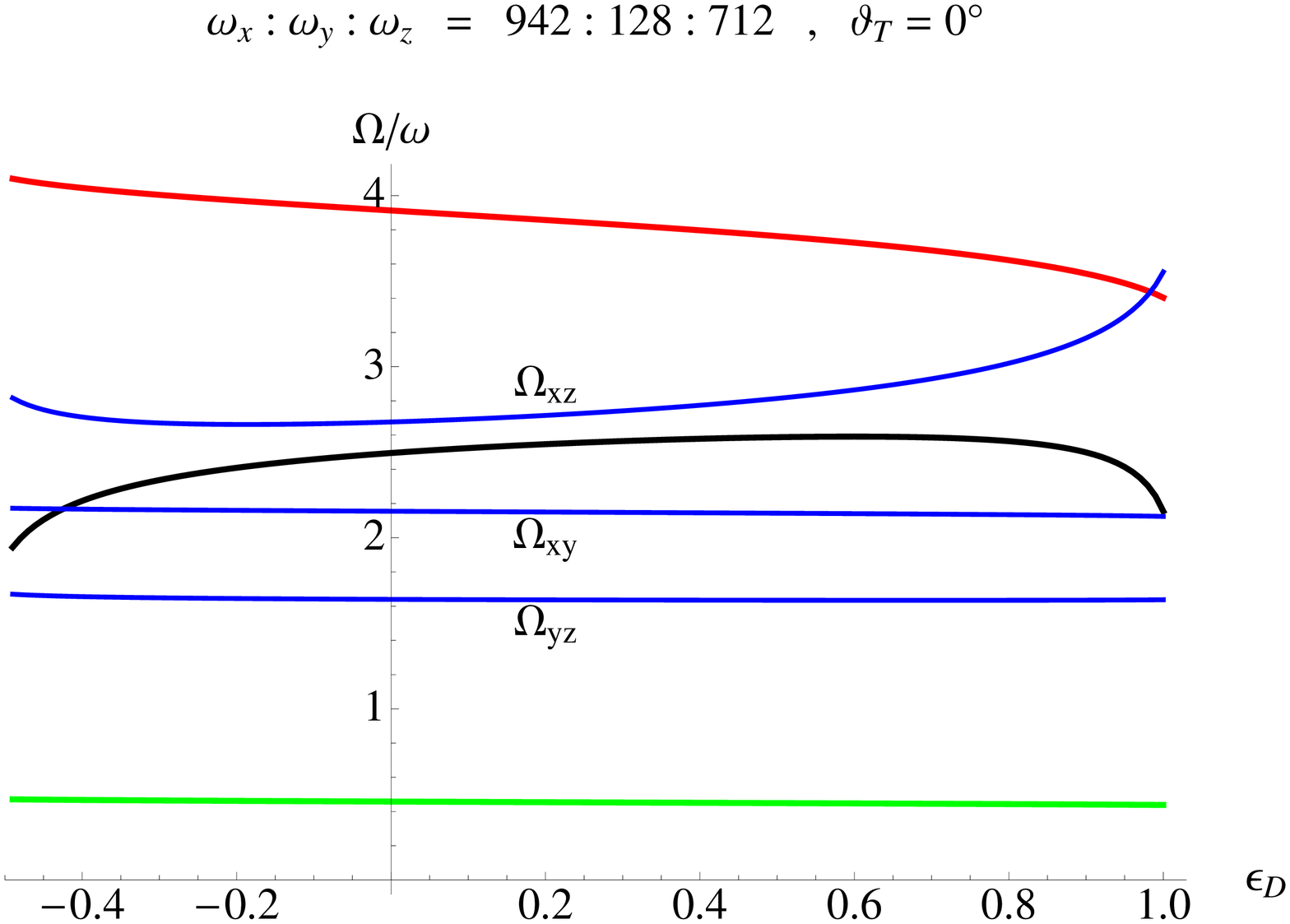}\caption{(Color online)
{}Dependence of eigenfrequencies of small amplitude collective modes combining
isotropic $s$-wave and quadrupolar $d$-wave basis elements vs. dipole
interaction strength $\varepsilon_{D}$ for dipolar BEC cloud confined in a
harmonic trap with reversed anisotropy ratio $\omega_{x}:\omega_{y}:\omega
_{z}=942:128:$ $712$ in the highly symmetric case $\vartheta_{T}=0$. Displayed
are three scissors modes (blue lines) and three hybridized modes that are
combinations of $s$-wave, $d_{x^{2}-y^{2}}$-and $d_{z^{2}}$-basis elements
(red line, black line and green line).}%
\label{Fig.14}%
\end{figure}

There exists fair agreement between our exact analytical results and the
numerical results obtained in Ref.\cite{Givanazzi}, which are based on the
method of solving Newton equations of motion for time dependent Thomas-Fermi
radii. As is evident from
(\ref{coupling of modes to rotation, shear and curvature of the trap}),
small amplitude fluctuations of the Thomas-Fermi radii are described in our
approach by the dilatation amplitudes $\widehat{\zeta}_{a}\left(  t\right)  $.
However, in the highly symmetric case $\vartheta_{T}=0$, these dilatation
amplitudes only couple to the diagonal basis elements of the tensor
$\widehat{\rho}_{ab}$ :
\begin{equation}
\widehat{\rho}_{aa}(\Omega)=\frac{\widehat{\zeta}_{a}\left(  \Omega\right)
}{\lambda_{a}}-\widehat{\eta}_{aa}^{\left(  1\right)  }(\Omega)
\label{coupling to hybridized modes s-wave and d-wave}%
\end{equation}
In the highly symmetric case $\vartheta_{T}=0$ no coupling of the dilation
amplitudes $\widehat{\zeta}_{a}\left(  \Omega\right)  $ to the off diagonal
elements $a\neq b$ of the tensor $\widehat{\rho}_{ab}$ exists, as is evident
from (\ref{scissors modes II}). To ease comparison of our results with the
results presented in Ref.(\cite{Griesmaier et al}) we also plot in Fig.11 and
Fig.12 the relative change of the collective mode frequencies $\frac
{\Omega-\Omega^{\left(  0\right)  }}{\Omega}$ vs. $\varepsilon_{D}$ for the
three hybridized modes displayed in Fig.9 and Fig.10 that couple via the
dilatation amplitudes $\widehat{\zeta}_{a}$ to the time dependent Thomas-Fermi radii.

\begin{figure}[h]
\centering
\includegraphics[width=0.90\textwidth]{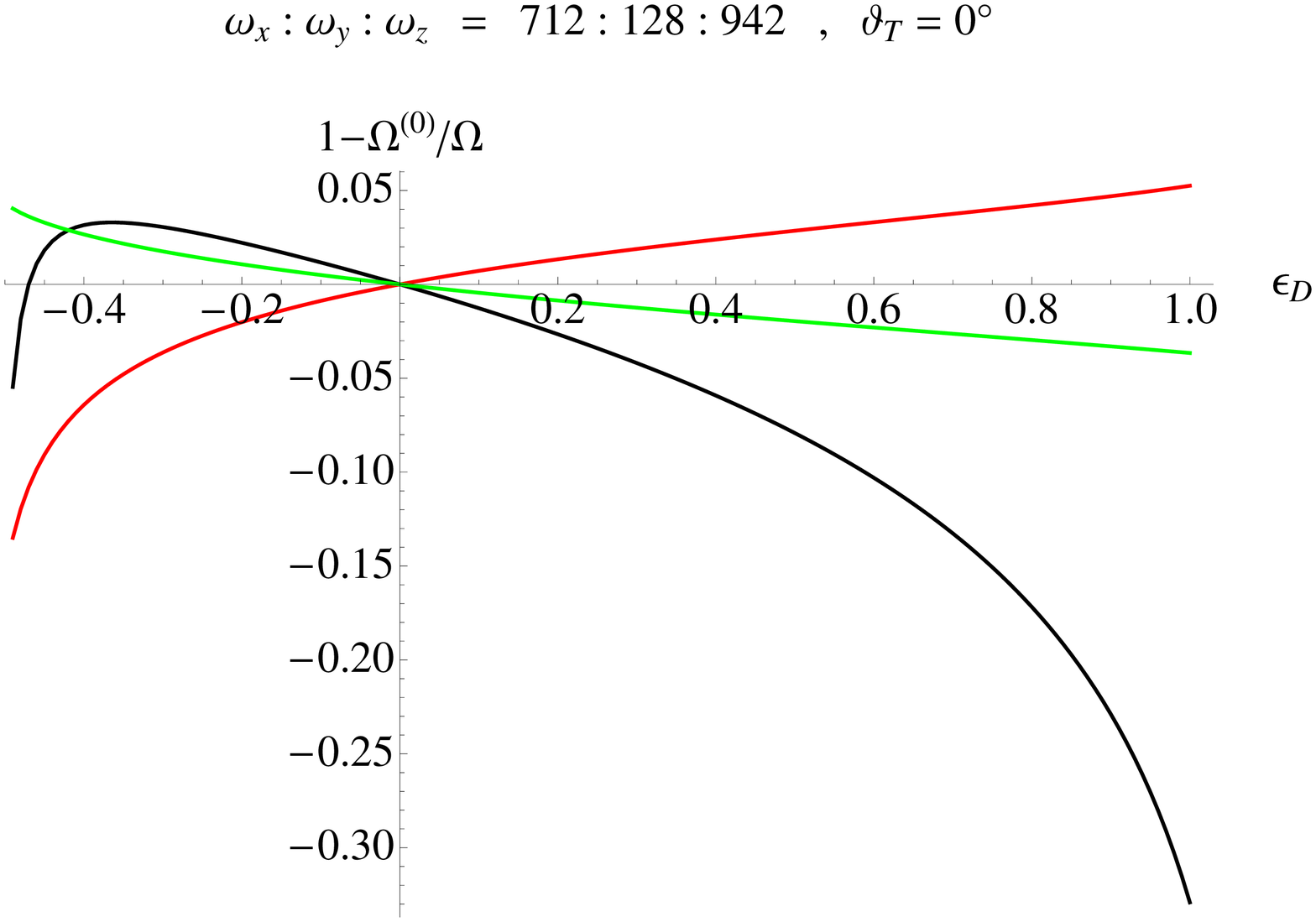}\caption{(Color online) Relative
change $\frac{\Omega-\Omega^{\left(  0\right)  }}{\Omega}$ vs. $\varepsilon
_{D}$ for the three hybridized collective modes as displayed in
Fig.~\ref{Fig.13} for the highly symmetric case $\vartheta_{T}=0$. }%
\label{Fig.15}%
\end{figure}

\begin{figure}[h]
\centering
\includegraphics[width=0.90\textwidth]{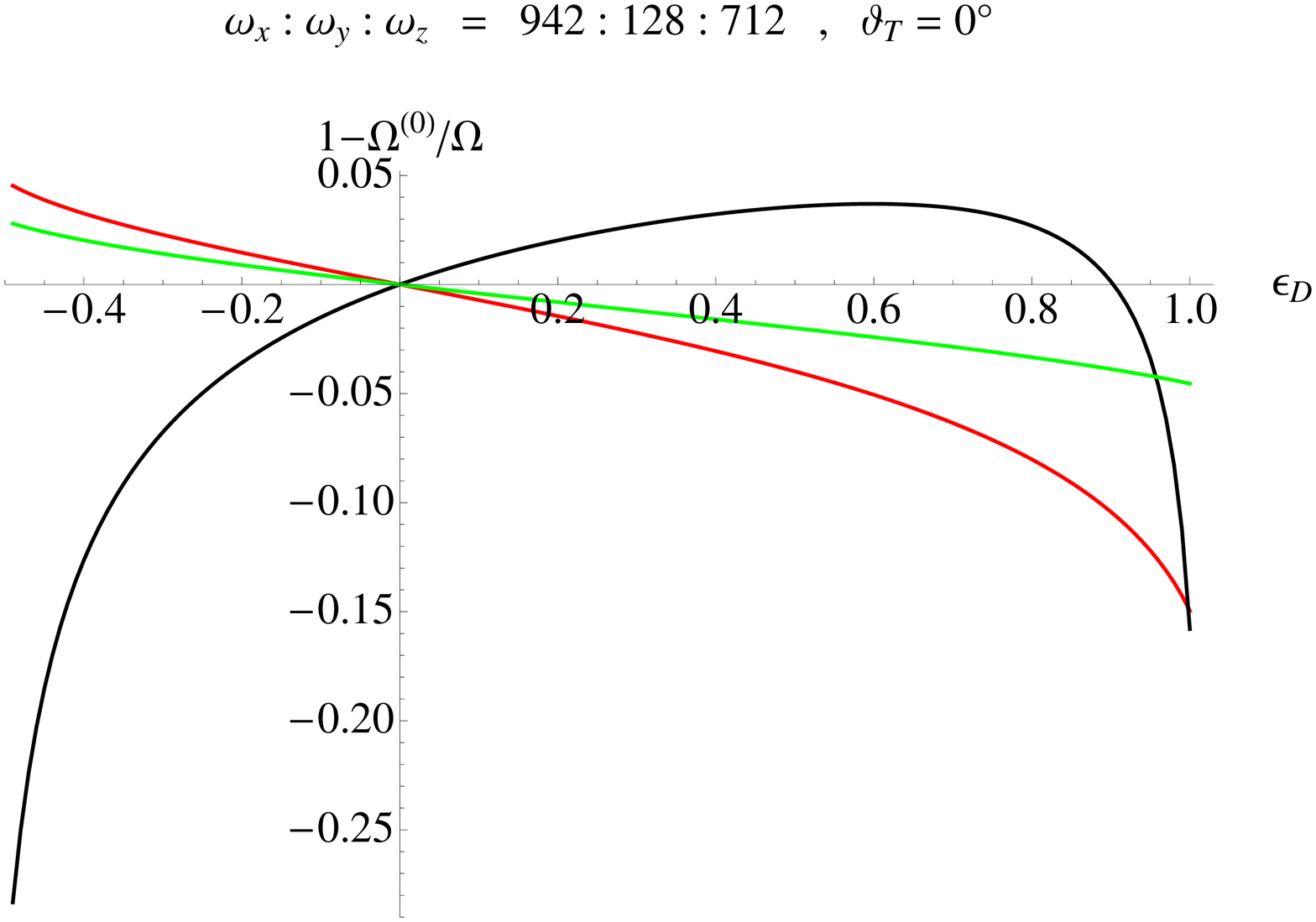}\caption{(Color online) Relative
change $\frac{\Omega-\Omega^{\left(  0\right)  }}{\Omega}$ vs. $\varepsilon
_{D}$ for the three hybridized collective modes as displayed in
Fig.~\ref{Fig.14} for the highly symmetric case $\vartheta_{T}=0$. }%
\end{figure}

It should be pointed out that a purely diagonal shear movement of the dipolar
BEC cloud at constant Thomas-Fermi radii, $\widehat{\zeta}_{a}=0$, as
described by the diagonal elements $\widehat{\eta}_{aa}^{\left(  1\right)  }$
of the tensor $\widehat{\eta}_{ab}^{\left(  1\right)  }$ spanning the
(solenoidal) displacement vectorfield
(\ref{expansion displacement vectorfield I}), may also excite these hybridized
modes coupling to $\widehat{\rho}_{aa}$. This degeneracy is a special property
of any quantum degenerate BEC groundstate with an ellipsoidal shaped density profile.

\subsection{Spectrum of Low-Lying Excitations for the Case $\vartheta_{T}\neq 0$.}

Sudden changes of the trap potential may excite various collective modes of a
dipolar BEC cloud. If the polarizing external magnetic field $\mathbf{B}$ is
not in alignment with the principal axis $\mathbf{e}_{z,T}$ of the trap, so
that $\mathbf{e}_{z,T}$ includes a finite angle $\vartheta_{T}\neq0$ with
$\mathbf{B}$ in the $xz$-plane (see Fig.\ref{Fig. 1}), the $s$-wave and
$d$-wave symmetry parts of the collective density oscillations combine to a
quadruplet and a doublet of modes. It is found from
(\ref{collective mode eigenvalue problem}), that the modes with mixed
$d_{x^{2}-y^{2}}$ , $d_{z^{2}}$, $d_{xz}$- and $s$-wave symmetry, consisting
of a linear combination of the three diagonal amplitudes $\widehat{\rho}_{xx}$
, $\widehat{\rho}_{yy}$ , $\widehat{\rho}_{zz}$ and one off-diagonal amplitude
$\widehat{\rho}_{xz}$, combine together to a quadruplet (see Fig.13), and the
modes with mixed $d_{xy}$- and $d_{yz}$-symmetry combine together to a doublet
of scissors modes (see Fig.14). From (\ref{scissors modes II}) it is evident,
that for $\varepsilon_{D}\neq0$ an infinitesimal rotation around the principal
axis $\mathbf{e}_{y,T}$ of a harmonic tri-axial trap may then excite via its
coupling to the $\widehat{\rho}_{xz}$-components of the eigenvectors all
$\ $four modes of the mentioned quadruplet of small amplitude oscillations of
the density simultaneously.

Likewise, a rotation around the principal axis $\mathbf{e}_{z,T}$ (or
$\mathbf{e}_{x,T}$ ) of the tri-axial harmonic trap may excite via the
coupling to the off diagonal amplitudes $\widehat{\rho}_{yz}\ $and
$\widehat{\rho}_{xy}$ the mentioned doublet of scissors modes simultaneously.
Alternatively, these scissors modes can also be excited by transversal shear
movements of the anisotropic harmonic trap, thus creating an excitation of the
BEC cloud that may be described by a (solenoidal) displacement vectorfield
(\ref{expansion displacement vectorfield I}) that is spanned by the
\emph{symmetric} off diagonal elements $\ $of the tensor $\widehat{\eta}%
_{ab}^{\left(  1\right)  }$.

A sudden change of the curvature of the trap potential, as described by the
dilatation amplitudes $\widehat{\zeta}_{a}$ in
(\ref{coupling to hybridized modes s-wave and d-wave}),excites in the geometry
under consideration the modes of the quadruplet, but never the scissors modes
of the doublet with mixed $d_{xy}$- and $d_{yz}$-symmetry.

The results displayed in Fig.13 and Fig.14 reveal, that a tri-axial harmonic
trap with trap frequencies $\omega_{x}=\omega_{1}$ , $\omega_{y}=\omega_{2}$ ,
$\omega_{z}=\omega_{3}$ , say $\omega_{1}>\omega_{2}>\omega_{3}$ , shows a
characteristic shift of the eigenfrequencies of these quadruplet- and
doublet-collective modes compared to a trap with \emph{reversed} trap
frequencies, i.e. a harmonic trap with $\omega_{x}=\omega_{3}$ , $\omega
_{y}=\omega_{2}$ , $\omega_{z}=\omega_{1}$.

\begin{figure}[h]
\centering
\includegraphics[width=0.90\textwidth]{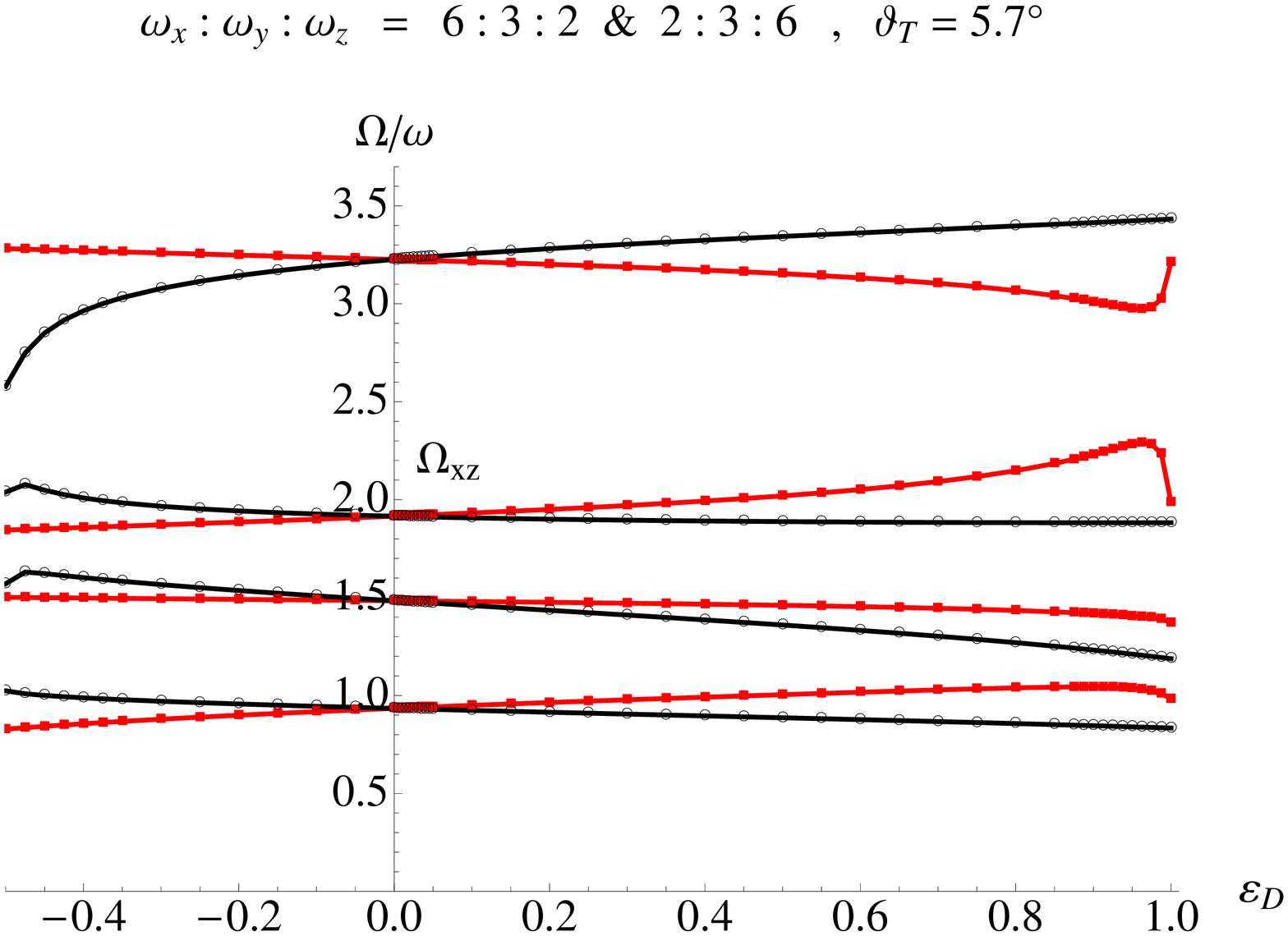}\caption{(Color online) Dependence
on dipole interaction strength $\varepsilon_{D}$ of eigenfrequencies $\Omega$ of
small amplitude density oscillations corresponding to $4\times4$-block in (\ref{collective mode eigenvalue problem}) when the BEC cloud is confined in tri-axial harmonic
anisotropic trap for $\omega_{x}:\omega_{y}:\omega_{z}$ $=6:3:2$ (red line)
and $\omega_{x}:\omega_{y}:\omega_{z}$ $=2:3:6$ (black line), choosing a trap
orientation angle $\vartheta_{T}=5.7^{\circ}$. All frequencies normalized to
geometric mean $\omega=\left(  \omega_{x}\omega_{y}\omega_{z}\right)
^{\frac{1}{3}}$.}%
\label{Fig.17}%
\end{figure}

\begin{figure}[h]
\centering
\includegraphics[width=0.90\textwidth]{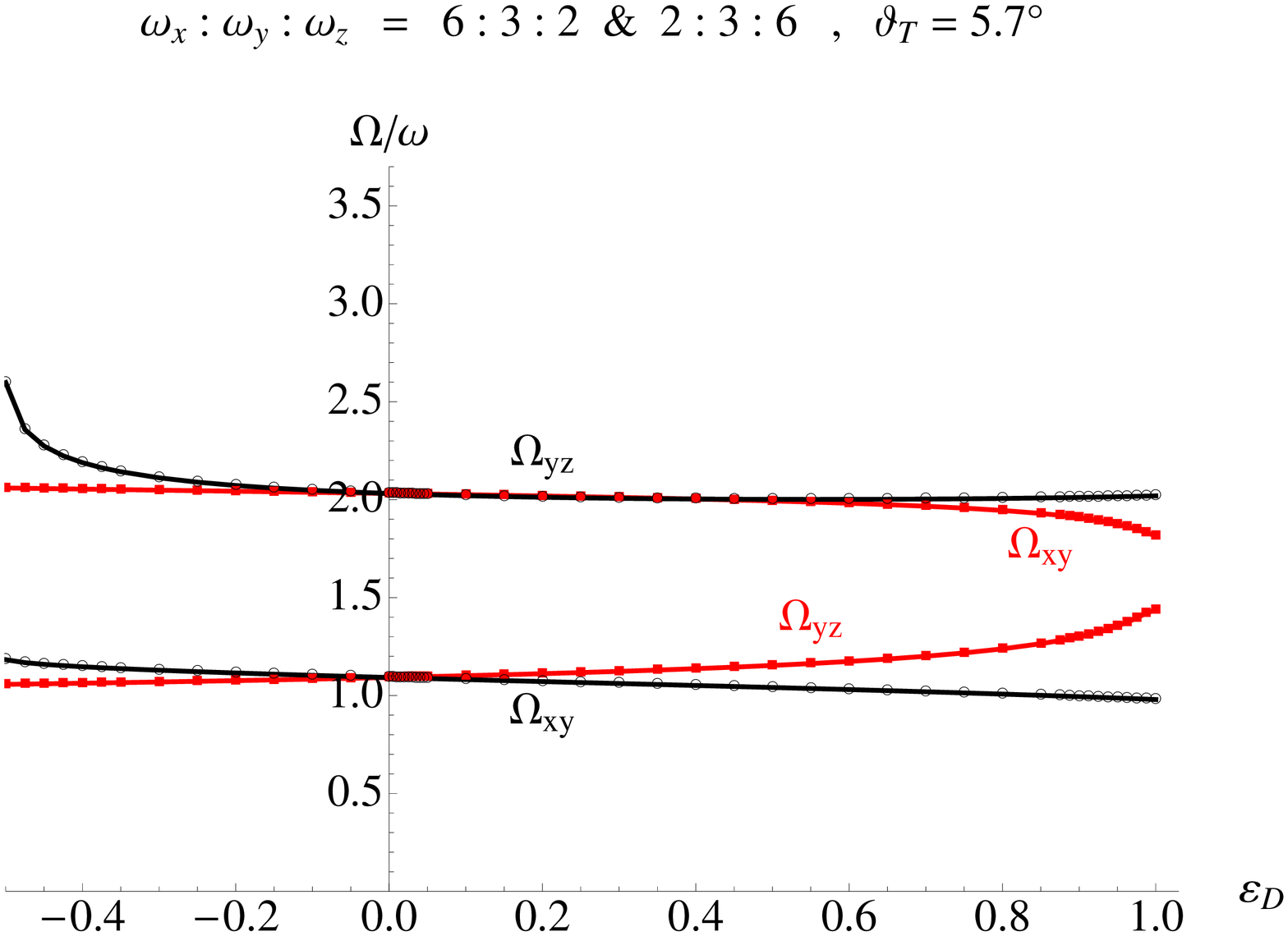}\caption{(Color online) Dependence
on dipole interaction strength $\varepsilon_{D}$ of eigenfrequencies $\Omega$ of
small amplitude density oscillations corresponding to $2\times2$-block in (\ref{collective mode eigenvalue problem}) when the BEC cloud is confined in tri-axial harmonic
anisotropic trap for $\omega_{x}:\omega_{y}:\omega_{z}$ $=6:3:2$ (red line)
and $\omega_{x}:\omega_{y}:\omega_{z}$ $=2:3:6$ (black line), choosing a trap
orientation angle $\vartheta_{T}=5.7^{\circ}$. All frequencies normalized to
geometric mean $\omega=\left(  \omega_{x}\omega_{y}\omega_{z}\right)
^{\frac{1}{3}}$.}%
\end{figure}From measurements of these characteristic shifts of the collective
mode frequencies of the quadruplet- and doublet-collective modes of a dipolar
BEC cloud for two such \emph{mutually reciprocal} tri-axial traps the strength
of the interaction parameter $\varepsilon_{D}$ could be determined
accurately\cite{Givanazzi}. Knowing the mass $m^{\star}$ and the magnetic
dipole moment $\left\vert \left\langle \mathbf{M}\right\rangle \right\vert $
of a single atom one then obtains immediately from (\ref{dipole strength})
the isotropic $s$-wave scattering length of the atoms \cite{Griesmaier et
al}:
\begin{equation}
a_{s}=\frac{\mu_{0}\left\vert \left\langle \mathbf{M}\right\rangle \right\vert
^{2}}{\frac{12\pi\hbar^{2}}{m^{\star}}\varepsilon_{D}}
\label{s-wave scattering length from measurement of eps_D}%
\end{equation}

The experiment suggested here consists in preparing a quantum degenerate spin
polarized dipolar BEC cloud confined in a harmonic trap with tri-axial
symmetry, so that the principal axis $\mathbf{e}_{z,T}$ of the trap is first
orientated colinear to the spin polarizing magnetic field $\mathbf{B}$, i.e.
at the beginning of the experiment $\vartheta_{T}=0=\vartheta_{0}$ (see
Fig.\ref{Fig. 1} ). Then, say at time $t=0$, the trap orientation angle
$\vartheta_{T}$ is changed suddenly to a new value, by making a rotation
around the principal axis $\mathbf{e}_{y,T}$ of the trap by a constant small
rotation angle, say $\vartheta_{T}=5.7^{\circ}$ , the value chosen in Fig.13
and Fig.14. A dipolar BEC cloud excited in this manner will then oscillate not
around the old cloud orientation angle $\vartheta_{0}=0$ , but around a new
cloud orientation angle $\vartheta_{0}\left(  \vartheta_{T}\right)  $, which
is via the selfconsistency equations (\ref{selfconsistent IV a}%
),(\ref{selfconsistent IV b}), (\ref{selfconsistent IV c}) not only dependent
on the strength of the dipole interaction parameter $\varepsilon_{D}$ , but
also on the chosen trap orientation angle $\vartheta_{T}$. The principal axis
$\mathbf{e}_{z,0}$ of the new equilibrium BEC cloud confined in a harmonic
trap with trap orientation angle $\vartheta_{T}$ then includes with the fixed
magnetic field $\mathbf{B}$ a finite angle $\vartheta_{0}$, that is smaller or
larger than $\vartheta_{T}$, depending on the anisotropy ratio of the trap
(see Fig.2 ). It follows from what has been said that the eigenfrequencies
$\Omega$ of the collective modes of the density fluctuations, that can be
excited in this manner, are functions of $\varepsilon_{D}$ and the trap
orientation angle $\vartheta_{T}$.

\begin{figure}[h]
\centering
\includegraphics[width=0.90\textwidth]{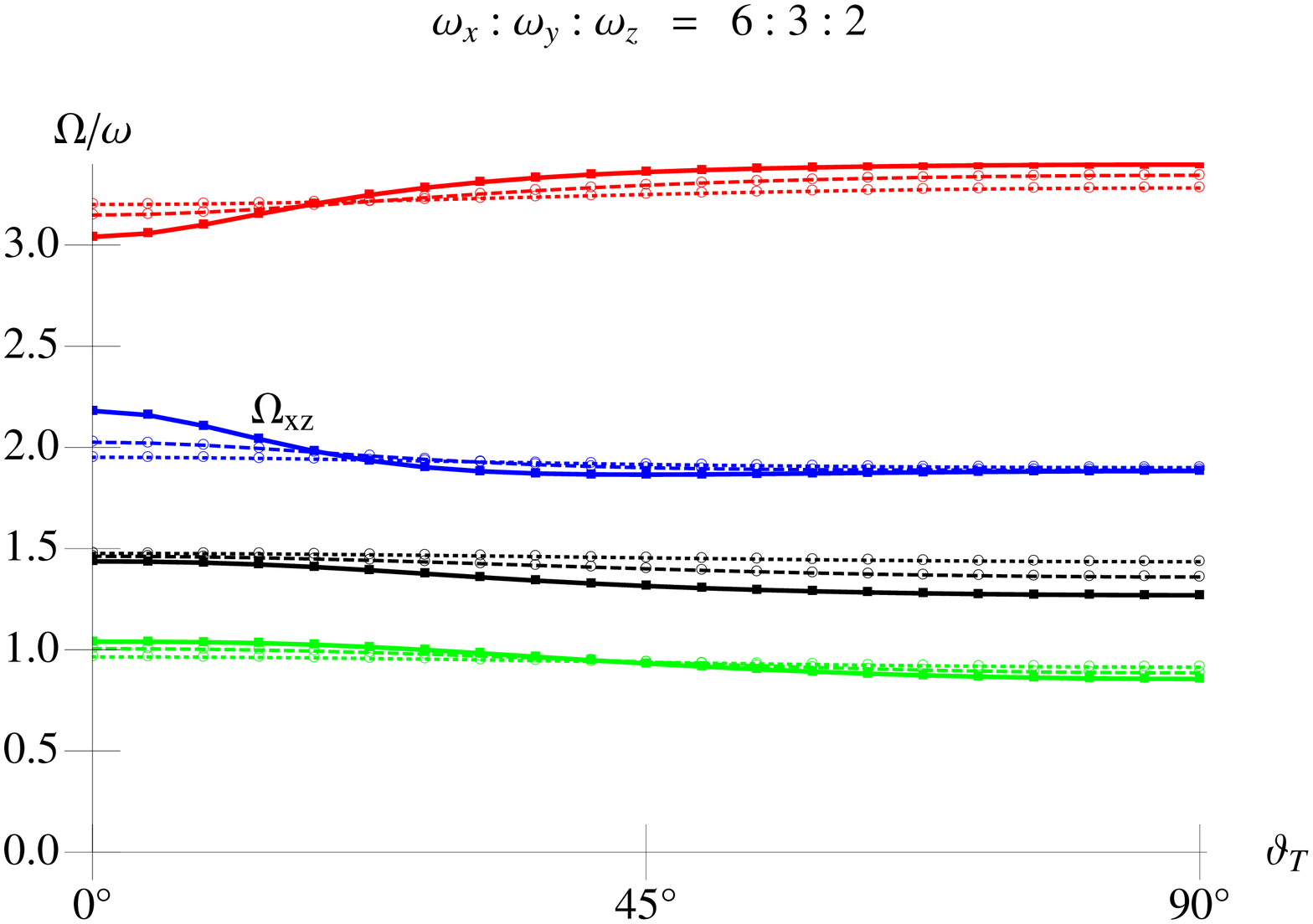}%
\caption{(Color online)
Dependence on trap orientation angle $\vartheta_{T}$ of eigenfrequencies $\Omega$ of
small amplitude density oscillations corresponding to $4\times4$-block in (\ref{collective mode eigenvalue problem}) when the BEC cloud is confined in tri-axial harmonic
anisotropic trap for $\omega_{x}:\omega_{y}:\omega_{z}$
$=6:3:2$. All frequencies normalized to geometric mean $\omega=\left(
\omega_{x}\omega_{y}\omega_{z}\right)  ^{\frac{1}{3}}$. The strength of the
dipole interaction is $\varepsilon_{D}=0.2$ (dotted line) , $\varepsilon_{D}=0.5$ (dashed line) , $\varepsilon_{D}=0.8$ (solid line).}%
\label{CollectiveModeFrequencyVSThetaTrap-4x4-Block}
\end{figure}

\begin{figure}[h]
\centering
\includegraphics[width=0.90\textwidth]{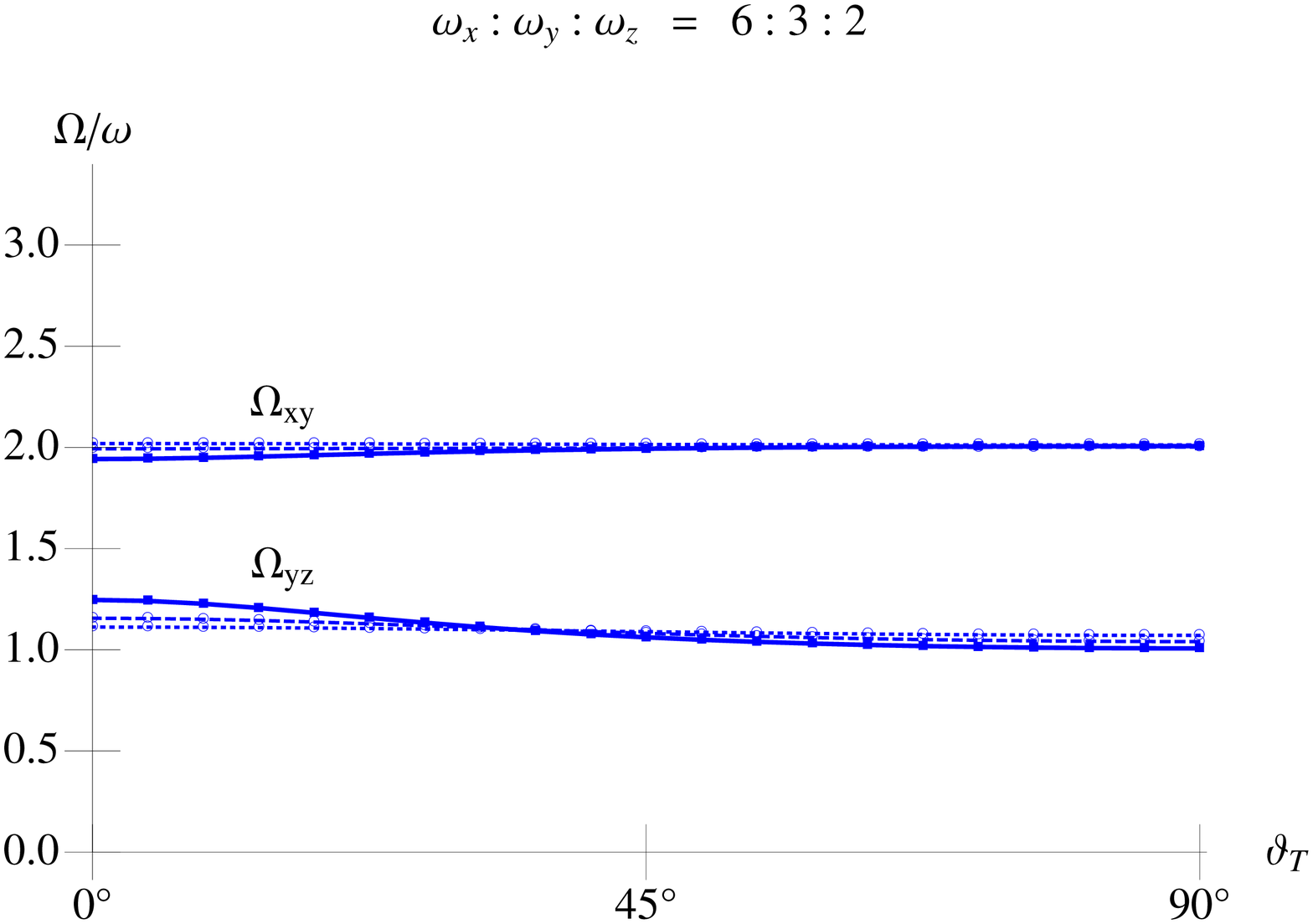}%
\caption{(Color online) Dependence on trap orientation angle $\vartheta_{T}$ of eigenfrequencies $\Omega$ of
small amplitude density oscillations corresponding to $2\times2$-block in (\ref{collective mode eigenvalue problem}) when the BEC cloud is confined in tri-axial harmonic
anisotropic trap for $\omega_{x}:\omega_{y}:\omega_{z}$
$=6:3:2$. All frequencies normalized to geometric mean $\omega=\left(
\omega_{x}\omega_{y}\omega_{z}\right)  ^{\frac{1}{3}}$. The strength of the
dipole interaction is $\varepsilon_{D}=0.2$ (dotted line) , $\varepsilon_{D}=0.5$ (dashed line) , $\varepsilon_{D}=0.8$ (solid line).}%
\label{CollectiveModeFrequencyVSThetaTrap-2x2-Block}
\end{figure}

In Fig.\ref{CollectiveModeFrequencyVSThetaTrap-4x4-Block} and Fig.\ref{CollectiveModeFrequencyVSThetaTrap-2x2-Block} the dependence of the collective mode frequencies of a
dipolar BEC cloud on the trap orientation angle $\vartheta_{T}$ is shown for
three values of the interaction strength parameter $\varepsilon_{D}$.

\clearpage

\clearpage

If a dipolar BEC cloud is confined in a harmonic trap with tri-axial symmetry,
so that the spin polarizing magnetic field $\mathbf{B}$ is orientated in a
completely general fashion, i.e. $\mathbf{B}$ its not orientated parallel to
any symmetry plane of the trap, the quantum degenerate groundstate of the BEC
cloud is then characterized by three Euler angles determining the orientation
of the ellipsoid $\mathbb{D}_{TF}$ relative to the axes of the trap. A study
of such spin polarized dipolar BEC clouds we shall present in a separate
publication~\cite{schopohl}, together with a discussion of the octupolar modes
of density oscillations, which can be described by fluctuation amplitudes
$\eta_{a;bc}^{\left(  2\right)  }\left(  t\right)  $ associated with a
solenoidal vectorfield with a quadratic spatial variation, i.e. $\eta
_{a}\left(  \mathbf{r},t\right)  =\sum_{b,.c}\eta_{a;bc}^{\left(  2\right)
}\left(  t\right)  r_{b}r_{c}$.

\section{Conclusions}

We have studied the groundstate and the low-lying collective modes of a dipolar
Bose-Einstein Condensate for the case that the external magnetic field is not
necessarily oriented parallel to one of the principal axes of the harmonic
anisotropic trap. In particular, we have determined the eigenfrequencies of
six low-lying collective modes that combine respectivly, to a quadruplet and doublet of atom density oscillations with mixed $s$- and $d$-wave symmetry, and obtained analytical expressions for them.
We have found the following results: the mode frequencies depend on the dipole
interaction parameter in a characteristic way that could be used to measure
the $s$-wave scattering length of the atoms accurately. In the special case
that the harmonic trap is \emph{spherical} we find the remarkable result that
the eigenfrequency of the isotropic breather mode does \emph{not} depend on
the dipole interaction strength, even though the shape of the condensate does.
Thus, this mode could be used as a reference frequency for the other
collective modes that depend on the dipole interaction strength. A rigorous
sum rule argument shows that this feature of the breather mode is a
consequence of the scaling property (\ref{homogeneous function of degree -3})
of the interaction potential in a dipolar BEC,  and the Thomas-Fermi approximation.

\begin{acknowledgments}
We thank J\'{o}zsef Fort\'{a}gh for inspiring discussions.
\end{acknowledgments}

\newpage

\appendix{}

\section{Index Integrals}
\label{appendixA}

Consider the index integrals%
\begin{align}
a,b,c  &  \in\left\{  x,y,z\right\} \label{index integrals}\\
I_{a}\left(  \lambda_{x},\lambda_{y},\lambda_{z}\right)   &  =\lambda
_{x}\lambda_{y}\lambda_{z}\int_{0}^{\infty}\frac{du}{\sqrt{\left(  \lambda
_{x}^{2}+u\right)  \left(  \lambda_{y}^{2}+u\right)  \left(  \lambda_{z}%
^{2}+u\right)  }}\frac{1}{\left(  \lambda_{a}^{2}+u\right)  }\nonumber\\
I_{ab}\left(  \lambda_{x},\lambda_{y},\lambda_{z}\right)   &  =\lambda
_{x}\lambda_{y}\lambda_{z}\int_{0}^{\infty}\frac{du}{\sqrt{\left(  \lambda
_{x}^{2}+u\right)  \left(  \lambda_{y}^{2}+u\right)  \left(  \lambda_{z}%
^{2}+u\right)  }}\frac{1}{\left(  \lambda_{a}^{2}+u\right)  \left(
\lambda_{b}^{2}+u\right)  }\nonumber\\
I_{abc}\left(  \lambda_{x},\lambda_{y},\lambda_{z}\right)   &  =\lambda
_{x}\lambda_{y}\lambda_{z}\int_{0}^{\infty}\frac{du}{\sqrt{\left(  \lambda
_{x}^{2}+u\right)  \left(  \lambda_{y}^{2}+u\right)  \left(  \lambda_{z}%
^{2}+u\right)  }}\frac{1}{\left(  \lambda_{a}^{2}+u\right)  \left(
\lambda_{b}^{2}+u\right)  \left(  \lambda_{c}^{2}+u\right)  }\nonumber
\end{align}
These integrals are symmetric under permutations of the indices $a,b,c\in
\left\{  x,y,z\right\}  $. It is also evident that%
\begin{align}
I_{a}-I_{b}  &  =-\left(  \lambda_{a}^{2}-\lambda_{b}^{2}\right)
I_{ab}\label{identity index integrals I}\\
I_{ac}-I_{bc}  &  =-\left(  \lambda_{a}^{2}-\lambda_{b}^{2}\right)
I_{abc}\nonumber
\end{align}
Also index integrals $I_{ab}$ and $I_{abc}$ are connected by a derivative
operation:%
\begin{equation}
\left(  \frac{1}{\lambda_{c}}-\frac{\partial}{\partial\lambda_{c}}\right)
I_{ab}=\left(  1+2\delta_{ac}+2\delta_{bc}\right)  \lambda_{c}I_{abc}
\label{derivative index integrals}%
\end{equation}

Let us note the identity
\begin{align}
&  \left(  -2\frac{d}{du}\right)  \left[  \frac{1}{\sqrt{\left(  \lambda
_{x}^{2}+u\right)  \left(  \lambda_{y}^{2}+u\right)  \left(  \lambda_{z}%
^{2}+u\right)  }}\right] \label{identity index integrals IIa}\\
&  =\frac{1}{\sqrt{\left(  \lambda_{x}^{2}+u\right)  \left(  \lambda_{y}%
^{2}+u\right)  \left(  \lambda_{z}^{2}+u\right)  }}\sum_{a\in\left\{
x,y,z\right\}  }\frac{1}{\lambda_{a}^{2}+u}\nonumber
\end{align}
Likewise%
\begin{align}
& \label{identity index integrals II b}\\
&  \left(  -2\frac{d}{du}\right)  \left[  \frac{1}{\sqrt{\left(  \lambda
_{x}^{2}+u\right)  \left(  \lambda_{y}^{2}+u\right)  \left(  \lambda_{z}%
^{2}+u\right)  }}\frac{1}{\lambda_{b}^{2}+u}\right] \nonumber\\
&  =\frac{1}{\sqrt{\left(  \lambda_{x}^{2}+u\right)  \left(  \lambda_{y}%
^{2}+u\right)  \left(  \lambda_{z}^{2}+u\right)  }}\frac{1}{\lambda_{b}^{2}%
+u}\left[  \frac{2}{\lambda_{b}^{2}+u}+\sum_{a\in\left\{  x,y,z\right\}
}\frac{1}{\lambda_{a}^{2}+u}\right] \nonumber
\end{align}
and%
\begin{align}
& \label{identity index integrals IIc}\\
&  \left(  -2\frac{d}{du}\right)  \left[  \frac{1}{\sqrt{\left(  \lambda
_{x}^{2}+u\right)  \left(  \lambda_{y}^{2}+u\right)  \left(  \lambda_{z}%
^{2}+u\right)  }}\frac{1}{\lambda_{b}^{2}+u}\frac{1}{\lambda_{c}^{2}+u}\right]
\nonumber\\
&  =\frac{1}{\sqrt{\left(  \lambda_{x}^{2}+u\right)  \left(  \lambda_{y}%
^{2}+u\right)  \left(  \lambda_{z}^{2}+u\right)  }}\frac{1}{\lambda_{b}^{2}%
+u}\frac{1}{\lambda_{c}^{2}+u}\left[  \frac{2}{\lambda_{b}^{2}+u}+\frac
{2}{\lambda_{c}^{2}+u}+\sum_{a\in\left\{  x,y,z\right\}  }\frac{1}{\lambda
_{a}^{2}+u}\right] \nonumber
\end{align}
Upon integration with respect to the variable $u$ from $0$ to $\infty$ there
follow now several useful identities:%
\begin{align}
2  &  =\sum_{a\in\left\{  x,y,z\right\}  }I_{a}%
\label{identity index integrals III}\\
\frac{2}{\lambda_{b}^{2}}  &  =2I_{bb}+\sum_{a\in\left\{  x,y,z\right\}
}I_{ba}\nonumber\\
\frac{2}{\lambda_{b}^{2}\lambda_{c}^{2}}  &  =2I_{bbc}+2I_{bcc}+\sum
_{a\in\left\{  x,y,z\right\}  }I_{bca}\nonumber
\end{align}
Using these relations various useful algebraic connections between the
integral $I_{a}$ , $I_{ab}$ , also between $I_{ab}$ and $I_{abc}$ become
evident \cite{Chandrasekhar}:%
\begin{align}
a,b,c  &  \in\left\{  x,y,z\right\} \label{identity index integrals IV}\\
a  &  \neq b\neq c\nonumber\\
& \nonumber\\
3I_{aa}\lambda_{a}^{2}+I_{ab}\lambda_{b}^{2}+I_{ac}\lambda_{c}^{2}  &
=3I_{a}\nonumber\\
& \nonumber\\
5I_{aaa}\lambda_{a}^{2}+I_{aab}\lambda_{b}^{2}+I_{aac}\lambda_{c}^{2}  &
=5I_{aa}\nonumber\\
& \nonumber\\
3I_{aab}\lambda_{a}^{2}+3I_{abb}\lambda_{b}^{2}+I_{abc}\lambda_{c}^{2}  &
=5I_{ab}\nonumber
\end{align}

Making the substitution%
\begin{equation}
u\rightarrow u=\Lambda^{2}\cdot u^{\prime} \label{scaling subsitution}%
\end{equation}
we obtain useful scaling relations%
\begin{align}
I_{a}\left(  \lambda_{x},\lambda_{y},\lambda_{z}\right)   &  =I_{a}\left(
\frac{\lambda_{x}}{\Lambda},\frac{\lambda_{y}}{\Lambda},\frac{\lambda_{z}%
}{\Lambda}\right)  \equiv\overline{I}_{a}\label{scaling index integrals}\\
I_{ab}\left(  \lambda_{x},\lambda_{y},\lambda_{z}\right)   &  =\frac
{1}{\Lambda^{2}}I_{ab}\left(  \frac{\lambda_{x}}{\Lambda},\frac{\lambda_{y}%
}{\Lambda},\frac{\lambda_{z}}{\Lambda}\right)  \equiv\frac{1}{\Lambda^{2}%
}\overline{I}_{ab}\nonumber\\
I_{abc}\left(  \lambda_{x},\lambda_{y},\lambda_{z}\right)   &  =\frac
{1}{\Lambda^{4}}I_{abc}\left(  \frac{\lambda_{x}}{\Lambda},\frac{\lambda_{y}%
}{\Lambda},\frac{\lambda_{z}}{\Lambda}\right)  \equiv\frac{1}{\Lambda^{4}%
}\overline{I}_{abc}\nonumber
\end{align}
In our calculations we find it convenient to choose $\Lambda=\lambda_{z}>0$.

The task to calculate a double index integrals $I_{ab}$ can be reduced to
calculating simpler single index integrals $I_{a}$. This is enabled by using%

\begin{equation}
\overline{I}_{zz}=\frac{2-\overline{I}_{zx}-\overline{I}_{zy}}{3}
\label{identity double index integral IIa}%
\end{equation}
, an immediate consequence of (\ref{identity index integrals III}).
Provided $\lambda_{a}\neq\lambda_{z}$ , the integrals $\overline{I}_{za}$ can
be reduced to calculating the simpler integrals $\overline{I}_{z}$ and
$\overline{I}_{a}$ using the identity:
\begin{align}
a  &  \in\left\{  x,y\right\} \label{identity double index integral IIb}\\
\overline{I}_{za}  &  =-\frac{\overline{I}_{z}-\overline{I}_{a}}%
{1-\frac{\lambda_{a}^{2}}{\lambda_{z}^{2}}}\nonumber
\end{align}
So for $\lambda_{a}\neq\lambda_{z}$ all double index integrals $\overline
{I}_{za}$ can be reduced to single index integrals $\overline{I}_{a}$. Carlson
\cite{Carlson} has provided an elegant efficient algorithm based on the well
known method of the arithmetic-geometric mean to calculate the single index
integral $\overline{I}_{a}$ directly, which method we highly recommend because
of its accuracy and speed \cite{numerical recipes}.

For $\lambda_{a}=\lambda_{z}$ the right hand side becomes formally undefined.
However, in this case we may calculate the integrals $\overline{I}_{zx}$ and
$\overline{I}_{zy}$ in closed form:%
\begin{align}
& \label{identity double index integral IIIa}\\
\lim_{\lambda_{a}\rightarrow\lambda_{z}}\overline{I}_{za}  &  =\lim
_{\lambda_{a}\rightarrow\lambda_{z}}\frac{\lambda_{x}}{\lambda_{z}}%
\frac{\lambda_{y}}{\lambda_{z}}\int_{0}^{\infty}\frac{du}{\sqrt{\left(
\frac{\lambda_{x}^{2}}{\lambda_{z}^{2}}+u\right)  \left(  \frac{\lambda
_{y}^{2}}{\lambda_{z}^{2}}+u\right)  \left(  1+u\right)  }}\frac{1}{\left(
1+u\right)  \left(  \frac{\lambda_{a}^{2}}{\lambda_{z}^{2}}+u\right)
}\nonumber\\
& \nonumber\\
\lim_{\lambda_{x}\rightarrow\lambda_{z}}\overline{I}_{zx}  &  =\frac
{\lambda_{y}}{\lambda_{z}}\int_{0}^{\infty}\frac{du}{\sqrt{\frac{\lambda
_{y}^{2}}{\lambda_{z}^{2}}+u}}\frac{1}{\left(  1+u\right)  ^{3}}\equiv
I(\frac{\lambda_{y}}{\lambda_{z}})\nonumber\\
\lim_{\lambda_{y}\rightarrow\lambda_{z}}\overline{I}_{zy}  &  =\frac
{\lambda_{x}}{\lambda_{z}}\int_{0}^{\infty}\frac{du}{\sqrt{\frac{\lambda
_{x}^{2}}{\lambda_{z}^{2}}+u}}\frac{1}{\left(  1+u\right)  ^{3}}\equiv
I(\frac{\lambda_{x}}{\lambda_{z}})\nonumber\\
& \nonumber\\
I(q)  &  =q\int_{0}^{\infty}\frac{du}{\sqrt{q^{2}+u}}\frac{1}{\left(
1+u\right)  ^{3}}=\frac{q}{4\left(  q^{2}-1\right)  ^{2}}\left(
2q^{3}-5q+3\frac{\mbox{arccosh}\left(  q\right)  }{\sqrt{q^{2}-1}}\right)
\nonumber
\end{align}
In the isotropic case $\lambda_{z}=\lambda_{x}=\lambda_{y}$ :
\begin{equation}
\lim_{\lambda_{x}\rightarrow\lambda_{z}}\lim_{\lambda_{y}\rightarrow
\lambda_{z}}\overline{I}_{zy}=\lim_{\lambda_{y}\rightarrow\lambda_{z}}%
\lim_{\lambda_{x}\rightarrow\lambda_{z}}\overline{I}_{zx}=I(1)=\frac{2}{5}
\label{identity double index integral IIIb}%
\end{equation}
\newpage

\section{Sum Rule}
\label{appendixB}

The sum rule%
\begin{equation}
\sum_{a}w_{aa}\left(  t\right)  =5w_{00}\left(  t\right)  \label{sum rule w00}%
\end{equation}
follows directly from the defining equations
(\ref{diagonal coefficients spatial derivative term of wave equation} ) and
the properties of the triple index integrals:%
\begin{align}
& \label{sum of diagonal coefficients w_aa}\\
\sum_{a}w_{aa}\left(  t\right)   &  =\left\{
\begin{array}
[c]{c}%
\frac{3}{2}\varepsilon_{D}\cdot\left[
\begin{array}
[c]{c}%
\lambda_{x}^{2}\left[
\begin{array}
[c]{c}%
\cos^{2}\left(  \vartheta_{0}\right)  \left(  3I_{xxz}+I_{xyz}+3I_{xzz}\right)
\\
+3\sin^{2}\left(  \vartheta_{0}\right)  \left(  5I_{xxx}+I_{xxy}%
+I_{xxz}\right)
\end{array}
\right]  \rho_{xx}(t)\\
+\lambda_{y}^{2}\left[
\begin{array}
[c]{c}%
\cos^{2}\left(  \vartheta_{0}\right)  \left(  I_{xyz}+3I_{yyz}+3I_{yzz}\right)
\\
+\sin^{2}\left(  \vartheta_{0}\right)  \left(  3I_{xxy}+3I_{xyy}%
+I_{xyz}\right)
\end{array}
\right]  \rho_{yy}(t)\\
+\lambda_{z}^{2}\left[
\begin{array}
[c]{c}%
3\cos^{2}\left(  \vartheta_{0}\right)  \left(  I_{xzz}+I_{yzz}+5I_{zzz}\right)
\\
+\sin^{2}\left(  \vartheta_{0}\right)  \left(  3I_{xxz}+I_{xyz}+3I_{xzz}%
\right)
\end{array}
\right]  \rho_{zz}(t)\\
+\sin\left(  2\vartheta_{0}\right)  \lambda_{z}\lambda_{x}\left(
3I_{xxz}+I_{xyz}+3I_{xzz}\right)  \ \rho_{xz}(t)
\end{array}
\right] \\
\\
+5\left(  1-\varepsilon_{D}\right)  \left[  \frac{1}{\lambda_{x}^{2}}\rho
_{xx}(t)+\frac{1}{\lambda_{y}^{2}}\rho_{yy}(t)+\frac{1}{\lambda_{z}^{2}}%
\rho_{zz}(t)\right] \\
\mathbf{+}\left[  5\frac{1-\varepsilon_{D}}{2}\left(  \frac{1}{\lambda_{x}%
^{2}}+\frac{1}{\lambda_{y}^{2}}+\frac{1}{\lambda_{z}^{2}}\right)
+\frac{9\varepsilon_{D}}{2}\left(  \frac{\cos^{2}\left(  \vartheta_{0}\right)
}{\lambda_{z}^{2}}+\frac{\sin^{2}\left(  \vartheta_{0}\right)  }{\lambda
_{x}^{2}}\right)  \right]  \rho_{00}(t)\\
+9\varepsilon_{D}\left[  \frac{\sin^{2}\left(  \vartheta_{0}\right)  }%
{\lambda_{x}^{2}}\rho_{xx}(t)+\frac{\cos^{2}\left(  \vartheta_{0}\right)
}{\lambda_{z}^{2}}\rho_{zz}(t)+\frac{\sin\left(  2\vartheta_{0}\right)
}{2\lambda_{x}\lambda_{z}}\rho_{xz}(t)\right]
\end{array}
\right\} \nonumber
\end{align}
There holds the following identity for the index integrals $I_{abc}$:%
\begin{align}
a,b  &  \in\left\{  x,y,z\right\} \label{identity triple index integrals II}\\
2I_{aab}+2I_{abb}+I_{abx}+I_{aby}+I_{abz}  &  =\frac{2}{\lambda_{a}^{2}%
\lambda_{b}^{2}}\nonumber
\end{align}
Taking into account that the index integrals $I_{abc}$ are invariant under
permutations of their indices $a,b,c$ $\in\left\{  x,y,z\right\}  $ it follows
for the linear combinations encountered in
(\ref{sum of diagonal coefficients w_aa}):%
\[
3I_{xxz}+I_{xyz}+3I_{xzz}=2I_{xxz}+2I_{xzz}+I_{xzx}+I_{xzy}+I_{xzz}=\frac
{2}{\lambda_{x}^{2}\lambda_{z}^{2}}%
\]%
\[
5I_{xxx}+I_{xxy}+I_{xxz}=2I_{xxx}+2I_{xxx}+I_{xxx}+I_{xxy}+I_{xxz}=\frac
{2}{\lambda_{x}^{2}\lambda_{x}^{2}}=\frac{2}{\lambda_{x}^{4}}%
\]%
\[
I_{xyz}+3I_{yyz}+3I_{yzz}=2I_{yyz}+2I_{yzz}+I_{yzx}+I_{yzy}+I_{yzz}=\frac
{2}{\lambda_{y}^{2}\lambda_{z}^{2}}%
\]%
\[
3I_{xxy}+3I_{xyy}+I_{xyz}=2I_{xxy}+2I_{xyy}+I_{xyx}+I_{xyy}+I_{xyz}=\frac
{2}{\lambda_{x}^{2}\lambda_{y}^{2}}%
\]%
\[
I_{xzz}+I_{yzz}+5I_{zzz}=2I_{zzz}+2I_{zzz}+I_{zzx}+I_{zzy}+I_{zzz}=\frac
{2}{\lambda_{z}^{2}\lambda_{z}^{2}}=\frac{2}{\lambda_{z}^{4}}%
\]
Making use of these identities, and taking into account (\ref{Q00 II}), we
in deed see that%
\begin{align}
& \label{sum w_aa}\\
\sum_{a\in\left\{  x,y,z\right\}  }w_{aa}\left(  t\right)   &  =\left\{
\begin{array}
[c]{c}%
3\varepsilon_{D}\cdot\left[
\begin{array}
[c]{c}%
\left[  \cos^{2}\left(  \vartheta_{0}\right)  \frac{1}{\lambda_{z}^{2}}%
+3\sin^{2}\left(  \vartheta_{0}\right)  \frac{1}{\lambda_{x}^{2}}\right]
\rho_{xx}(t)\\
+\left[  \cos^{2}\left(  \vartheta_{0}\right)  \frac{1}{\lambda_{z}^{2}}%
+\sin^{2}\left(  \vartheta_{0}\right)  \frac{1}{\lambda_{x}^{2}}\right]
\rho_{yy}(t)\\
+\left[  3\cos^{2}\left(  \vartheta_{0}\right)  \frac{1}{\lambda_{z}^{2}}%
+\sin^{2}\left(  \vartheta_{0}\right)  \frac{1}{\lambda_{x}^{2}}\right]
\rho_{zz}(t)\\
+\sin\left(  2\vartheta_{0}\right)  \frac{1}{\lambda_{x}\lambda_{z}}%
\ \rho_{xz}(t)
\end{array}
\right] \\
+5\left(  1-\varepsilon_{D}\right)  \left[  \frac{1}{\lambda_{x}^{2}}\rho
_{xx}(t)+\frac{1}{\lambda_{y}^{2}}\rho_{yy}(t)+\frac{1}{\lambda_{z}^{2}}%
\rho_{zz}(t)\right] \\
\mathbf{+}\left[  5\frac{1-\varepsilon_{D}}{2}\left(  \frac{1}{\lambda_{x}%
^{2}}+\frac{1}{\lambda_{y}^{2}}+\frac{1}{\lambda_{z}^{2}}\right)
+\frac{9\varepsilon_{D}}{2}\left(  \frac{\cos^{2}\left(  \vartheta_{0}\right)
}{\lambda_{z}^{2}}+\frac{\sin^{2}\left(  \vartheta_{0}\right)  }{\lambda
_{x}^{2}}\right)  \right]  \rho_{00}(t)\\
+9\varepsilon_{D}\left[  \frac{\sin^{2}\left(  \vartheta_{0}\right)  }%
{\lambda_{x}^{2}}\rho_{xx}(t)+\frac{\cos^{2}\left(  \vartheta_{0}\right)
}{\lambda_{z}^{2}}\rho_{zz}(t)+\frac{\sin\left(  2\vartheta_{0}\right)
}{2\lambda_{x}\lambda_{z}}\rho_{xz}(t)\right]
\end{array}
\right\} \nonumber\\
& \nonumber\\
&  =\left\{
\begin{array}
[c]{c}%
3\varepsilon_{D}\left[  \cos^{2}\left(  \vartheta_{0}\right)  \frac{1}%
{\lambda_{z}^{2}}+\sin^{2}\left(  \vartheta_{0}\right)  \frac{1}{\lambda
_{x}^{2}}\right]  \left[  \rho_{xx}(t)+\rho_{yy}(t)+\rho_{zz}(t)\right] \\
+5\left(  1-\varepsilon_{D}\right)  \left[  \frac{1}{\lambda_{x}^{2}}\rho
_{xx}(t)+\frac{1}{\lambda_{y}^{2}}\rho_{yy}(t)+\frac{1}{\lambda_{z}^{2}}%
\rho_{zz}(t)\right] \\
\mathbf{+}\left[  5\frac{1-\varepsilon_{D}}{2}\left(  \frac{1}{\lambda_{x}%
^{2}}+\frac{1}{\lambda_{y}^{2}}+\frac{1}{\lambda_{z}^{2}}\right)
+\frac{9\varepsilon_{D}}{2}\left(  \frac{\cos^{2}\left(  \vartheta_{0}\right)
}{\lambda_{z}^{2}}+\frac{\sin^{2}\left(  \vartheta_{0}\right)  }{\lambda
_{x}^{2}}\right)  \right]  \rho_{00}(t)\\
+3\varepsilon_{D}\cdot\left(  3+2\right)  \left[  \frac{\sin^{2}\left(
\vartheta_{0}\right)  }{\lambda_{x}^{2}}\rho_{xx}(t)+\frac{\cos^{2}\left(
\vartheta_{0}\right)  }{\lambda_{z}^{2}}\rho_{zz}(t)+\frac{\sin\left(
2\vartheta_{0}\right)  }{2\lambda_{x}\lambda_{z}}\rho_{xz}(t)\right]
\end{array}
\right\} \nonumber\\
& \nonumber\\
&  =5\cdot\left\{
\begin{array}
[c]{c}%
\left(  1-\varepsilon_{D}\right)  \left[  \frac{1}{\lambda_{x}^{2}}\rho
_{xx}\left(  t\right)  +\frac{1}{\lambda_{y}^{2}}\rho_{yy}\left(  t\right)
+\frac{1}{\lambda_{z}^{2}}\rho_{zz}\left(  t\right)  \right] \\
\\
+3\varepsilon_{D}\left[  \frac{\cos^{2}\left(  \vartheta_{0}\right)  }%
{\lambda_{z}^{2}}\rho_{zz}\left(  t\right)  +\frac{\sin^{2}\left(
\vartheta_{0}\right)  }{\lambda_{x}^{2}}\rho_{xx}\left(  t\right)  +\frac
{\sin\left(  2\vartheta_{0}\right)  }{2\lambda_{x}\lambda_{z}}\rho_{xz}\left(
t\right)  \right] \\
\\
\mathbf{+}\left[  \frac{1-\varepsilon_{D}}{2}\left(  \frac{1}{\lambda_{x}^{2}%
}+\frac{1}{\lambda_{y}^{2}}+\frac{1}{\lambda_{z}^{2}}\right)  +\frac
{3\varepsilon_{D}}{2}\left(  \frac{\cos^{2}\left(  \vartheta_{0}\right)
}{\lambda_{z}^{2}}+\frac{\sin^{2}\left(  \vartheta_{0}\right)  }{\lambda
_{x}^{2}}\right)  \right]  \rho_{00}\left(  t\right)
\end{array}
\right\} \nonumber\\
& \nonumber\\
&  =5w_{00}\left(  t\right) \nonumber
\end{align}

\bigskip\newpage

\section{Explicit Expressions for $G_{ab}(t)-F_{ab}(t)$ and Matrix Elements
$C_{ab,cd}$ .}
\label{appendixC}

We collect here explicit expressions for the linear combinations
$G_{ab}(t)-F_{ab}(t)$ that occur in
(\ref{diagonal coefficients spatial derivative term of wave equation}),%

\begin{align}
& \label{Gaa-Faa}\\
G_{xx}(t)-F_{xx}(t)  &  =\frac{1}{\lambda_{x}^{2}}\left\{
\begin{array}
[c]{c}%
3\lambda_{x}^{2}\left[  \cos^{2}\left(  \vartheta_{0}\right)  I_{xxz}%
+5\sin^{2}\left(  \vartheta_{0}\right)  I_{xxx}\right]  \rho_{xx}\left(
t\right) \\
+\lambda_{y}^{2}\left[  \cos^{2}\left(  \vartheta_{0}\right)  I_{xyz}%
+3\sin^{2}\left(  \vartheta_{0}\right)  I_{xxy}\right]  \rho_{yy}\left(
t\right) \\
+3\lambda_{z}^{2}\left[  \cos^{2}\left(  \vartheta_{0}\right)  I_{xzz}%
+\sin^{2}\left(  \vartheta_{0}\right)  I_{xxz}\right]  \rho_{zz}\left(
t\right) \\
+3\sin\left(  2\vartheta_{0}\right)  I_{xxz}\lambda_{x}\lambda_{z}\rho
_{xz}\left(  t\right)
\end{array}
\right\} \nonumber
\end{align}%
\[
G_{yy}(t)-F_{yy}(t)=\frac{1}{\lambda_{y}^{2}}\left\{
\begin{array}
[c]{c}%
\lambda_{x}^{2}\left[  \cos^{2}\left(  \vartheta_{0}\right)  I_{xyz}+3\sin
^{2}\left(  \vartheta_{0}\right)  I_{xxy}\right]  \rho_{xx}\left(  t\right) \\
+3\lambda_{y}^{2}\left[  \cos^{2}\left(  \vartheta_{0}\right)  I_{yyz}%
+\sin^{2}\left(  \vartheta_{0}\right)  I_{xyy}\right]  \rho_{yy}\left(
t\right) \\
+\lambda_{z}^{2}\left[  3\cos^{2}\left(  \vartheta_{0}\right)  I_{yzz}%
+\sin^{2}\left(  \vartheta_{0}\right)  I_{xyz}\right]  \rho_{zz}\left(
t\right) \\
+\sin\left(  2\vartheta_{0}\right)  I_{xyz}\cdot\lambda_{x}\lambda_{z}%
\rho_{xz}\left(  t\right)
\end{array}
\right\}
\]%
\[
G_{zz}(t)-F_{zz}(t)=\frac{1}{\lambda_{z}^{2}}\left\{
\begin{array}
[c]{c}%
3\lambda_{x}^{2}\left[  \cos^{2}\left(  \vartheta_{0}\right)  I_{xzz}+\sin
^{2}\left(  \vartheta_{0}\right)  I_{xxz}\right]  \rho_{xx}\left(  t\right) \\
+\lambda_{y}^{2}\left[  3\cos^{2}\left(  \vartheta_{0}\right)  I_{yzz}%
+\sin^{2}\left(  \vartheta_{0}\right)  I_{xyz}\right]  \rho_{yy}\left(
t\right) \\
+3\lambda_{z}^{2}\left[  5\cos^{2}\left(  \vartheta_{0}\right)  I_{zzz}%
+\sin^{2}\left(  \vartheta_{0}\right)  I_{xzz}\right]  \rho_{zz}\left(
t\right) \\
+3\sin\left(  2\vartheta_{0}\right)  I_{xzz}\lambda_{x}\lambda_{z}\rho
_{xz}\left(  t\right)
\end{array}
\right\}
\]
, and also in
(\ref{off diagonal coefficients of derivative term of wave equation}):%
\[
G_{xz}(t)-F_{xz}(t)=\left\{
\begin{array}
[c]{c}%
3\left(  \frac{1}{\lambda_{x}^{2}}+\frac{1}{\lambda_{z}^{2}}\right)
\lambda_{x}\lambda_{z}\left[  \cos^{2}\left(  \vartheta_{0}\right)
I_{xzz}+\sin^{2}\left(  \vartheta_{0}\right)  I_{xxz}\right]  \rho_{xz}\left(
t\right) \\
+\sin\left(  2\vartheta_{0}\right)  \left[
\begin{array}
[c]{c}%
3\left(  1+\frac{\lambda_{x}^{2}}{\lambda_{z}^{2}}\right)  I_{xxz}\rho
_{xx}\left(  t\right)  +\left(  \frac{\lambda_{y}^{2}}{\lambda_{x}^{2}}%
+\frac{\lambda_{y}^{2}}{\lambda_{z}^{2}}\right)  I_{xyz}\rho_{yy}\left(
t\right) \\
+3\left(  \frac{\lambda_{z}^{2}}{\lambda_{x}^{2}}+1\right)  I_{xzz}\rho
_{zz}\left(  t\right)
\end{array}
\right]
\end{array}
\right\}
\]%
\begin{align*}
G_{yz}(t)-F_{yz}(t)  &  =\left(  \frac{1}{\lambda_{y}^{2}}+\frac{1}%
{\lambda_{z}^{2}}\right)  \left[
\begin{array}
[c]{c}%
\sin\left(  2\vartheta_{0}\right)  \lambda_{x}\lambda_{y}I_{xyz}\rho
_{xy}\left(  t\right)  \ \\
+\lambda_{y}\lambda_{z}\left[  3\cos^{2}\left(  \vartheta_{0}\right)
I_{yzz}+\sin^{2}\left(  \vartheta_{0}\right)  I_{xyz}\right]  \rho_{yz}\left(
t\right)  \
\end{array}
\right] \\
& \\
G_{xy}(t)-F_{xy}(t)  &  =\left(  \frac{1}{\lambda_{x}^{2}}+\frac{1}%
{\lambda_{y}^{2}}\right)  \left[
\begin{array}
[c]{c}%
\lambda_{x}\lambda_{y}\left[  \cos^{2}\left(  \vartheta_{0}\right)
I_{xyz}+3\sin^{2}\left(  \vartheta_{0}\right)  I_{xxy}\right]  \rho
_{xy}\left(  t\right)  \ \\
+\lambda_{y}\lambda_{z}\sin\left(  2\vartheta_{0}\right)  I_{xyz}\rho
_{yz}\left(  t\right)  \ \
\end{array}
\right]
\end{align*}
The matrix elements $C_{ab,cd}$ occurring in the eigenvalue problem
(\ref{collective mode eigenvalue problem}) are explicitely given by%

\begin{align}
& \label{coefficients Cxx,ab}\\
\frac{2n_{0}g^{\left(  s\right)  }}{m^{\star}}C_{xx,xx}  &  =\omega_{y}%
^{2}\frac{\lambda_{y}^{2}}{\lambda_{x}^{2}}\frac{3\left(  1-\varepsilon
_{D}\right)  +\frac{9}{2}\varepsilon_{D}\cdot\frac{\lambda_{x}^{4}}%
{\lambda_{z}^{4}}\left[  \cos^{2}\left(  \vartheta_{0}\right)  \overline
{I}_{xxz}+5\sin^{2}\left(  \vartheta_{0}\right)  \overline{I}_{xxx}\right]
}{1-\varepsilon_{D}+\frac{3\varepsilon_{D}}{2}\frac{\lambda_{y}^{2}}%
{\lambda_{z}^{2}}\left[  \cos^{2}\left(  \vartheta_{0}\right)  \overline
{I}_{zy}+\sin^{2}\left(  \vartheta_{0}\right)  \overline{I}_{xy}\right]
}\nonumber\\
& \nonumber\\
\frac{2n_{0}g^{\left(  s\right)  }}{m^{\star}}C_{xx,yy}  &  =\omega_{y}%
^{2}\frac{\lambda_{y}^{2}}{\lambda_{x}^{2}}\frac{\left(  1-\varepsilon
_{D}\right)  +\frac{3}{2}\varepsilon_{D}\cdot\frac{\lambda_{x}^{2}}%
{\lambda_{z}^{2}}\frac{\lambda_{y}^{2}}{\lambda_{z}^{2}}\left[  \cos
^{2}\left(  \vartheta_{0}\right)  \overline{I}_{xyz}+3\sin^{2}\left(
\vartheta_{0}\right)  \overline{I}_{xxy}\right]  }{1-\varepsilon_{D}%
+\frac{3\varepsilon_{D}}{2}\frac{\lambda_{y}^{2}}{\lambda_{z}^{2}}\left[
\cos^{2}\left(  \vartheta_{0}\right)  \overline{I}_{zy}+\sin^{2}\left(
\vartheta_{0}\right)  \overline{I}_{xy}\right]  }\nonumber\\
& \nonumber\\
\frac{2n_{0}g^{\left(  s\right)  }}{m^{\star}}C_{xx,zz}  &  =\omega_{y}%
^{2}\frac{\lambda_{y}^{2}}{\lambda_{x}^{2}}\frac{\left(  1-\varepsilon
_{D}\right)  +\frac{9}{2}\varepsilon_{D}\cdot\frac{\lambda_{x}^{2}}%
{\lambda_{z}^{2}}\left[  \cos^{2}\left(  \vartheta_{0}\right)  \overline
{I}_{xzz}+\sin^{2}\left(  \vartheta_{0}\right)  \overline{I}_{xxz}\right]
}{1-\varepsilon_{D}+\frac{3\varepsilon_{D}}{2}\frac{\lambda_{y}^{2}}%
{\lambda_{z}^{2}}\left[  \cos^{2}\left(  \vartheta_{0}\right)  \overline
{I}_{zy}+\sin^{2}\left(  \vartheta_{0}\right)  \overline{I}_{xy}\right]
}\nonumber\\
& \nonumber\\
\frac{2n_{0}g^{\left(  s\right)  }}{m^{\star}}C_{xx,xz}  &  =\omega_{y}%
^{2}\frac{\frac{9}{2}\varepsilon_{D}\sin\left(  2\vartheta_{0}\right)
\frac{\lambda_{x}}{\lambda_{z}}\frac{\lambda_{y}^{2}}{\lambda_{z}^{2}%
}\overline{I}_{xxz}}{1-\varepsilon_{D}+\frac{3\varepsilon_{D}}{2}\frac
{\lambda_{y}^{2}}{\lambda_{z}^{2}}\left[  \cos^{2}\left(  \vartheta
_{0}\right)  \overline{I}_{zy}+\sin^{2}\left(  \vartheta_{0}\right)
\overline{I}_{xy}\right]  }\nonumber
\end{align}%
\begin{align}
& \label{coefficients Cyy,ab}\\
\frac{2n_{0}g^{\left(  s\right)  }}{m^{\star}}C_{yy,xx}  &  =\omega_{y}%
^{2}\frac{\left(  1-\varepsilon_{D}\right)  +\frac{3}{2}\varepsilon_{D}%
\cdot\frac{\lambda_{x}^{2}}{\lambda_{z}^{2}}\frac{\lambda_{y}^{2}}{\lambda
_{z}^{2}}\left[  \cos^{2}\left(  \vartheta_{0}\right)  \overline{I}%
_{xyz}+3\sin^{2}\left(  \vartheta_{0}\right)  \overline{I}_{xxy}\right]
}{1-\varepsilon_{D}+\frac{3\varepsilon_{D}}{2}\frac{\lambda_{y}^{2}}%
{\lambda_{z}^{2}}\left[  \cos^{2}\left(  \vartheta_{0}\right)  \overline
{I}_{zy}+\sin^{2}\left(  \vartheta_{0}\right)  \overline{I}_{xy}\right]
}\nonumber\\
& \nonumber\\
\frac{2n_{0}g^{\left(  s\right)  }}{m^{\star}}C_{yy,yy}  &  =\omega_{y}%
^{2}\frac{3\left(  1-\varepsilon_{D}\right)  +\frac{9}{2}\varepsilon_{D}%
\cdot\frac{\lambda_{y}^{4}}{\lambda_{z}^{4}}\left[  \cos^{2}\left(
\vartheta_{0}\right)  \overline{I}_{yyz}+\sin^{2}\left(  \vartheta_{0}\right)
\overline{I}_{xyy}\right]  }{1-\varepsilon_{D}+\frac{3\varepsilon_{D}}{2}%
\frac{\lambda_{y}^{2}}{\lambda_{z}^{2}}\left[  \cos^{2}\left(  \vartheta
_{0}\right)  \overline{I}_{zy}+\sin^{2}\left(  \vartheta_{0}\right)
\overline{I}_{xy}\right]  }\nonumber\\
& \nonumber\\
\frac{2n_{0}g^{\left(  s\right)  }}{m^{\star}}C_{yy,zz}  &  =\omega_{y}%
^{2}\frac{\left(  1-\varepsilon_{D}\right)  +\frac{3}{2}\varepsilon_{D}%
\cdot\frac{\lambda_{y}^{2}}{\lambda_{z}^{2}}\left[  3\cos^{2}\left(
\vartheta_{0}\right)  \overline{I}_{yzz}+\sin^{2}\left(  \vartheta_{0}\right)
\overline{I}_{xyz}\right]  }{1-\varepsilon_{D}+\frac{3\varepsilon_{D}}{2}%
\frac{\lambda_{y}^{2}}{\lambda_{z}^{2}}\left[  \cos^{2}\left(  \vartheta
_{0}\right)  \overline{I}_{zy}+\sin^{2}\left(  \vartheta_{0}\right)
\overline{I}_{xy}\right]  }\nonumber\\
& \nonumber\\
\frac{2n_{0}g^{\left(  s\right)  }}{m^{\star}}C_{yy,xz}  &  =\omega_{y}%
^{2}\frac{\frac{3}{2}\varepsilon_{D}\sin\left(  2\vartheta_{0}\right)
\frac{\lambda_{x}}{\lambda_{z}}\frac{\lambda_{y}^{2}}{\lambda_{z}^{2}%
}\overline{I}_{xyz}}{1-\varepsilon_{D}+\frac{3\varepsilon_{D}}{2}\frac
{\lambda_{y}^{2}}{\lambda_{z}^{2}}\left[  \cos^{2}\left(  \vartheta
_{0}\right)  \overline{I}_{zy}+\sin^{2}\left(  \vartheta_{0}\right)
\overline{I}_{xy}\right]  }\nonumber
\end{align}%
\begin{align}
& \label{coefficients Czz,ab}\\
\frac{2n_{0}g^{\left(  s\right)  }}{m^{\star}}C_{zz,xx}  &  =\omega_{y}%
^{2}\frac{\lambda_{y}^{2}}{\lambda_{z}^{2}}\frac{\left(  1-\varepsilon
_{D}\right)  +\frac{9}{2}\varepsilon_{D}\cdot\frac{\lambda_{x}^{2}}%
{\lambda_{z}^{2}}\left[  \cos^{2}\left(  \vartheta_{0}\right)  \overline
{I}_{xzz}+\sin^{2}\left(  \vartheta_{0}\right)  \overline{I}_{xxz}\right]
}{1-\varepsilon_{D}+\frac{3\varepsilon_{D}}{2}\frac{\lambda_{y}^{2}}%
{\lambda_{z}^{2}}\left[  \cos^{2}\left(  \vartheta_{0}\right)  \overline
{I}_{zy}+\sin^{2}\left(  \vartheta_{0}\right)  \overline{I}_{xy}\right]
}\nonumber\\
& \nonumber\\
\frac{2n_{0}g^{\left(  s\right)  }}{m^{\star}}C_{zz,yy}  &  =\omega_{y}%
^{2}\frac{\lambda_{y}^{2}}{\lambda_{z}^{2}}\frac{\left(  1-\varepsilon
_{D}\right)  +\frac{3}{2}\varepsilon_{D}\cdot\frac{\lambda_{y}^{2}}%
{\lambda_{z}^{2}}\left[  3\cos^{2}\left(  \vartheta_{0}\right)  \overline
{I}_{yzz}+\sin^{2}\left(  \vartheta_{0}\right)  \overline{I}_{xyz}\right]
}{1-\varepsilon_{D}+\frac{3\varepsilon_{D}}{2}\frac{\lambda_{y}^{2}}%
{\lambda_{z}^{2}}\left[  \cos^{2}\left(  \vartheta_{0}\right)  \overline
{I}_{zy}+\sin^{2}\left(  \vartheta_{0}\right)  \overline{I}_{xy}\right]
}\nonumber\\
& \nonumber\\
\frac{2n_{0}g^{\left(  s\right)  }}{m^{\star}}C_{zz,zz}  &  =\omega_{y}%
^{2}\frac{\lambda_{y}^{2}}{\lambda_{z}^{2}}\frac{3\left(  1-\varepsilon
_{D}\right)  +\frac{9}{2}\varepsilon_{D}\cdot\left[  5\cos^{2}\left(
\vartheta_{0}\right)  \overline{I}_{zzz}+\sin^{2}\left(  \vartheta_{0}\right)
\overline{I}_{xzz}\right]  }{1-\varepsilon_{D}+\frac{3\varepsilon_{D}}{2}%
\frac{\lambda_{y}^{2}}{\lambda_{z}^{2}}\left[  \cos^{2}\left(  \vartheta
_{0}\right)  \overline{I}_{zy}+\sin^{2}\left(  \vartheta_{0}\right)
\overline{I}_{xy}\right]  }\nonumber\\
& \nonumber\\
\frac{2n_{0}g^{\left(  s\right)  }}{m^{\star}}C_{zz,xz}  &  =\omega_{y}%
^{2}\frac{\frac{9}{2}\varepsilon_{D}\sin\left(  2\vartheta_{0}\right)
\frac{\lambda_{x}}{\lambda_{z}}\frac{\lambda_{y}^{2}}{\lambda_{z}^{2}%
}\overline{I}_{xzz}}{1-\varepsilon_{D}+\frac{3\varepsilon_{D}}{2}\frac
{\lambda_{y}^{2}}{\lambda_{z}^{2}}\left[  \cos^{2}\left(  \vartheta
_{0}\right)  \overline{I}_{zy}+\sin^{2}\left(  \vartheta_{0}\right)
\overline{I}_{xy}\right]  }\nonumber
\end{align}%
\begin{align}
& \label{coefficients Cxz,ab}\\
\frac{2n_{0}g^{\left(  s\right)  }}{m^{\star}}C_{xz,xx}  &  =\omega_{y}%
^{2}\left(  1+\frac{\lambda_{x}^{2}}{\lambda_{z}^{2}}\right)  \frac{\frac
{9}{2}\varepsilon_{D}\cdot\sin\left(  2\vartheta_{0}\right)  \frac{\lambda
_{x}}{\lambda_{z}}\frac{\lambda_{y}^{2}}{\lambda_{z}^{2}}\overline{I}_{xxz}%
}{1-\varepsilon_{D}+\frac{3\varepsilon_{D}}{2}\frac{\lambda_{y}^{2}}%
{\lambda_{z}^{2}}\left[  \cos^{2}\left(  \vartheta_{0}\right)  \overline
{I}_{zy}+\sin^{2}\left(  \vartheta_{0}\right)  \overline{I}_{xy}\right]
}\nonumber\\
& \nonumber\\
\frac{2n_{0}g^{\left(  s\right)  }}{m^{\star}}C_{xz,yy}  &  =\omega_{y}%
^{2}\left(  \frac{\lambda_{y}^{2}}{\lambda_{x}^{2}}+\frac{\lambda_{y}^{2}%
}{\lambda_{z}^{2}}\right)  \frac{\frac{3}{2}\varepsilon_{D}\cdot\sin\left(
2\vartheta_{0}\right)  \frac{\lambda_{x}}{\lambda_{z}}\frac{\lambda_{y}^{2}%
}{\lambda_{z}^{2}}\overline{I}_{xyz}}{1-\varepsilon_{D}+\frac{3\varepsilon
_{D}}{2}\frac{\lambda_{y}^{2}}{\lambda_{z}^{2}}\left[  \cos^{2}\left(
\vartheta_{0}\right)  \overline{I}_{zy}+\sin^{2}\left(  \vartheta_{0}\right)
\overline{I}_{xy}\right]  }\nonumber\\
& \nonumber\\
\frac{2n_{0}g^{\left(  s\right)  }}{m^{\star}}C_{xz,zz}  &  =\omega_{y}%
^{2}\left(  1+\frac{\lambda_{z}^{2}}{\lambda_{x}^{2}}\right)  \frac{\frac
{9}{2}\varepsilon_{D}\cdot\sin\left(  2\vartheta_{0}\right)  \frac{\lambda
_{x}}{\lambda_{z}}\frac{\lambda_{y}^{2}}{\lambda_{z}^{2}}\overline{I}_{xzz}%
}{1-\varepsilon_{D}+\frac{3\varepsilon_{D}}{2}\frac{\lambda_{y}^{2}}%
{\lambda_{z}^{2}}\left[  \cos^{2}\left(  \vartheta_{0}\right)  \overline
{I}_{zy}+\sin^{2}\left(  \vartheta_{0}\right)  \overline{I}_{xy}\right]
}\nonumber\\
& \nonumber\\
\frac{2n_{0}g^{\left(  s\right)  }}{m^{\star}}C_{xz,xz}  &  =\omega_{y}%
^{2}\left(  \frac{\lambda_{y}^{2}}{\lambda_{x}^{2}}+\frac{\lambda_{y}^{2}%
}{\lambda_{z}^{2}}\right)  \frac{\left(  1-\varepsilon_{D}\right)  +\frac
{9}{2}\varepsilon_{D}\cdot\frac{\lambda_{x}^{2}}{\lambda_{z}^{2}}\left[
\cos^{2}\left(  \vartheta_{0}\right)  \overline{I}_{xzz}+\sin^{2}\left(
\vartheta_{0}\right)  \overline{I}_{xxz}\right]  }{1-\varepsilon_{D}%
+\frac{3\varepsilon_{D}}{2}\frac{\lambda_{y}^{2}}{\lambda_{z}^{2}}\left[
\cos^{2}\left(  \vartheta_{0}\right)  \overline{I}_{zy}+\sin^{2}\left(
\vartheta_{0}\right)  \overline{I}_{xy}\right]  }\nonumber
\end{align}%
\begin{align}
& \label{Coefficients Cyz,ab}\\
\frac{2n_{0}g^{\left(  s\right)  }}{m^{\star}}C_{yz,yz}  &  =\omega_{y}%
^{2}\left(  1+\frac{\lambda_{y}^{2}}{\lambda_{z}^{2}}\right)  \frac{\left(
1-\varepsilon_{D}\right)  +\frac{3}{2}\varepsilon_{D}\frac{\lambda_{y}^{2}%
}{\lambda_{z}^{2}}\left[  3\cos^{2}\left(  \vartheta_{0}\right)  \overline
{I}_{yzz}+\sin^{2}\left(  \vartheta_{0}\right)  \overline{I}_{xyz}\right]
}{1-\varepsilon_{D}+\frac{3\varepsilon_{D}}{2}\frac{\lambda_{y}^{2}}%
{\lambda_{z}^{2}}\left[  \cos^{2}\left(  \vartheta_{0}\right)  \overline
{I}_{zy}+\sin^{2}\left(  \vartheta_{0}\right)  \overline{I}_{xy}\right]
}\nonumber\\
& \nonumber\\
\frac{2n_{0}g^{\left(  s\right)  }}{m^{\star}}C_{yz,xy}  &  =\omega_{y}%
^{2}\left(  1+\frac{\lambda_{y}^{2}}{\lambda_{z}^{2}}\right)  \frac{\frac
{3}{2}\varepsilon_{D}\sin\left(  2\vartheta_{0}\right)  \frac{\lambda_{x}%
}{\lambda_{z}}\frac{\lambda_{y}^{2}}{\lambda_{z}^{2}}\overline{I}_{xyz}%
}{1-\varepsilon_{D}+\frac{3\varepsilon_{D}}{2}\frac{\lambda_{y}^{2}}%
{\lambda_{z}^{2}}\left[  \cos^{2}\left(  \vartheta_{0}\right)  \overline
{I}_{zy}+\sin^{2}\left(  \vartheta_{0}\right)  \overline{I}_{xy}\right]
}\nonumber
\end{align}%
\begin{align}
& \label{coefficients Cxy,ab}\\
\frac{2n_{0}g^{\left(  s\right)  }}{m^{\star}}C_{xy,yz}  &  =\omega_{y}%
^{2}\left(  1+\frac{\lambda_{y}^{2}}{\lambda_{x}^{2}}\right)  \frac{\frac
{3}{2}\varepsilon_{D}\sin\left(  2\vartheta_{0}\right)  \frac{\lambda_{x}%
}{\lambda_{z}}\frac{\lambda_{y}^{2}}{\lambda_{z}^{2}}\overline{I}_{xyz}%
}{1-\varepsilon_{D}+\frac{3\varepsilon_{D}}{2}\frac{\lambda_{y}^{2}}%
{\lambda_{z}^{2}}\left[  \cos^{2}\left(  \vartheta_{0}\right)  \overline
{I}_{zy}+\sin^{2}\left(  \vartheta_{0}\right)  \overline{I}_{xy}\right]
}\nonumber\\
& \nonumber\\
\frac{2n_{0}g^{\left(  s\right)  }}{m^{\star}}C_{xy,xy}  &  =\omega_{y}%
^{2}\left(  1+\frac{\lambda_{y}^{2}}{\lambda_{x}^{2}}\right)  \frac{\left(
1-\varepsilon_{D}\right)  +\frac{3}{2}\varepsilon_{D}\frac{\lambda_{x}^{2}%
}{\lambda_{z}^{2}}\frac{\lambda_{y}^{2}}{\lambda_{z}^{2}}\left[  \cos
^{2}\left(  \vartheta_{0}\right)  \overline{I}_{xyz}+3\sin^{2}\left(
\vartheta_{0}\right)  \overline{I}_{xxy}\right]  }{1-\varepsilon_{D}%
+\frac{3\varepsilon_{D}}{2}\frac{\lambda_{y}^{2}}{\lambda_{z}^{2}}\left[
\cos^{2}\left(  \vartheta_{0}\right)  \overline{I}_{zy}+\sin^{2}\left(
\vartheta_{0}\right)  \overline{I}_{xy}\right]  }\nonumber
\end{align}
Here, the quantities $\overline{I}_{ab}$ and $\overline{I}_{abc}$ denote
(scaled) double- and triple index integrals, as explained in
(\ref{scaling index integrals}).

\section{Coupled Monopole-Quadrupole Modes of Density Oscillations}
\label{appendixD}

For completeness, we discuss here the coupled small amplitude
monopole-quadrupole oscillations of density for a BEC confined in a harmonic
trap with cylindrical (uniaxial) symmetry, restricting to the case of zero
dipole-dipole interaction, $\varepsilon_D=0$. Setting $\omega_{z}\neq\ \omega_{y}=\omega
_{x}=\omega_{\perp}$ in the eigenvalue problem
(\ref{collective modes for eps_D=0}) we easily find analytical expressions
for three eigenmodes. First
\begin{align}
\Omega_{x^{2}-y^{2}}^{\left(  0\right)  }  &  =\sqrt{2}\omega_{\perp
}\label{cylindrical trap eps_D=0 collective mode d_(x^2-y^2)-wave}\\
& \nonumber\\
\left[
\begin{array}
[c]{c}%
\widehat{\rho}_{xx}\left(  \Omega_{x^{2}-y^{2}}^{\left(  0\right)  }\right) \\
\widehat{\rho}_{yy}\left(  \Omega_{x^{2}-y^{2}}^{\left(  0\right)  }\right) \\
\widehat{\rho}_{zz}\left(  \Omega_{x^{2}-y^{2}}^{\left(  0\right)  }\right)
\end{array}
\right]   &  =\left[
\begin{array}
[c]{c}%
1\\
-1\\
0
\end{array}
\right] \nonumber
\end{align}
It follows directly from (\ref{density fluctuation II}) that this eigenmode
corresponds for all anisotropy ratios to a density fluctuation $\delta
n_{\Omega}\left(  \mathbf{r},t\right)  $ with pure $d_{x^{2}-y^{2}}%
$-symmetry:
\begin{align*}
\Omega &  =\Omega_{x^{2}-y^{2}}^{\left(  0\right)  }\\
\delta n_{\Omega}\left(  \mathbf{r},t\right)   &  =2n_{0}\cos\left(  \Omega
t+\delta_{\Omega}\right)  \frac{r_{x}^{2}-r_{y}^{2}}{\left[  \lambda_{\perp
}^{\left(  0\right)  }\right]  ^{2}}%
\end{align*}
The second and third eigenmodes $\widehat{\rho}_{aa}\left(  \Omega_{\pm
\ \ }^{\left(  0\right)  }\right)  $ with eigenfrequencies $\Omega_{\pm
\ \ }^{\left(  0\right)  }$ form a doublet consisting of a combination of
basis elements with $s$-wave and $d_{z^{2}}$-wave symmetry. We obtain as a
function of the anisotropy ratio $\nu=\frac{\omega_{z}}{\omega_{\perp}}$ the
following exact results for the eigenfrequencies and the eigenvectors:
\begin{align}
\Omega_{+\ \ }^{\left(  0\right)  }  &  =\omega_{\perp}\left[  \frac
{4+3\nu^{2}+\sqrt{16-16\nu^{2}+9\nu^{4}}}{2}\right]  ^{\frac{1}{2}}\nonumber\\
& \label{cylindrical trap eps_D=0 collective mode s-wave}\\
\left[
\begin{array}
[c]{c}%
\widehat{\rho}_{xx}\left(  \Omega_{+\ \ }^{\left(  0\right)  }\right) \\
\widehat{\rho}_{yy}\left(  \Omega_{+\ \ }^{\left(  0\right)  }\right) \\
\widehat{\rho}_{zz}\left(  \Omega_{+\ \ }^{\left(  0\right)  }\right)
\end{array}
\right]   &  =\left[
\begin{array}
[c]{c}%
\frac{4-3\nu^{2}+\sqrt{16-16\nu^{2}+9\nu^{4}}}{4\nu^{2}}\\
\\
\frac{4-3\nu^{2}+\sqrt{16-16\nu^{2}+9\nu^{4}}}{4\nu^{2}}\\
\\
1
\end{array}
\right] \nonumber
\end{align}
For $\nu\rightarrow\infty$ this mode becomes quasi one-dimensional
\begin{align*}
\nu &  >>1\\
& \\
\Omega_{+\ \ }^{\left(  0\right)  }  &  =\sqrt{3}\omega_{z}\left(  1+\frac
{1}{9\nu^{2}}+...\right) \\
& \\
\left[
\begin{array}
[c]{c}%
\widehat{\rho}_{xx}\left(  \Omega_{+\ \ }^{\left(  0\right)  }\right) \\
\widehat{\rho}_{yy}\left(  \Omega_{+\ \ }^{\left(  0\right)  }\right) \\
\widehat{\rho}_{zz}\left(  \Omega_{+\ \ }^{\left(  0\right)  }\right)
\end{array}
\right]   &  =\left[
\begin{array}
[c]{c}%
\frac{1}{3\nu^{2}}+...\\
\frac{1}{3\nu^{2}}+...\\
1
\end{array}
\right]
\end{align*}
, while for $\nu\rightarrow0$ it is quasi two-dimensional:
\begin{align*}
\Omega_{+\ \ }^{\left(  0\right)  }  &  =2\omega_{\perp}\left(  1+\frac{1}%
{16}\nu^{2}+....\right) \\
& \\
\left[
\begin{array}
[c]{c}%
\widehat{\rho}_{xx}\left(  \Omega_{+\ \ }^{\left(  0\right)  }\right) \\
\widehat{\rho}_{yy}\left(  \Omega_{+\ \ }^{\left(  0\right)  }\right) \\
\widehat{\rho}_{zz}\left(  \Omega_{+\ \ }^{\left(  0\right)  }\right)
\end{array}
\right]   &  =\left[
\begin{array}
[c]{c}%
\frac{2}{\nu^{2}}-\frac{5}{4}+...\\
\frac{2}{\nu^{2}}-\frac{5}{4}+...\\
1
\end{array}
\right]
\end{align*}
For an anisotropy ratio $\nu$ $\simeq1\ $(slightly deformed sphere) the
associated density fluctuation $\delta n_{\Omega}\left(  \mathbf{r},t\right)
$ is of the \emph{breather} type, i.e. a strongly weighted isotropic $s$-wave
part is combined with only a small admixture of quadrupolar $d_{z^{2}}$-wave
symmetry:
\begin{align}
\Omega &  =\Omega_{+\ \ }^{\left(  0\right)  }%
\label{cylindrical trap eps_D=0 density fluctuation s -wave}\\
\delta n_{\Omega}\left(  \mathbf{r},t\right)   &  =2n_{0}\cos\left(  \Omega
t+\delta_{\Omega}\right)  \left[
\begin{array}
[c]{c}%
\left(  \frac{4-3\nu^{2}+\sqrt{16-16\nu^{2}+9\nu^{4}}}{2\nu^{2}}+\frac{1}%
{2}\right)  \frac{r_{x}^{2}+r_{y}^{2}}{\left[  \lambda_{\perp}^{\left(
0\right)  }\right]  ^{2}}\\
\\
+\left(  \frac{4-3\nu^{2}+\sqrt{16-16\nu^{2}+9\nu^{4}}}{4\nu^{2}}+\frac{3}%
{2}\right)  \frac{r_{z}^{2}}{\left[  \lambda_{z}^{\left(  0\right)  }\right]
^{2}}\\
\\
-\left(  \frac{4-3\nu^{2}+\sqrt{16-16\nu^{2}+9\nu^{4}}}{4\nu^{2}}+\frac{1}%
{2}\right)
\end{array}
\right] \nonumber
\end{align}

The other eigenmode of the doublet is characterized by:%

\begin{align}
\Omega_{-\ \ }^{\left(  0\right)  }  &  =\omega_{\perp}\left(  \frac
{4+3\nu^{2}-\sqrt{16-16\nu^{2}+9\nu^{4}}}{2}\right)  ^{\frac{1}{2}}\nonumber\\
& \label{cylindrical trap eps_D=0 collective mode d_(z^2)-wave}\\
\left[
\begin{array}
[c]{c}%
\widehat{\rho}_{xx}\left(  \Omega_{-\ \ }^{\left(  0\right)  }\right) \\
\widehat{\rho}_{yy}\left(  \Omega_{-\ \ }^{\left(  0\right)  }\right) \\
\widehat{\rho}_{zz}\left(  \Omega_{-\ \ }^{\left(  0\right)  }\right)
\end{array}
\right]   &  =\left[
\begin{array}
[c]{c}%
\frac{4-3\nu^{2}-\sqrt{16-16\nu^{2}+9\nu^{4}}}{4\nu^{2}}\\
\\
\frac{4-3\nu^{2}-\sqrt{16-16\nu^{2}+9\nu^{4}}}{4\nu^{2}}\\
\\
1
\end{array}
\right] \nonumber
\end{align}
For $\nu\rightarrow\infty$ this mode behaves asymptotically like
\begin{align*}
\Omega_{-\ \ }^{\left(  0\right)  }  &  =\sqrt{\frac{10}{3}}\omega_{\perp
}\left(  1-\frac{1}{9\nu^{2}}+...\right) \\
& \\
\left[
\begin{array}
[c]{c}%
\widehat{\rho}_{xx}\left(  \Omega_{-\ \ }^{\left(  0\right)  }\right) \\
\widehat{\rho}_{yy}\left(  \Omega_{-\ \ }^{\left(  0\right)  }\right) \\
\widehat{\rho}_{zz}\left(  \Omega_{-\ \ }^{\left(  0\right)  }\right)
\end{array}
\right]   &  =\left[
\begin{array}
[c]{c}%
-\frac{3}{2}+\frac{5}{3\nu^{2}}+...\\
\\
-\frac{3}{2}+\frac{5}{3\nu^{2}}+...\\
\\
1
\end{array}
\right]  \
\end{align*}
For $\nu\rightarrow0\ $ we find
\begin{align*}
\Omega_{-\ \ }^{\left(  0\right)  }  &  =\sqrt{\frac{5}{2}}\omega_{z}\ \left(
1-\frac{\nu^{2}}{16}+...\right) \\
& \\
\left[
\begin{array}
[c]{c}%
\widehat{\rho}_{xx}\left(  \Omega_{-\ \ }^{\left(  0\right)  }\right) \\
\widehat{\rho}_{yy}\left(  \Omega_{-\ \ }^{\left(  0\right)  }\right) \\
\widehat{\rho}_{zz}\left(  \Omega_{-\ \ }^{\left(  0\right)  }\right)
\end{array}
\right]   &  =\left[
\begin{array}
[c]{c}%
-\frac{1}{4}\ -\frac{5}{32}\nu^{2}+...\\
\\
-\frac{1}{4}\ -\frac{5}{32}\nu^{2}+...\\
\\
1
\end{array}
\right]  \
\end{align*}
For an anisotropy ratio $\nu$ $\simeq1$ (slightly deformed sphere) this mode
describes a density fluctuation with a strongly weighted $d_{z^{2}}$- wave
part and only a small admixture of isotropic $s$-wave symmetry:%
\begin{align}
\Omega &  =\Omega_{-\ \ }^{\left(  0\right)  }%
\label{cylindrical trap eps_D=0 density fluctuation d_z^2 -wave}\\
\delta n_{\Omega}\left(  \mathbf{r},t\right)   &  =2n_{0}\cos\left(  \Omega
t+\delta_{\Omega}\right)  \left[
\begin{array}
[c]{c}%
\left(  \frac{4-3\nu^{2}-\sqrt{16-16\nu^{2}+9\nu^{4}}}{2\nu^{2}}+\frac{1}%
{2}\right)  \frac{r_{x}^{2}+r_{y}^{2}}{\left[  \lambda_{\perp}^{\left(
0\right)  }\right]  ^{2}}\\
\\
+\left(  \frac{4-3\nu^{2}-\sqrt{16-16\nu^{2}+9\nu^{4}}}{4\nu^{2}}+\frac{3}%
{2}\right)  \frac{r_{z}^{2}}{\lambda_{z}^{2}}\\
\\
-\left(  \frac{4-3\nu^{2}-\sqrt{16-16\nu^{2}+9\nu^{4}}}{4\nu^{2}}+\frac{1}%
{2}\right)
\end{array}
\right] \nonumber
\end{align}
The derived frequencies for the coupled monopole-quadrupole oscillations of a
BEC without dipole-dipole interaction, i.e. $\varepsilon_{D}=0$, that is
confined inside a harmonic trap with uniaxial (cylindrical) symmetry, coincide
with well known results first derived by Stringari~\cite{Stringari II} using a
different method, that enabled him also to derive all the higher lying frequencies.

\end{document}